\documentstyle[thmsa,12pt,sw20lart]{article}

\input tcilatex
\QQQ{Language}{
British English
}

\begin{document}

\section{ Introduction}

In several papers[1-5], we have developed the Land\'e interpretation of
quantum mechanics[6-9] to devise new methods of treating spin systems. By
these methods, we have not only derived the matrix treatment of spin from
probability amplitudes but we have also obtained new generalized forms of
spin operators, their eigenvectors, and of spin states. These results have
been for isolated spin systems. In this paper, we extend the formalism to
the case of systems of compounded angular momentum. While addition of
angular momentum might seem to be adequately treated by standard methods, we
show that new results and insights are the reward of applying the new
methods to this subject.

Our task in this paper is to derive explicit joint probability amplitudes
for measurements of spin projections of subsystems compounded to give new
systems, and then to use them to derive the matrix treatment of the
compounded systems. We here present a new method for calculating the
probability amplitudes and we present results which are more generalized
than any we have seen to date for these probability amplitudes. The theory
is applied to the cases of total spin $0$ and $1$ resulting from the
addition of two spins of spin $1/2$ each. For spin $0$, we obtain one matrix
representation of the system. For spin $1$ however, there are two matrix
representations.

Our considerations lead us to the conclusion that the Clebsch-Gordan
coefficients can be generalized. We give these new generalized forms for the
particular Clebsch-Gordan coefficients appearing in the cases treated here.

As an example of an application, the results we obtain for ${\bf S}=0$ and $%
{\bf S}=1$ are used to investigate joint measurements of the kind used to
study entangled systems. We confirm standard results and find a reason for
the correlations in the results for the ${\bf S}=0$ state.

This paper is organized according to the following plan. Section $2$, which
follows, is an exposition of basic theory. We there give the main equation
from the work of Land\'e which our considerations are based upon. In Section 
$2.2$, we explain the basic features of our approach. In Section $3$, this
approach is applied to the general problem of adding two angular momenta.
After explaining the notation in Section $3.1$, we give in Section $3.2$ the
basic formulas for the matrix treatment, as derived from probability
amplitudes for the vectors and the operators.

In Section $4$, we pose some of the questions which measurements on a
compounded-spin system are designed to answer. We begin our answer of these
questions in Section $5$. After explaining all the possibilities in Section $%
5.1$, we give general formulas for probability amplitudes in Section $5.2$.
Section $5.3$ lists the results of measurements on two uncoupled spin-$1/2$
systems. These are combined into the results of joint measurements on such
systems in Section $5.4$.

We derive the probability amplitudes for measurements on the singlet state
in Section $5.5$. Section $5.6$ gives the probability amplitudes for the
triplet state. Section $6$ is devoted to the calculation of probabilities:
thus, Section $6.1$ contains the singlet-state probability amplitudes, and
Section $6.2$ the triplet-state probability amplitudes.

According to our treatment, the Clebsch-Gordan coefficients can be
generalized. This is discussed in Section $7$, and the generalized
Clebsch-Gordan coefficients for both singlet and triplet states are given.

In Section $8$, we shift to matrix mechanics. We derive the matrix-mechanics
treatment of the singlet state in Section $8.2$ and of the triplet state in
Section $8.3$.

Section $9$ applies the new results to the calculation of the expectation
values for joint measurements on the singlet states and triplet states.
Section $10$ presents the Discussion and Conclusion, which closes the paper.

\section{Theory}

\subsection{Preliminary Results}

We begin by reminding ourselves of some basic features of the Land\'e
approach [6-9]. Let a quantum system have the observables $A$, $B$ and $C$.
The eigenvalues of $A$ are $A_1$, $A_2,$....; the eigenvalues of $B$ are $%
B_1 $,$B_2$, ..; and the eigenvalues of $C$ are $C_1$, $C_2$,... If the
system is in the state corresponding to the eigenvalue $A_i$ of $A$,
measurement of $C$ yields any of the eigenvalues $C_1$,$C_2,...$ with
probability amplitudes $\psi (A_i;C_n)$. Measurement of $B$ yields any of
the eigenvalues $B_1$, $B_2$,.. with probability amplitudes $\chi (A_i;B_n)$%
. Finally, measurement of $C$ when the system is in the state corresponding
to the eigenvalue $B_i$ of $B$ yields eigenvalues of $C$ with probability
amplitudes $\phi (B_i;C_n).$ Then the probability amplitudes are connected
by[6-9]

\begin{equation}
\psi (A_i;C_n)=\dsum_q\chi (A_i;B_q)\phi (B_q;C_n),  \label{one1}
\end{equation}
which we shall call the Land\'e formula. The probability amplitudes satisfy
the Hermiticity condition

\begin{equation}
\psi (A_i;C_n)=\psi ^{*}(C_n;A_i).  \label{two2}
\end{equation}

The expansions for the $\phi $'s and $\chi $'s are [1] 
\begin{equation}
\phi (B_l;C_k)=\dsum\limits_i\chi (B_l;A_i)\psi (A_i;C_k)  \label{three3}
\end{equation}
and 
\begin{equation}
\chi (A_i;B_m)=\dsum\limits_k\psi (A_i;C_k)\phi (C_k;B_m).  \label{four4}
\end{equation}

\subsection{Further Results}

The basis of our treatment is the expansion Eq. (\ref{one1}) and the
interpretation of the wave function due to Land\'e. According to Land\'e,
the wave function or the eigenfunction is to be interpreted as a probability
amplitude which connects well-defined initial and final states. Thus the
solution $u_E({\bf r})$ of the time-independent Schr\"odinger equation for a
particular system is a probability amplitude that connects the initial state
defined by the energy eigenvalue $E$ to the final state corresponding to the
position eigenvalue ${\bf r}$. Therefore $\left| u_E({\bf r)}\right| ^2$ $d%
{\bf r}$ is the probability that if the system is initially in the state
corresponding to the energy $E$, a measurement of its position yields the
value ${\bf r}$ in the volume element $d{\bf r}$. By the same token, the
spherical harmonic $Y_{lm}(\theta ,\varphi )$ is a probability amplitude
connecting an initial state defined by the quantum numbers $l$ and $m$ to a
final state characterized by the eigenvalues $(\theta ,\varphi )$. Hence, $%
\left| Y_{lm}(\theta ,\varphi )\right| ^2d\Omega $ is the probability that
if the square of the angular momentum is initially $l(l+1)\hbar ^2$ and its $%
z$ component is $m\hbar $, a measurement of the angular position gives $%
(\theta ,\varphi )$ in $d\Omega .$

This approach can be very fruitfully applied to the treatment of spin, as we
have shown[1-5]. We have utilized it to develop a way to derive generalized
probability amplitudes, operators and vector states for spin and have
illustrated the method by applying it to the cases of spin $1/2$[1,2,4,5]
and spin $1$ [3]. We shall denote the spin probability amplitudes by the
functions $\chi (m^{(\widehat{{\bf a}})};m^{(\widehat{{\bf b}})}).$ Here $%
\widehat{{\bf a}}$ is a direction vector with respect to which the spin
projection is initially known and $\widehat{{\bf b}}$ is the direction
vector with respect to which we subsequently measure it. Hence, $m^{(%
\widehat{{\bf a}})}$, for example, represents the projection $m^{(\widehat{%
{\bf a}})}\hbar $ in the direction $\widehat{{\bf a}}$. Thus $\left| \chi
((+\frac 12)^{(\widehat{{\bf a}})};(+\frac 12)^{(\widehat{{\bf b}})})\right|
^2$ is the probability that if the spin projection is initially up with
respect to the unit vector $\widehat{{\bf a}}$, a new measurement finds it
up with respect to the vector $\widehat{{\bf b}}$. Similarly, $\left| \chi
((-\frac 12)^{(\widehat{{\bf a}})};(+\frac 12)^{(\widehat{{\bf b}})})\right|
^2$ is the probability of finding the spin projection to be up with respect
to $\widehat{{\bf b}}$ if it was initially down with respect to $\widehat{%
{\bf a}}$. For spin $1/2$, there are two other probability amplitudes $\chi
((+\frac 12)^{(\widehat{{\bf a}})};(-\frac 12)^{(\widehat{{\bf b}})})$ and $%
\chi ((-\frac 12)^{(\widehat{{\bf a}})};(-\frac 12)^{(\widehat{{\bf b}})}$ $%
),$ whose interpretation is obvious.

By expanding the probability amplitudes using Eq. (\ref{one1}), we obtain
the matrix treatment of spin. We have demonstrated this for spin $1/2$ and
spin $1$. Thus, by means of this expansion, the treatment of spin is made
analogous to that of orbital angular momentum, since the matrix treatment of
orbital angular momentum is achieved with the aid of an expansion in terms
of the spherical harmonics. The main difference between the two cases
derives from the fact that spin is distinguished from most other dynamical
variables by the circumstance that the eigenvalues corresponding to both
initial and final states are discrete. In contrast, orbital angular momentum
as described by the spherical harmonics is characterized by a discrete
initial eigenvalue spectrum and a continuous final spectrum.

\section{General Addition of Angular Momentum}

\subsection{Re-interpretation and Notation}

We now wish to generalize the new method of deriving probability amplitudes
to systems compounded of other systems. We are specifically interested in
the singlet and triplet states obtained by adding two spins of value $1/2$.
We want to derive the probability amplitudes pertaining to measurements on
such systems and we want to show that the matrix treatment of such systems
can be derived from first principles. To this end, we first look at the
general problem of obtaining a matrix treatment of a compounded
angular-momentum system.

Our approach to the problem is based on the re-interpretation of standard
quantities and equations in terms of the Land\'e approach. Let the angular
momenta of two subsystems $1$ and $2$ be ${\bf J}_1$ and ${\bf J}_2.$ Let
these angular momenta be added to give the total angular momentum ${\bf J}$.
Let the simultaneous eigenfunction of ${\bf J}^2$, ${\bf J}_z,{\bf J}_1^2$
and ${\bf J}_2^2$ be $\Psi _{jMj_1j_2}(\theta _1,\varphi _1,\theta
_2,\varphi _2).$ Here $(\theta _1,\varphi _1)$ are the angular variables
pertaining to ${\bf J}_1$, while $(\theta _2,\varphi _2) $ are the variables
for ${\bf J}_2$. In the spirit of the Land\'e interpretation of quantum
mechanics, this function is a probability amplitude which connects an
initial state characterized by the eigenvalues $j(j+1)\hbar ^2,$ $M\hbar ,$ $%
j_1(j_1+1)\hbar ^2$and $j_2(j_2+1)\hbar ^2$ with a final state characterized
by the eigenvalues $\theta _1,\varphi _{1,}\theta _2$ and $\varphi _2.$ For
this reason, $\left| \Psi _{jMj_1j_2}(\theta _1,\varphi _1,\theta _2,\varphi
_2)\right| ^2$ $d\Omega _1d\Omega _2$ is the probability that starting from
the initial state characterised by the eigenvalues $j(j+1)\hbar ^2,$ $M\hbar
,$ $j_1(j_1+1)\hbar ^2$and $j_2(j_2+1)\hbar ^2$, a measurement of angular
position gives $(\theta _1,\varphi _1,\theta _2,\varphi _2)$ in $d\Omega
_1d\Omega _2.$

Let the eigenfunctions for isolated subsystem $1$ be $\phi
_{j_1m_1}^{(1)}(\theta _1,\varphi _1)$ and those for isolated subsystem $2$
be $\phi _{j_2m_2}^{(2)}(\theta _2,\varphi _2).$ According to the new
formalism, the wave function for subsystem $1$ gives the probability
amplitude that if the initial eigenstate of the system has the quantum
numbers $j_1$ and $m_1$ a measurement of angular position gives $(\theta
_1,\varphi _1).$ A corresponding interpretation holds for subsystem 2. To
bring out this interpretation, we rewrite these probability amplitudes as 
\begin{equation}
\phi _1(j_1,m_1;\theta _1,\varphi _1)=\phi _{j_1m_1}^{(1)}(\theta _1,\varphi
_1)  \label{five5a}
\end{equation}
and

\begin{equation}
\phi _2(j_2,m_2;\theta _2,\varphi _2)=\phi _{j_2m_2}^{(2)}(\theta _2,\varphi
_2).  \label{five5b}
\end{equation}
We denote their product by 
\begin{equation}
\Phi (j_1,m_1,j_2,m_2;\theta _1,\varphi _1,\theta _2,\varphi _2)=\phi
_1(j_1,m_1;\theta _1,\varphi _1)\phi _2(j_2,m_2;\theta _2,\varphi _2).
\label{six6}
\end{equation}

The eigenfunction for the state obtained by adding the two angular momenta $%
{\bf J}_1$ and ${\bf J}_2$ is[10]

\begin{equation}
\Psi _{jMj_1j_2}(\theta _1,\varphi _1,\theta _2,\varphi
_2)=\dsum\limits_{m_1}C(j_1j_2j;m_1m_2M)\phi _{j_1m_1}^{(1)}(\theta
_1,\varphi _1)\phi _{j_2m_2}^{(2)}(\theta _2,\varphi _2),  \label{seven7}
\end{equation}
where we have used the notation in Rose[10] for the Clebsch-Gordan
coefficients $C(j_1j_2j;m_1m_2M).$ In order to interpret this eigenfunction
in terms of the Land\'e approach, we have to change the notation
appropriately. Hence we rewrite Eq. (\ref{seven7}) as

\begin{eqnarray}
&&\Psi (j,M,j_1,j_2;\theta _1,\varphi _1,\theta _2,\varphi
_2)=\dsum\limits_{m_1}C(j_1j_2j;m_1m_2M)  \nonumber \\
&&\times \phi _1(j_1,m_1;\theta _1,\varphi _1)\phi _2(j_2,m_2;\theta
_2,\varphi _2).  \label{eight8}
\end{eqnarray}
This eigenfunction is the probability amplitude that if the initial state is
characterized by the quantum numbers $(j,M,j_1,j_2),$ a measurement of
angular position gives $(\theta _1,\varphi _1,\theta _2,\varphi _2).$

Now if we compare the basic equation Eq. (\ref{one1}) and the expansion Eq. (%
\ref{eight8}), the Clebsch-Gordan coefficients are immediately seen to be
probability amplitudes. Thus $C(j_1j_2j;m_1m_2M)$ is the probability
amplitude that if the compound system is in a state corresponding to the
quantum numbers $(j,M,j_1,j_2)$, a measurement of the $z$ components of the
spins of systems $1$ and $2$ gives $m_1\hbar $ and $m_2\hbar $ respectively,
while a measurement of the squares of the spins gives $j_1(j_1+1)\hbar ^2$
and $j_2(j_2+1)\hbar ^2$ respectively. To emphasize the fact that the
Clebsch-Gordan coefficients are just probability amplitudes, and in order to
cast Eq. (\ref{eight8}) in terms of the fundamental expansion Eq. (\ref{one1}%
), we set 
\begin{equation}
\eta (j_1,j_2,j,M;j_1,j_2,m_1,m_2)=C(j_1j_2j;m_1m_2M).  \label{nine9}
\end{equation}
Thus, with the aid of Eq. (\ref{six6}) we express Eq. (\ref{eight8}) in the
form

\begin{eqnarray}
&&\Psi (j,j_1,j_2,M;\theta _1,\varphi _1,\theta _2,\varphi
_2)=\dsum\limits_{m_1}\eta (j_1,j_2,j,M;j_1,j_2,m_1,m_2)  \nonumber \\
&&\times \Phi (j_1,j_2,m_1,m_2;\theta _1,\varphi _1,\theta _2,\varphi _2).
\label{ten10}
\end{eqnarray}
We can suppress the indices $j_1$ and $j_2$ because for given subsystems 1
and 2, they are fixed. However, they may, of course, give different values
of $j$ within the range $\left| j_1-j_2\right| \leq j\leq j_1+j_2.$ In
addition, we can suppress the index $m_2$ because of the constraint $%
m_1+m_2=M$. We then have 
\begin{equation}
\eta (j,M;m_1)=\eta (j_1,j_2,j,M;j_1,j_2,m_1,m_2)  \label{ten10a}
\end{equation}
so that

\begin{equation}
\Psi (j,M;\theta _1,\varphi _1,\theta _2,\varphi _2)=\dsum\limits_{m_1}\eta
(j,M;m_1)\Phi (m_1;\theta _1,\varphi _1,\theta _2,\varphi _2).  \label{el11}
\end{equation}

We see immediately that the standard expression for the wave function of a
system of compounded angular momentum is of the form Eq. (\ref{one1}). This
justifies the interpretation we have given to the Clebsch-Gordan
coefficients.

In the expansion, $\Psi $ and the $\eta ^{\prime }s$ are not known, while
the $\Phi ^{\prime }s$ are known. If we are adding two orbital angular
momenta, $\Phi $ is the product of two spherical harmonics. When we are
adding spins, $\Phi $ is the product of two spin probability amplitudes. If
orbital angular momentum and spin are being compounded, then $\Phi $ is the
product of a spherical harmonic and a spin probability amplitude.

\subsection{General Expressions For Operators And Vectors}

In this section, we show how to obtain the matrix treatment of added angular
momentum. Thus, we obtain the general formulas for the vector states and
operators. We achieve this by going through the definition of the
expectation value. Consider the case where we are measuring values of the
observable $R(\theta _1,\varphi _1,\theta _2,\varphi _2)$. The expectation
value of $R$ is

\begin{equation}
\left\langle R\right\rangle =\int \left| \Psi (j,M;\theta _1,\varphi
_1,\theta _2,\varphi _2)\right| ^2R(\theta _1,\varphi _1,\theta _2,\varphi
_2)d\Omega _1d\Omega _2.  \label{tw12}
\end{equation}

Using

\begin{equation}
\Psi (j,M;\theta _1,\varphi _1,\theta _2,\varphi _2)=\dsum\limits_{m_1}\eta
(j,M;m_1)\Phi (m_1;\theta _1,\varphi _1,\theta _2,\varphi _2)  \label{th13}
\end{equation}
and

\begin{equation}
\Psi ^{*}(j,M;\theta _1,\varphi _1,\theta _2,\varphi
_2)=\dsum\limits_{m_1^{\prime }}\eta ^{*}(j,M;m_1^{\prime })\Phi
^{*}(m_1^{\prime };\theta _1,\varphi _1,\theta _2,\varphi _2),  \label{fo14}
\end{equation}
we obtain

\begin{equation}
\left\langle R\right\rangle =\sum_{m_1}\sum_{m_1^{\prime }}\eta
^{*}(j,M;m_1^{\prime })R_{m_1^{\prime }m_1}\eta (j,M;m_1),  \label{fi15}
\end{equation}
where 
\begin{eqnarray}
R_{m_1^{\prime }m_1} &=&\int \Phi ^{*}(m_1^{\prime };\theta _1,\varphi
_1,\theta _2,\varphi _2)R(\theta _1,\varphi _1,\theta _2,\varphi _2) 
\nonumber \\
&&\times \Phi (m_1;\theta _1,\varphi _1,\theta _2,\varphi _2)d\Omega
_1d\Omega _2.  \label{si16}
\end{eqnarray}
Hence

\begin{equation}
\left\langle R\right\rangle =[\Psi ]^{\dagger }[R][\Psi ],  \label{se17}
\end{equation}
where

\begin{equation}
\lbrack \Psi ]=\left( 
\begin{array}{c}
\eta (j,M;(m_1)_1) \\ 
\eta (j,M;(m_1)_2) \\ 
.. \\ 
\eta (j,M;(m_1)_N)
\end{array}
\right)  \label{ei18}
\end{equation}
and 
\begin{equation}
\lbrack R]=\left( 
\begin{array}{cccc}
R_{11} & R_{22} & ... & R_{1N} \\ 
R_{21} & R_{22} & ... & R_{2N} \\ 
... & ... & ... & ... \\ 
R_{N1} & R_{N2} & ... & R_{NN}
\end{array}
\right) .  \label{ni19}
\end{equation}
Here $N$ is the number of combinations of $m_1$ and $m_2$ such that $%
m_1+m_2=M$. We note also that $(m_1)_i$ are individual values of $m_1,$
given labels from $1$ to $N$.

We see that the matrix representation of the state is a row vector whose
elements are the Clebsch-Gordan coefficients.

Actually, the matrix representation presented here is not the only one
possible. Others can be realized. They arise if we seek the generalized form
of the probability amplitude Eq. (\ref{el11}).

The standard treatment of angular momentum addition assumes that the initial
direction of projection of the total spin is the $z$ direction. But in the
generalized treatment we shall give here, this direction is arbitrary. In
that case, Eq. (\ref{el11}) is replaced by 
\begin{equation}
\Psi (j,M^{(\widehat{{\bf a}})};\theta _1,\varphi _1,\theta _2,\varphi
_2)=\dsum\limits_{m_1,m_2}\eta (j,M^{(\widehat{{\bf a}})};m_1,m_2)\Phi
(m_1,m_2;\theta _1,\varphi _1,\theta _2,\varphi _2),  \label{ni19d}
\end{equation}
where the generalized probability amplitude $\Psi (j,M^{(\widehat{{\bf a}}%
)};\theta _1,\varphi _1,\theta _2,\varphi _2)$ is such that the initial
projection of the total spin is with respect to the direction $\widehat{{\bf %
a}}$. By the same token, $\eta (j,M^{(\widehat{{\bf a}})};m_1,m_2) $ now
contains the quantum number $M^{(\widehat{{\bf a}})}$, which refers to the
direction $\widehat{{\bf a}}$ instead of the $z$ direction. This quantity is
no longer a Clebsch-Gordan coefficient, but as we shall see later, it can be
expressed in terms of the Clebsch-Gordan coefficients, by means of the
expansion Eq. (\ref{two2}). But this new expansion can be used to realize a
different matrix representation. We observe that since $\eta (j,M^{(\widehat{%
{\bf a}})};m_1,m_2)$ is no longer a Clebsch-Gordan coefficient, the
condition $m_1+m_2=M$ no longer necessarily holds, so that in principle the
summation in Eq. (\ref{ni19d}) is over both $m_1$ and $m_2$.

\section{Measurements on Systems of Compounded Spin}

The theory we have outlined above is most easily applied to spin systems. We
consider the case of two such systems compounded to give one system. The
total spin of the compounded system is

\begin{equation}
{\bf S}={\bf S}_1+{\bf S}_2,  \label{tw20}
\end{equation}
where ${\bf S}_1$ is the spin of subsystem $1$ and ${\bf S}_2$ is the spin
of subsystem $2$. We may ask the following questions regarding these systems.

(a) Suppose we focus attention on one of the subsystems. If the spin
projection of that subsystem is initially known in a given direction, what
is the probability of obtaining a given value of the spin projection in a
new direction? The answer is easily obtained since the subsystem can be
treated as if isolated. When the subsystem is of spin $1/2$, or of spin $1$,
the probability amplitudes that answer this question for the general case
are as given in Refs. [1,2,4,5] and [3] respectively. If the spin is not $%
1/2 $ or $1$, we can derive the generalized probability amplitudes by the
method illustrated in these references. If we are satisfied with a less
general answer, we can use standard expressions [11] for the probability
amplitudes. As shown in Refs. [1-5], the standard expressions can be
obtained from the generalized expressions by setting the direction in which
we initially know the spin projection to be the $z$ direction.

(b) Suppose we focus attention on the compound system. If its spin
projection is initially known along a given direction, what is the
probability of obtaining a given value of the projection along a new
direction? The answer depends only on the total spin of the compound system.
It is not necessary to know what individual spins have been combined to give
the particular value of the compound spin. If the compound spin is $0$, then
of course there is only one projection and the question is trivial because
the required probability is unity. If the compound spin is $1/2$ or $1$, the
probability amplitudes are the ones derived in Refs. [1,2,4,5] and in [3]
respectively. If the compound spin is neither $1/2$ nor $1$, it is necessary
to use the method outlined in Refs. [1-5] to derive these probability
amplitudes. Again, if we are satisfied with less generality, we can use the
standard formulas for that value of spin.

(c) Finally, suppose we focus attention on the compound system. If the
projection of the total spin is initially known along the direction $%
\widehat{{\bf a}}$, what is the probability that a measurement of the spin
projection of subsystem $1$ along the vector $\widehat{{\bf c}}_1$ gives a
specified value while a simultaneous measurement of the spin projection of
subsystem $2$ along the direction $\widehat{{\bf c}}_2$ gives a certain
value? This is the question we address here. We show that the probability
amplitude represented by the functions for the compound spin are correspond
to precisely this measurement.

\section{Probability Amplitudes}

\subsection{Possible Results of Measurements on Subsystems}

The compounded system has states characterized by simultaneous eigenvalues
of $S^2$, $S_1^2$, $S_2^2$ and $S_z$. Since $s_1$ and $s_2$ are fixed, we
omit them in the labelling of the states. The basic quantity we wish to
obtain is the probability amplitude for obtaining a specified value of $%
S_{1z}$ along the vector $\widehat{{\bf c}}_1$ and a specified value of $%
S_{2z}$ along the vector $\widehat{{\bf c}}_2$ starting from a state
characterised by the specified value $S_z$ along the vector $\widehat{{\bf a}%
}.$ We shall make our investigation concrete by considering the case $%
s_1=s_2=1/2$. The reason for this is that this is the simplest non-trivial
case to which we may apply the new theory we shall develop here.

For $s_1=s_2=1/2$, there are only two possible values of the spin projection
for each subsystem. If the spin projection is found up (down) with respect
to $\widehat{{\bf c}}_1$ or $\widehat{{\bf c}}_2$, we label this outcome by $%
(+\frac 12)^{(\widehat{{\bf c}}_1)}$ $((-\frac 12)^{(\widehat{{\bf c}}_1)})$
or $(+\frac 12)^{(\widehat{{\bf c}}_2)}$ $((-\frac 12)^{(\widehat{{\bf c}}%
_2)}),$ respectively. Thus the possible combinations of the results of the
measurements are as follows:

(1 ) spin $1$ up with respect to $\widehat{{\bf c}}_1$ and spin $2$ up with
respect to $\widehat{{\bf c}}_2$ (denoted by $((+\frac 12)^{(\widehat{{\bf c}%
}_1)},(+\frac 12)^{(\widehat{{\bf c}}_2)})$);

(2) spin $1$ up with respect to $\widehat{{\bf c}}_1$ and spin $2$ down with
respect to $\widehat{{\bf c}}_2$ (denoted by $((+\frac 12)^{(\widehat{{\bf c}%
}_1)},(-\frac 12)^{(\widehat{{\bf c}}_2)})$);

(3) spin $1$ down with respect to $\widehat{{\bf c}}_1$ and spin $2$ up with
respect to $\widehat{{\bf c}}_2$ (denoted by $((-\frac 12)^{(\widehat{{\bf c}%
}_1)},(+\frac 12)^{(\widehat{{\bf c}}_2)})$);

(4) spin $1$ down with respect to $\widehat{{\bf c}}_1$ and spin $2$ down
with respect to $\widehat{{\bf c}}_2$ (denoted by $((-\frac 12)^{(\widehat{%
{\bf c}}_1)},(-\frac 12)^{(\widehat{{\bf c}}_2)})$).

We shall denote a particular such outcome by $((m_1)_u^{(\widehat{{\bf c}}%
_1)},(m_2)_v^{(\widehat{{\bf c}}_2)})$, with $u,v=1,2$. Thus, 
\begin{equation}
((m_1)_1^{(\widehat{{\bf c}}_1)},(m_2)_1^{(\widehat{{\bf c}}_2)})=((+\tfrac
12)^{(\widehat{{\bf c}}_1)},(+\tfrac 12)^{(\widehat{{\bf c}}_2)}),
\label{tw20a}
\end{equation}

\begin{equation}
((m_1)_1^{(\widehat{{\bf c}}_1)},(m_2)_2^{(\widehat{{\bf c}}_2)})=((+\tfrac
12)^{(\widehat{{\bf c}}_1)},(-\tfrac 12)^{(\widehat{{\bf c}}_2)}),
\label{tw20b}
\end{equation}
\begin{equation}
((m_1)_2^{(\widehat{{\bf c}}_1)},(m_2)_1^{(\widehat{{\bf c}}_2)})=((-\tfrac
12)^{(\widehat{{\bf c}}_1)},(+\tfrac 12)^{(\widehat{{\bf c}}_2)})
\label{tw20c}
\end{equation}
and

\begin{equation}
((m_1)_2^{(\widehat{{\bf c}}_1)},(m_2)_2^{(\widehat{{\bf c}}_2)})=((-\tfrac
12)^{(\widehat{{\bf c}}_1)},(-\tfrac 12)^{(\widehat{{\bf c}}_2)}).
\label{tw20d}
\end{equation}

Corresponding to each initial state characterised by the quantum number $%
M_i^{(\widehat{{\bf a}})}$ for the spin projection along $\widehat{{\bf a}}$
of the compound system, there are four possible probability amplitudes.
These possibilities are denoted by $\Psi (s,M_i^{(\widehat{{\bf a}}%
)};(+\tfrac 12)^{(\widehat{{\bf c}}_1)},(+\tfrac 12)^{(\widehat{{\bf c}}%
_2)}),$ $\Psi (s,M_i^{(\widehat{{\bf a}})};(+\tfrac 12)^{(\widehat{{\bf c}}%
_1)},(-\tfrac 12)^{(\widehat{{\bf c}}_2)})$, $\Psi (s,M_i^{(\widehat{{\bf a}}%
)};(-\tfrac 12)^{(\widehat{{\bf c}}_1)},(+\tfrac 12)^{(\widehat{{\bf c}}%
_2)}) $ and $\Psi (s,M_i^{(\widehat{{\bf a}})};(-\tfrac 12)^{(\widehat{{\bf c%
}}_1)},(-\tfrac 12)^{(\widehat{{\bf c}}_2)}).$ When we mean to denote these
probability amplitudes in a general way, we shall use the shorthand $\Psi
(s,M_i^{(\widehat{{\bf a}})};(m_1)_u^{(\widehat{{\bf c}}_1)},(m_2)_v^{(%
\widehat{{\bf c}}_2)})$.

\subsection{General Formulas For Probability Amplitudes}

We now seek the explicit forms of these probability amplitudes. The basic
method is to use the Land\'e expansion, Eq. (\ref{one1}), to express the
required probability amplitudes in terms of known probability amplitudes.
Consider the probability amplitude $\Psi (s,M_i^{(\widehat{{\bf a}}%
)};(m_1)_u^{(\widehat{{\bf c}}_1)},(m_2)_v^{(\widehat{{\bf c}}_2)}).$
Referring to Eq. (\ref{one1}), and assuming that the total spin $s$ is
fixed, we first identify the observable $A$ as the spin projection $S_z$
along $\widehat{{\bf a}}$. The observable $C$ corresponds to the
combinations of final spin projections of the subsystems. For the
intermediate states belonging to the observable $B$, in terms of which we
expand $\Psi (s,M_i^{(\widehat{{\bf a}})};(m_1)_u^{(\widehat{{\bf c}}%
_1)},(m_2)_v^{(\widehat{{\bf c}}_2)})$, we take the set of states resulting
from measuring components of the compound spin ${\bf S}$ along the $z$ axis.
The reason for this choice of intermediate states will become clear later.
Thus, we have

\begin{eqnarray}
&&\Psi (s,M_i^{(\widehat{{\bf a}})};(m_1)_u^{(\widehat{{\bf c}}%
_1)},(m_2)_v^{(\widehat{{\bf c}}_2)})=\sum_j\chi (s,M_i^{(\widehat{{\bf a}}%
)};s,M_j^{(\widehat{{\bf k}})})  \nonumber \\
&&\times \xi (s,M_j^{(\widehat{{\bf k}})};(m_1)_u^{(\widehat{{\bf c}}%
_1)},(m_2)_v^{(\widehat{{\bf c}}_2)}),  \label{tw21}
\end{eqnarray}
where $\chi (s,M_i^{(\widehat{{\bf a}})};s,M_j^{(\widehat{{\bf k}})})$ is
the probability amplitude for finding that the spin projection along the $z$
direction $\widehat{{\bf k}}$ is $M_j^{(\widehat{{\bf k}})}\hbar $ when
initially, the spin projection is $M_i^{(\widehat{{\bf a}})}\hbar $ along
the direction $\widehat{{\bf a}}.$ Since $s_1=s_2=1/2$, it follows that $%
s=0,1$. For this reason the probability amplitudes $\chi $ belong to $s=0$
or $s=1$ and are therefore known. If $s=0$, then there is only one value $%
M^{(\widehat{{\bf a}})}=0$; hence $\chi (s,0^{(\widehat{{\bf a}})};s,0^{(%
\widehat{{\bf k}})})=e^{i\delta }$, where $\delta $ is a real number which
we shall set equal to zero. For $s=1$, the probability amplitudes $\chi $
are given in Ref. [3].

The function $\xi (s,M_j^{(\widehat{{\bf k}})};(m_1)_u^{(\widehat{{\bf c}}%
_1)},(m_2)_v^{(\widehat{{\bf c}}_2)})$ is the probability amplitude that if
the state of the compound system is initially characterized by the
eigenvalue $M_j^{(\widehat{{\bf k}})}\hbar $, a measurement of the spin
projections of subsystems $1$ and $2$ yields the projections $(m_1)_u\hbar $%
and $(m_2)_v\hbar $ with respect to the direction vectors $\widehat{{\bf c}}%
_1$ and $\widehat{{\bf c}}_2$ respectively. However, the functions $\xi $
are not known. But they can be obtained by using the Land\'e expansion once
more. Thus, we set

\begin{eqnarray}
&&\xi (s,M_j^{(\widehat{{\bf k}})};(m_1)_u^{(\widehat{{\bf c}}_1)},(m_2)_v^{(%
\widehat{{\bf c}}_2)})=\sum_{\alpha ,\alpha ^{\prime }}\eta (s,M_j^{(%
\widehat{{\bf k}})};(m_1)_\alpha ^{(\widehat{{\bf k}})},(m_2)_{\alpha
^{\prime }}^{(\widehat{{\bf k}})})  \nonumber \\
&&\times \psi ((m_1)_\alpha ^{(\widehat{{\bf k}})},(m_2)_{\alpha ^{\prime
}}^{(\widehat{{\bf k}})};(m_1)_u^{(\widehat{{\bf c}}_1)},(m_2)_v^{(\widehat{%
{\bf c}}_2)}).  \label{tw22}
\end{eqnarray}

In this expansion, the values $((m_1)_\alpha ^{(\widehat{{\bf k}}%
)},(m_2)_{\alpha ^{\prime }}^{(\widehat{{\bf k}})})$ are given by Eqs. (\ref
{tw20a}) - (\ref{tw20d}) with $\widehat{{\bf c}}_1=\widehat{{\bf c}}_2=%
\widehat{{\bf k}}.$ Thus for example, $\eta (s,M_j^{(\widehat{{\bf k}}%
)};(+\tfrac 12)^{(\widehat{{\bf k}})},(+\tfrac 12)^{(\widehat{{\bf k}})})$
is the probability amplitude for the spin projection of subsystem $1$ to be
found up with respect to the $z$ direction and for that of subsystem $2 $ to
be found up with respect to $z$ direction upon measurement, when the initial
compound state is characterized by the spin projection being $M_j^{(\widehat{%
{\bf k}})}\hbar $ along the $z$ direction$.$ Thus, according to the
interpretation of Section $3.1$, the $\eta (s,M_j^{(\widehat{{\bf k}}%
)};(m_1)_\alpha ^{(\widehat{{\bf k}})},(m_2)_{\alpha ^{\prime }}^{(\widehat{%
{\bf k}})})$ must be Clebsch-Gordan coefficients. Hence, when systems $1$
and $2$ are spin-$1/2$ systems, we have

\begin{equation}
\eta (s,M_j^{(\widehat{{\bf k}})};(\pm \tfrac 12)^{(\widehat{{\bf k}})},(\pm
\tfrac 12)^{(\widehat{{\bf k}})})=C(\tfrac 12\tfrac 12s;\pm \tfrac 12,\pm
\tfrac 12,M_j^{(\widehat{{\bf k}})}).  \label{tw23}
\end{equation}

On the other hand, the function $\psi ((m_1)_\alpha ^{(\widehat{{\bf k}}%
)},(m_2)_{\alpha ^{\prime }}^{(\widehat{{\bf k}})};(m_1)_u^{(\widehat{{\bf c}%
}_1)},(m_2)_v^{(\widehat{{\bf c}}_2)})$ is the probability amplitude that if
the spin projection of subsystem $1$ is $(m_1)_\alpha \hbar $along the $z$
axis, and that of subsystem $2$ is $(m_2)_{\alpha ^{\prime }}\hbar $along
the $z$ axis, a measurement of the spin projection of subsystem $1$ along $%
\widehat{{\bf c}}_1$ finds $(m_1)_u\hbar $ while a measurement of the spin
projection of subsystem $2$ along $\widehat{{\bf c}}_2$ finds $(m_2)_v\hbar $%
. Thus, it follows that

\begin{eqnarray}
&&\psi ((m_1)_\alpha ^{(\widehat{{\bf k}})},(m_2)_{\alpha ^{\prime }}^{(%
\widehat{{\bf k}})};(m_1)_u^{(\widehat{{\bf c}}_1)},(m_2)_v^{(\widehat{{\bf c%
}}_2)})=\phi _1((m_1)_\alpha ^{(\widehat{{\bf k}})};(m_1)_u^{(\widehat{{\bf c%
}}_1)})  \nonumber \\
&&\times \phi _2((m_2)_{\alpha ^{\prime }}^{(\widehat{{\bf k}})};(m_2)_v^{(%
\widehat{{\bf c}}_2)}),  \label{tw24}
\end{eqnarray}
where the $\phi _1$ are the probability amplitudes for the measurement of
spin projections of subsystem $1$ from one direction to another. By the same
token, the $\phi _2$ are the probability amplitudes for the measurement of
spin projections of subsystem $2$ from one direction to another. The
probability amplitudes $\phi _1$ and $\phi _2$ come in a variety of forms,
depending on the choice of phase made when they are being derived[1,4].

\subsection{Results of Measurements on Isolated Spin-1/2 Systems}

The derivation of the probability amplitudes $\phi _1((m_1)_\alpha ^{(%
\widehat{{\bf k}})};(m_1)_u^{(\widehat{{\bf c}}_1)}$ or $\phi
_2((m_2)_{\alpha ^{\prime }}^{(\widehat{{\bf k}})};(m_2)_v^{(\widehat{{\bf c}%
}_2)})$ has already been done for two different choices of phase[1,4].
Consider either subsystem $1$ or $2$. Let the spin projection be initially
in the direction of the vector $\widehat{{\bf d}}$, whose polar angles are $%
(\theta ,\varphi ).$ We subsequently measure it in the direction of the
vector $\widehat{{\bf e}}$, whose polar angles are $(\theta ^{\prime
},\varphi ^{\prime })$. In Ref. [1], we found the following probability
amplitudes for the measurements: 
\begin{eqnarray}
\phi ((+\tfrac 12)^{(\widehat{{\bf d}})};(+\tfrac 12)^{(\widehat{{\bf e}})})
&=&\cos \theta /2\cos \theta ^{\prime }/2+e^{i(\varphi -\varphi ^{\prime
})}\sin \theta /2\sin \theta ^{\prime }/2,  \label{tw25} \\
&&\ \ .  \nonumber
\end{eqnarray}

\begin{equation}
\phi ((+\tfrac 12)^{(\widehat{{\bf d}})};(-\tfrac 12)^{(\widehat{{\bf e}}%
)})=\cos \theta /2\sin \theta ^{\prime }/2-e^{i(\varphi -\varphi ^{\prime
})}\sin \theta /2\cos \theta ^{\prime }/2,  \label{tw26}
\end{equation}

\begin{equation}
\phi ((-\tfrac 12)^{(\widehat{{\bf d}})};(+\tfrac 12)^{(\widehat{{\bf e}}%
)})=\sin \theta /2\cos \theta ^{\prime }/2-e^{i(\varphi -\varphi ^{\prime
})}\cos \theta /2\sin \theta ^{\prime }/2  \label{tw27}
\end{equation}
and

\begin{equation}
\phi ((-\tfrac 12)^{(\widehat{{\bf d}})};(-\tfrac 12)^{(\widehat{{\bf e}}%
)})=\sin \theta /2\sin \theta ^{\prime }/2+e^{i(\varphi -\varphi ^{\prime
})}\cos \theta /2\cos \theta ^{\prime }/2.  \label{tw28}
\end{equation}

It turns out however that these probability amplitudes will not serve here,
because they ultimately lead to incorrect results when employed. Since they
differ from other sets of probability amplitudes for the same measurements
only in phase, this means that in general we cannot use an arbitrary choice
of phase to develop our theory. That the probability amplitudes Eq. (\ref
{tw25}) - (\ref{tw28}) lead to the wrong results when used is deduced from
the following circumstance. The probability amplitudes Eq. (\ref{tw21}) must
reduce to the Clebsch-Gordan coefficients to within a phase factor if we set 
$\widehat{{\bf a}}=\widehat{{\bf c}}_1=\widehat{{\bf c}}_2=\widehat{{\bf k}}%
. $ (The argument leading to this requirement is given in Section $7$, where
the Clebsch-Gordan coefficients are discussed). When we use Eqs. (\ref{tw25}%
) - (\ref{tw28}) in the treatment that follows, we produce probability
amplitudes $\Psi (s,M_i^{(\widehat{{\bf a}})};(m_1)_u^{(\widehat{{\bf c}}%
_1)},(m_2)_v^{(\widehat{{\bf c}}_2)})$ that do not have this property. For
this reason, it is necessary to try other choices of phase until this
condition is satisfied. In fact the phase choice that gives the correct
results corresponds to the following probability amplitudes:

\begin{equation}
\phi ((+\tfrac 12)^{(\widehat{{\bf d}})};(+\tfrac 12)^{(\widehat{{\bf e}}%
)})=\cos \dfrac \theta 2\cos \dfrac{\theta ^{\prime }}2+e^{i(\varphi
-\varphi ^{\prime })}\sin \dfrac \theta 2\sin \dfrac{\theta ^{\prime }}2,
\label{tw29}
\end{equation}

\begin{equation}
\phi ((+\tfrac 12)^{(\widehat{{\bf d}})};(-\tfrac 12)^{(\widehat{{\bf e}}%
)})=-\cos \dfrac \theta 2\sin \dfrac{\theta ^{\prime }}2+e^{i(\varphi
-\varphi ^{\prime })}\sin \dfrac \theta 2\cos \dfrac{\theta ^{\prime }}2,
\label{th30}
\end{equation}

\begin{equation}
\phi ((-\tfrac 12)^{(\widehat{{\bf d}})};(+\tfrac 12)^{(\widehat{{\bf e}}%
)})=-\sin \dfrac \theta 2\cos \dfrac{\theta ^{\prime }}2+e^{i(\varphi
-\varphi ^{\prime })}\cos \dfrac \theta 2\sin \dfrac{\theta ^{\prime }}2
\label{th31}
\end{equation}
and 
\begin{equation}
\phi ((-\tfrac 12)^{(\widehat{{\bf d}})};(-\tfrac 12)^{(\widehat{{\bf e}}%
)})=\sin \dfrac \theta 2\sin \dfrac{\theta ^{\prime }}2+e^{i(\varphi
-\varphi ^{\prime })}\cos \dfrac \theta 2\cos \dfrac{\theta ^{\prime }}2.
\label{th32}
\end{equation}

Setting $\theta =\varphi =0$, so that $\widehat{{\bf d}}=\widehat{{\bf k}}$,
and letting $\theta ^{\prime }=\theta _1,$ $\varphi ^{\prime }=\varphi _1$
so that $\widehat{{\bf e}}=\widehat{{\bf c}}_1$ (whose polar angles are $%
(\theta _1,\varphi _1)$)$,$ we get

\begin{equation}
\phi _1((+\tfrac 12)^{(\widehat{{\bf k}})};(+\tfrac 12)^{(\widehat{{\bf c}}%
_1)})=\cos \dfrac{\theta _1}2,  \label{th33}
\end{equation}

\begin{equation}
\phi _1((+\tfrac 12)^{(\widehat{{\bf k}})};(-\tfrac 12)^{(\widehat{{\bf c}}%
_1)})=-\sin \dfrac{\theta _1}2,  \label{th34}
\end{equation}

\begin{equation}
\phi _1((-\tfrac 12)^{(\widehat{{\bf k}})};(+\tfrac 12)^{(\widehat{{\bf c}}%
_1)})=e^{-i\varphi _1}\sin \dfrac{\theta _1}2  \label{th35}
\end{equation}
and

\begin{equation}
\phi _1((-\tfrac 12)^{(\widehat{{\bf k}})};(-\tfrac 12)^{(\widehat{{\bf c}}%
_1)})=e^{-i\varphi _1}\cos \dfrac{\theta _1}2.  \label{th36}
\end{equation}
The same formulas apply for the $\phi _2$ except that the index $1$ is
everywhere replaced by the index $2$.

\subsection{Results of Joint Measurements on the Uncoupled Subsystems}

It is now a straightforward matter to obtain the $\psi ((m_1)_\alpha ^{(%
\widehat{{\bf k}})},(m_2)_{\alpha ^{\prime }}^{(\widehat{{\bf k}}%
)}:(m_1)_u^{(\widehat{{\bf c}}_1)},(m_2)_v^{(\widehat{{\bf c}}_2)}).$ Using
Eq. (\ref{tw24}), and Eqs. (\ref{th33}) - (\ref{th36}), we obtain

\begin{equation}
\psi ((+\tfrac 12)^{(\widehat{{\bf k}})},(+\tfrac 12)^{(\widehat{{\bf k}}%
)};(+\tfrac 12)^{(\widehat{{\bf c}}_1)},(+\tfrac 12)^{(\widehat{{\bf c}}%
_2)})=\cos \dfrac{\theta _1}2\cos \dfrac{\theta _2}2,  \label{th37}
\end{equation}

\begin{equation}
\psi ((+\tfrac 12)^{(\widehat{{\bf k}})},(+\tfrac 12)^{(\widehat{{\bf k}}%
)};(+\tfrac 12)^{(\widehat{{\bf c}}_1)},(-\tfrac 12)^{(\widehat{{\bf c}}%
_2)})=-\cos \dfrac{\theta _1}2\sin \dfrac{\theta _2}2,  \label{th38}
\end{equation}

\begin{equation}
\psi ((+\tfrac 12)^{(\widehat{{\bf k}})},(+\tfrac 12)^{(\widehat{{\bf k}}%
)};(-\tfrac 12)^{(\widehat{{\bf c}}_1)},(+\tfrac 12)^{(\widehat{{\bf c}}%
_2)})=-\sin \dfrac{\theta _1}2\cos \dfrac{\theta _2}2,  \label{th39}
\end{equation}

\begin{equation}
\psi ((+\tfrac 12)^{(\widehat{{\bf k}})},(+\tfrac 12)^{(\widehat{{\bf k}}%
)};(-\tfrac 12)^{(\widehat{{\bf c}}_1)},(-\tfrac 12)^{(\widehat{{\bf c}}%
_2)})=\sin \dfrac{\theta _1}2\sin \dfrac{\theta _2}2,  \label{fo40}
\end{equation}
\begin{equation}
\psi ((+\tfrac 12)^{(\widehat{{\bf k}})},(-\tfrac 12)^{(\widehat{{\bf k}}%
)};(+\tfrac 12)^{(\widehat{{\bf c}}_1)},(+\tfrac 12)^{(\widehat{{\bf c}}%
_2)})=\cos \dfrac{\theta _1}2\sin \dfrac{\theta _2}2e^{-i\varphi _2},
\label{fo41}
\end{equation}
\begin{equation}
\psi ((+\tfrac 12)^{(\widehat{{\bf k}})},(-\tfrac 12)^{(\widehat{{\bf k}}%
)};(+\tfrac 12)^{(\widehat{{\bf c}}_1)},(-\tfrac 12)^{(\widehat{{\bf c}}%
_2)})=\cos \dfrac{\theta _1}2\cos \dfrac{\theta _2}2e^{-i\varphi _2},
\label{fo41b}
\end{equation}
\begin{equation}
\psi ((+\tfrac 12)^{(\widehat{{\bf k}})},(-\tfrac 12)^{(\widehat{{\bf k}}%
)};(-\tfrac 12)^{(\widehat{{\bf c}}_1)},(+\tfrac 12)^{(\widehat{{\bf c}}%
_2)})=-\sin \dfrac{\theta _1}2\sin \dfrac{\theta _2}2e^{-i\varphi _2},
\label{fo42}
\end{equation}

\begin{equation}
\psi ((+\tfrac 12)^{(\widehat{{\bf k}})},(-\tfrac 12)^{(\widehat{{\bf k}}%
)};(-\tfrac 12)^{(\widehat{{\bf c}}_1)},(-\tfrac 12)^{(\widehat{{\bf c}}%
_2)})=-\sin \dfrac{\theta _1}2\cos \dfrac{\theta _2}2e^{-i\varphi _2},
\label{fo43}
\end{equation}

\begin{equation}
\psi ((-\tfrac 12)^{(\widehat{{\bf k}})},(+\tfrac 12)^{(\widehat{{\bf k}}%
)};(+\tfrac 12)^{(\widehat{{\bf c}}_1)},(+\tfrac 12)^{(\widehat{{\bf c}}%
_2)})=\sin \dfrac{\theta _1}2\cos \dfrac{\theta _2}2e^{-i\varphi _1},
\label{fo44}
\end{equation}

\begin{equation}
\psi ((-\tfrac 12)^{(\widehat{{\bf k}})},(+\tfrac 12)^{(\widehat{{\bf k}}%
)};(+\tfrac 12)^{(\widehat{{\bf c}}_1)},(-\tfrac 12)^{(\widehat{{\bf c}}%
_2)})=-\sin \dfrac{\theta _1}2\sin \dfrac{\theta _2}2e^{-i\varphi _1},
\label{fo45}
\end{equation}
\begin{equation}
\psi ((-\tfrac 12)^{(\widehat{{\bf k}})},(+\tfrac 12)^{(\widehat{{\bf k}}%
)};(-\tfrac 12)^{(\widehat{{\bf c}}_1)},(+\tfrac 12)^{(\widehat{{\bf c}}%
_2)})=\cos \dfrac{\theta _1}2\cos \dfrac{\theta _2}2e^{-i\varphi _1},
\label{fo46}
\end{equation}
\begin{equation}
\psi (-\tfrac 12)^{(\widehat{{\bf k}})},(+\tfrac 12)^{(\widehat{{\bf k}}%
)};(-\tfrac 12)^{(\widehat{{\bf c}}_1)},(-\tfrac 12)^{(\widehat{{\bf c}}%
_2)})=-\cos \dfrac{\theta _1}2\sin \dfrac{\theta _2}2e^{-i\varphi _1},
\label{fo49}
\end{equation}

\begin{equation}
\psi ((-\tfrac 12)^{(\widehat{{\bf k}})},(-\tfrac 12)^{(\widehat{{\bf k}}%
)};(+\tfrac 12)^{(\widehat{{\bf c}}_1)},(+\tfrac 12)^{(\widehat{{\bf c}}%
_2)})=\sin \dfrac{\theta _1}2\sin \dfrac{\theta _2}2e^{-i(\varphi _1+\varphi
_2)},  \label{fi50}
\end{equation}
\begin{equation}
\psi ((-\tfrac 12)^{(\widehat{{\bf k}})},(-\tfrac 12)^{(\widehat{{\bf k}}%
)};(+\tfrac 12)^{(\widehat{{\bf c}}_1)},(-\tfrac 12)^{(\widehat{{\bf c}}%
_2)})=\sin \dfrac{\theta _1}2\cos \dfrac{\theta _2}2e^{-i(\varphi _1+\varphi
_2)},  \label{fi51}
\end{equation}
\begin{equation}
\psi ((-\tfrac 12)^{(\widehat{{\bf k}})},(-\tfrac 12)^{(\widehat{{\bf k}}%
)};(-\tfrac 12)^{(\widehat{{\bf c}}_1)},(+\tfrac 12)^{(\widehat{{\bf c}}%
_2)})=\cos \dfrac{\theta _1}2\sin \dfrac{\theta _2}2e^{-i(\varphi _1+\varphi
_2)}  \label{fi52}
\end{equation}
and

\begin{equation}
\psi ((-\tfrac 12)^{(\widehat{{\bf k}})},(-\tfrac 12)^{(\widehat{{\bf k}}%
)};(-\tfrac 12)^{(\widehat{{\bf c}}_1)},(-\tfrac 12)^{(\widehat{{\bf c}}%
_2)})=\cos \dfrac{\theta _1}2\cos \dfrac{\theta _2}2e^{-i(\varphi _1+\varphi
_2)}.  \label{fi53}
\end{equation}

With these results, we are in a position to compute the probability
amplitudes for the compounded system.

\subsection{The Singlet-State Probability Amplitudes}

We first deal with the case $s=0$. Thus, there is only one projection,
characterised by $M^{(\widehat{{\bf a}})}=0.$ For this case therefore, there
is only the summation in Eq. (\ref{tw22}) to be carried out. The one in Eq. (%
\ref{tw21}) is redundant. Since 
\begin{equation}
\chi (s=0,M_i^{(\widehat{{\bf a}})}=0;s=0,M_j^{(\widehat{{\bf k}})}=0)=\chi
(0,0^{(\widehat{{\bf a}})};0,0^{(\widehat{{\bf k}})})=1,  \label{fi53a}
\end{equation}
we have 
\begin{eqnarray}
&&\Psi (0,0^{(\widehat{{\bf a}})};(m_1)_u^{(\widehat{{\bf c}}_1)},(m_2)_v^{(%
\widehat{{\bf c}}_2)})=\xi (0,0^{(\widehat{{\bf a}})};(m_1)_u^{(\widehat{%
{\bf c}}_1)},(m_2)_v^{(\widehat{{\bf c}}_2)})  \nonumber \\
\ &=&\sum_{\alpha ,\alpha ^{\prime }}\eta (0,0^{(\widehat{{\bf k}}%
)};(m_1)_\alpha ^{(\widehat{{\bf k}})},(m_2)_{\alpha ^{\prime }}^{(\widehat{%
{\bf k}})})  \nonumber \\
&&\times \psi ((m_1)_\alpha ^{(\widehat{{\bf k}})},(m_2)_{\alpha ^{\prime
}}^{(\widehat{{\bf k}})};(m_1)_u^{(\widehat{{\bf c}}_1)},(m_2)_v^{(\widehat{%
{\bf c}}_2)}).  \label{fi53b}
\end{eqnarray}

Since the $\eta $'s are Clebsch-Gordan coefficients, they are 
\begin{equation}
\eta (0,0^{(\widehat{{\bf k}})};(+\tfrac 12)^{(\widehat{{\bf k}})},(+\tfrac
12)^{(\widehat{{\bf k}})})=C(\tfrac 12\tfrac 120;\tfrac 12\tfrac 120)=0,
\label{fi53c}
\end{equation}

\begin{equation}
\eta (0,0^{(\widehat{{\bf k}})};(+\tfrac 12)^{(\widehat{{\bf k}})},(-\tfrac
12)^{(\widehat{{\bf k}})})=C(\tfrac 12\tfrac 120;\tfrac 12,-\tfrac
120)=\frac 1{\sqrt{2}},  \label{fi53d}
\end{equation}

\begin{equation}
\eta (0,0^{(\widehat{{\bf k}})};(-\tfrac 12)^{(\widehat{{\bf k}})},(+\tfrac
12)^{(\widehat{{\bf k}})})=C(\tfrac 12\tfrac 120;-\tfrac 12\tfrac
120)=-\frac 1{\sqrt{2}},  \label{fi53e}
\end{equation}
and 
\begin{equation}
\eta (0,0^{(\widehat{{\bf k}})};(-\tfrac 12)^{(\widehat{{\bf k}})},(-\tfrac
12)^{(\widehat{{\bf k}})})=C(\tfrac 12\tfrac 120;-\tfrac 12,-\tfrac 120)=0.
\label{fi53f}
\end{equation}
Here, we have obtained the values of the Clebsch-Gordan coefficients from
Rose[10].

Using the expressions for the $\psi $'s given by Eqs. (\ref{th37}) - (\ref
{fi53}), we obtain the probability amplitudes

\begin{equation}
\Psi (0,0^{(\widehat{{\bf a}})};(+\tfrac 12)^{(\widehat{{\bf c}}%
_1)},(+\tfrac 12)^{(\widehat{{\bf c}}_2)})=\frac 1{\sqrt{2}}[\cos \dfrac{%
\theta _1}2\sin \dfrac{\theta _2}2e^{-i\varphi _2}-\sin \dfrac{\theta _1}%
2\cos \dfrac{\theta _2}2e^{-i\varphi _1}],  \label{fi54}
\end{equation}
\begin{equation}
\Psi (0,0^{(\widehat{{\bf a}})};(+\tfrac 12)^{(\widehat{{\bf c}}%
_1)},(-\tfrac 12)^{(\widehat{{\bf c}}_2)})=\frac 1{\sqrt{2}}[\cos \dfrac{%
\theta _1}2\cos \dfrac{\theta _2}2e^{-i\varphi _2}+\sin \dfrac{\theta _1}%
2\sin \dfrac{\theta _2}2e^{-i\varphi _1}],  \label{fi55}
\end{equation}
\begin{equation}
\Psi (0,0^{(\widehat{{\bf a}})};(-\tfrac 12)^{(\widehat{{\bf c}}%
_1)},(+\tfrac 12)^{(\widehat{{\bf c}}_2)})=-\frac 1{\sqrt{2}}[\sin \dfrac{%
\theta _1}2\sin \dfrac{\theta _2}2e^{-i\varphi _2}+\cos \dfrac{\theta _1}%
2\cos \dfrac{\theta _2}2e^{-i\varphi _1}]  \label{fi56}
\end{equation}
and 
\begin{equation}
\Psi (0,0^{(\widehat{{\bf a}})};(-\tfrac 12)^{(\widehat{{\bf c}}%
_1)},(-\tfrac 12)^{(\widehat{{\bf c}}_2)})=\frac 1{\sqrt{2}}[\cos \dfrac{%
\theta _1}2\sin \dfrac{\theta _2}2e^{-i\varphi _1}-\sin \dfrac{\theta _1}%
2\cos \dfrac{\theta _2}2e^{-i\varphi _2}].  \label{fi57}
\end{equation}

\subsection{The Triplet State Probability Amplitudes}

Owing to the presence of two summations, the triplet case is a bit more
involved than the singlet case. The general formula for the probability
amplitudes is given by Eq. (\ref{tw21}). Each of the four probability
amplitudes has the form

\begin{eqnarray}
&&\Psi (1,M_i^{(\widehat{{\bf a}})};(m_1)_u^{(\widehat{{\bf c}}%
_1)},(m_2)_v^{(\widehat{{\bf c}}_2)})=\chi (1,M^{(\widehat{{\bf a}})};1,1^{(%
\widehat{{\bf k}})})\xi (1,1^{(\widehat{{\bf k}})};(m_1)_u^{(\widehat{{\bf c}%
}_1)},(m_2)_v^{(\widehat{{\bf c}}_2)})  \nonumber \\
&&+\chi (1,M^{(\widehat{{\bf a}})};1,0^{(\widehat{{\bf k}})})\xi (1,0^{(%
\widehat{{\bf k}})};(m_1)_u^{(\widehat{{\bf c}}_1)},(m_2)_v^{(\widehat{{\bf c%
}}_2)})  \nonumber \\
&&+\chi (1,M^{(\widehat{{\bf a}})};1,(-1)^{(\widehat{{\bf k}})})\xi
(1,(-1)^{(\widehat{{\bf k}})};(m_1)_u^{(\widehat{{\bf c}}_1)},(m_2)_v^{(%
\widehat{{\bf c}}_2)}).  \label{si64}
\end{eqnarray}

The probability amplitudes $\chi (1,M_i^{(\widehat{{\bf a}})};1,M_j^{(%
\widehat{{\bf k}})})$ connect spin projections measurements such that the
initial state corresponds to the vector $\widehat{{\bf a}}$, while the
result corresponds to the vector $\widehat{{\bf k}}$. Since $s=1$, there are
three values of $M^{(\widehat{{\bf a}})}$ and $M^{(\widehat{{\bf k}})}$:
these are $M^{(\widehat{{\bf a}})},M^{(\widehat{{\bf k}})}=0,\pm 1$. These
probability amplitudes $\chi $ are derived in Ref. [3].

\subsubsection{The $M^{(\widehat{{\bf a}})}=1$ State}

For $M^{(\widehat{{\bf a}})}=1$, the $\chi (1,M_i^{(\widehat{{\bf a}}%
)};1,M_j^{(\widehat{{\bf k}})})$ are

\begin{equation}
\chi (1,1^{(\widehat{{\bf a}})};1,1^{(\widehat{{\bf k}})})=\cos ^2\dfrac
\theta 2e^{-i\varphi },  \label{si65}
\end{equation}
\begin{equation}
\chi (1,1^{(\widehat{{\bf a}})};1,0^{(\widehat{{\bf k}})})=\frac 1{\sqrt{2}%
}\sin \theta  \label{si66}
\end{equation}
and

\begin{equation}
\chi (1,1^{(\widehat{{\bf a}})};1,(-1)^{(\widehat{{\bf k}})})=\sin ^2\dfrac
\theta 2e^{i\varphi }.  \label{si67}
\end{equation}

The probability amplitudes $\xi $ are given by Eq. (\ref{tw22}) and are:

\begin{eqnarray}
&&\xi (1,M_j^{(\widehat{{\bf k}})};(m_1)_u^{(\widehat{{\bf c}}_1)},(m_2)_v^{(%
\widehat{{\bf c}}_2)})=\eta (1,M_j^{(\widehat{{\bf k}})};(+\tfrac 12)^{(%
\widehat{{\bf k}})},(+\tfrac 12)^{(\widehat{{\bf k}})})  \nonumber \\
&&\times \psi ((+\tfrac 12)^{(\widehat{{\bf k}})},(+\tfrac 12)^{(\widehat{%
{\bf k}})};(m_1)_u^{(\widehat{{\bf c}}_1)},(m_2)_v^{(\widehat{{\bf c}}_2)}) 
\nonumber \\
&&\ +\eta (1,M_j^{(\widehat{{\bf k}})};(+\tfrac 12)^{(\widehat{{\bf k}}%
)},(-\tfrac 12)^{(\widehat{{\bf k}})})\psi ((+\tfrac 12)^{(\widehat{{\bf k}}%
)},(-\tfrac 12)^{(\widehat{{\bf k}})};(m_1)_u^{(\widehat{{\bf c}}%
_1)},(m_2)_v^{(\widehat{{\bf c}}_2)})  \nonumber \\
&&\ +\eta (1,M_j^{(\widehat{{\bf k}})};(-\tfrac 12)^{(\widehat{{\bf k}}%
)},(+\tfrac 12)^{(\widehat{{\bf k}})})\psi ((-\tfrac 12)^{(\widehat{{\bf k}}%
)},(+\tfrac 12)^{(\widehat{{\bf k}})};(m_1)_u^{(\widehat{{\bf c}}%
_1)},(m_2)_v^{(\widehat{{\bf c}}_2)})  \nonumber \\
&&\ +\eta (1,M_j^{(\widehat{{\bf k}})};(-\tfrac 12)^{(\widehat{{\bf k}}%
)},(-\tfrac 12)^{(\widehat{{\bf k}})})\psi ((-\tfrac 12)^{(\widehat{{\bf k}}%
)},(-\tfrac 12)^{(\widehat{{\bf k}})};(m_1)_u^{(\widehat{{\bf c}}%
_1)},(m_2)_v^{(\widehat{{\bf c}}_2)}).  \nonumber \\
&&  \label{si68}
\end{eqnarray}

The $\eta $'s are Clebsch-Gordan coefficients; thus setting $M^{(\widehat{%
{\bf k}})}=1$, we have

\begin{equation}
\eta (1,1^{(\widehat{{\bf k}})};(+\tfrac 12)^{(\widehat{{\bf k}})},(+\tfrac
12)^{(\widehat{{\bf k}})})=C(\tfrac 12\tfrac 121;\tfrac 12\tfrac 121)=1,
\label{si69}
\end{equation}
\begin{equation}
\eta (1,1^{(\widehat{{\bf k}})};(+\tfrac 12)^{(\widehat{{\bf k}})},(-\tfrac
12)^{(\widehat{{\bf k}})})=C(\tfrac 12\tfrac 121;\tfrac 12,-\tfrac 121)=0,
\label{se70}
\end{equation}

\begin{equation}
\eta (1,1^{(\widehat{{\bf k}})};(-\tfrac 12)^{(\widehat{{\bf k}})},(+\tfrac
12)^{(\widehat{{\bf k}})})=C(\tfrac 12\tfrac 121;-\tfrac 12\tfrac 121)=0
\label{se71}
\end{equation}
and

\begin{equation}
\eta (1,1^{(\widehat{{\bf k}})};(-\tfrac 12)^{(\widehat{{\bf k}})},(-\tfrac
12)^{(\widehat{{\bf k}})})=C(\tfrac 12\tfrac 121;-\tfrac 12,-\tfrac 121)=0.
\label{se72}
\end{equation}
Therefore, 
\begin{equation}
\xi (1,1^{(\widehat{{\bf k}})};(m_1)_u^{(\widehat{{\bf c}}_1)},(m_2)_v^{(%
\widehat{{\bf c}}_2)})=\psi ((+\tfrac 12)^{(\widehat{{\bf k}})},(+\tfrac
12)^{(\widehat{{\bf k}})};(m_1)_u^{(\widehat{{\bf c}}_1)},(m_2)_v^{(\widehat{%
{\bf c}}_2)}).  \label{se73}
\end{equation}
Setting $M^{(\widehat{{\bf k}})}=0$, we find 
\begin{equation}
\eta (1,0^{(\widehat{{\bf k}})};(+\tfrac 12)^{(\widehat{{\bf k}})},(+\tfrac
12)^{(\widehat{{\bf k}})})=C(\tfrac 12\tfrac 121;\tfrac 12\tfrac 120)=0,
\label{se74}
\end{equation}
\begin{equation}
\eta (1,0^{(\widehat{{\bf k}})};(+\tfrac 12)^{(\widehat{{\bf k}})},(-\tfrac
12)^{(\widehat{{\bf k}})})=C(\tfrac 12\tfrac 121;\tfrac 12,-\tfrac
120)=\frac 1{\sqrt{2}},  \label{se75}
\end{equation}

\begin{equation}
\eta (1,0^{(\widehat{{\bf k}})};(-\tfrac 12)^{(\widehat{{\bf k}})},(+\tfrac
12)^{(\widehat{{\bf k}})})=C(\tfrac 12\tfrac 121;-\tfrac 12\tfrac 120)=\frac
1{\sqrt{2}},  \label{se76}
\end{equation}
and 
\begin{equation}
\eta (1,0^{(\widehat{{\bf k}})};(-\tfrac 12)^{(\widehat{{\bf k}})},(-\tfrac
12)^{(\widehat{{\bf k}})})=C(\tfrac 12\tfrac 121;-\tfrac 12,-\tfrac 120)=0.
\label{se77}
\end{equation}
Thus 
\begin{eqnarray}
&&\xi (1,0^{(\widehat{{\bf k}})};(m_1)_u^{(\widehat{{\bf c}}_1)},(m_2)_v^{(%
\widehat{{\bf c}}_2)})=\frac 1{\sqrt{2}}[\psi ((+\tfrac 12)^{(\widehat{{\bf k%
}})},(-\tfrac 12)^{(\widehat{{\bf k}})};(m_1)_u^{(\widehat{{\bf c}}%
_1)},(m_2)_v^{(\widehat{{\bf c}}_2)})  \nonumber \\
&&+\psi ((-\tfrac 12)^{(\widehat{{\bf k}})},(+\tfrac 12)^{(\widehat{{\bf k}}%
)};(m_1)_u^{(\widehat{{\bf c}}_1)},(m_2)_v^{(\widehat{{\bf c}}_2)})].
\label{se78}
\end{eqnarray}
Finally, setting $M^{(\widehat{{\bf k}})}=-1$, we find 
\begin{equation}
\eta (1,(-1)^{(\widehat{{\bf k}})};(+\tfrac 12)^{(\widehat{{\bf k}}%
)},(+\tfrac 12)^{(\widehat{{\bf k}})})=C(\tfrac 12\tfrac 121;\tfrac 12\tfrac
12,-1)=0,  \label{se79}
\end{equation}
\begin{equation}
\eta (1,(-1)^{(\widehat{{\bf k}})};(+\tfrac 12)^{(\widehat{{\bf k}}%
)},(-\tfrac 12)^{(\widehat{{\bf k}})})=C(\tfrac 12\tfrac 121;\tfrac
12,-\tfrac 12,-1)=0,  \label{ei80}
\end{equation}

\begin{equation}
\eta (1,(-1)^{(\widehat{{\bf k}})};(-\tfrac 12)^{(\widehat{{\bf k}}%
)},(+\tfrac 12)^{(\widehat{{\bf k}})})=C(\tfrac 12\tfrac 121;-\tfrac
12\tfrac 12,-1)=0,  \label{ei81}
\end{equation}
and 
\begin{equation}
\eta (1,(-1)^{(\widehat{{\bf k}})};(-\tfrac 12)^{(\widehat{{\bf k}}%
)},(-\tfrac 12)^{(\widehat{{\bf k}})})=C(\tfrac 12\tfrac 121;-\tfrac
12,-\tfrac 12,-1)=1.  \label{ei82}
\end{equation}
Thus,

\begin{equation}
\xi (1,(-1)^{(\widehat{{\bf k}})};(m_1)_u^{(\widehat{{\bf c}}_1)},(m_2)_v^{(%
\widehat{{\bf c}}_2)})=\psi ((-\tfrac 12)^{(\widehat{{\bf k}})},(-\tfrac
12)^{(\widehat{{\bf k}})};(m_1)_u^{(\widehat{{\bf c}}_1)},(m_2)_v^{(\widehat{%
{\bf c}}_2)}).  \label{ei83}
\end{equation}

Using all these results, and recalling Eqs. (\ref{th37}) - (\ref{fi53}) for
the $\psi $, we find that for $M^{(\widehat{{\bf a}})}=1$, 
\begin{eqnarray}
&&\Psi (1,1^{(\widehat{{\bf a}})};(+\tfrac 12)^{(\widehat{{\bf c}}%
_1)},(+\tfrac 12)^{(\widehat{{\bf c}}_2)})=\cos ^2\dfrac \theta 2\cos \dfrac{%
\theta _1}2\cos \dfrac{\theta _2}2e^{-i\varphi }  \nonumber \\
&&+\sin ^2\dfrac \theta 2\sin \dfrac{\theta _1}2\sin \dfrac{\theta _2}%
2e^{i(\varphi -\varphi _1-\varphi _2)}  \nonumber \\
&&\ \ +\frac 12\sin \theta [\cos \dfrac{\theta _1}2\sin \dfrac{\theta _2}%
2e^{-i\varphi _2}+\sin \dfrac{\theta _1}2\cos \dfrac{\theta _2}2e^{-i\varphi
_1}],  \label{ei84}
\end{eqnarray}
\begin{eqnarray}
&&\Psi (1,1^{(\widehat{{\bf a}})};(+\tfrac 12)^{(\widehat{{\bf c}}%
_1)},(-\tfrac 12)^{(\widehat{{\bf c}}_2)})=-\cos ^2\dfrac \theta 2\cos 
\dfrac{\theta _1}2\sin \dfrac{\theta _2}2e^{-i\varphi }  \nonumber \\
&&+\sin ^2\dfrac \theta 2\sin \dfrac{\theta _1}2\cos \dfrac{\theta _2}%
2e^{i(\varphi -\varphi _1-\varphi _2)}  \nonumber \\
&&\ \ \ +\frac 12\sin \theta [\cos \dfrac{\theta _1}2\cos \dfrac{\theta _2}%
2e^{-i\varphi _2}-\sin \dfrac{\theta _1}2\sin \dfrac{\theta _2}2e^{-i\varphi
_1}],  \label{ei85}
\end{eqnarray}

\begin{eqnarray}
&&\Psi (1,1^{(\widehat{{\bf a}})};(-\tfrac 12)^{(\widehat{{\bf c}}%
_1)},(+\tfrac 12)^{(\widehat{{\bf c}}_2)})=-\cos ^2\dfrac \theta 2\sin 
\dfrac{\theta _1}2\cos \dfrac{\theta _2}2e^{-i\varphi }  \nonumber \\
&&+\sin ^2\dfrac \theta 2\cos \dfrac{\theta _1}2\sin \dfrac{\theta _2}%
2e^{i(\varphi -\varphi _1-\varphi _2)}  \nonumber \\
&&\ \ \ \ +\frac 12\sin \theta [\cos \dfrac{\theta _1}2\cos \dfrac{\theta _2}%
2e^{-i\varphi _1}-\sin \dfrac{\theta _1}2\sin \dfrac{\theta _2}2e^{-i\varphi
_2}]  \label{ei86}
\end{eqnarray}
and 
\begin{eqnarray}
&&\Psi (1,1^{(\widehat{{\bf a}})};(-\tfrac 12)^{(\widehat{{\bf c}}%
_1)},(-\tfrac 12)^{(\widehat{{\bf c}}_2)})=\cos ^2\dfrac \theta 2\sin \dfrac{%
\theta _1}2\sin \dfrac{\theta _2}2e^{-i\varphi }  \nonumber \\
&&+\sin ^2\dfrac \theta 2\cos \dfrac{\theta _1}2\cos \dfrac{\theta _2}%
2e^{i(\varphi -\varphi _1-\varphi _2)}  \nonumber \\
&&\ \ \ \ \ -\frac 12\sin \theta [\sin \dfrac{\theta _1}2\cos \dfrac{\theta
_2}2e^{-i\varphi _2}+\cos \dfrac{\theta _1}2\sin \dfrac{\theta _2}%
2e^{-i\varphi _1}].  \label{ei87}
\end{eqnarray}

\subsubsection{The $M^{(\widehat{{\bf a}})}=0$ State}

For $M^{(\widehat{{\bf a}})}=0$, the $\xi $'s remain the same as for $M^{(%
\widehat{{\bf a}})}=1$. However, the $\chi $'s are[3]

\begin{equation}
\chi (1,0^{(\widehat{{\bf a}})};1,1^{(\widehat{{\bf k}})})=-\frac 1{\sqrt{2}%
}\sin \theta e^{-i\varphi },  \label{ei88}
\end{equation}
\begin{equation}
\chi (1,0^{(\widehat{{\bf a}})};1,0^{(\widehat{{\bf k}})})=\cos \theta
\label{ei89}
\end{equation}
and

\begin{equation}
\chi (1,0^{(\widehat{{\bf a}})};1,(-1)^{(\widehat{{\bf k}})})=\frac 1{\sqrt{2%
}}\sin \theta e^{i\varphi }.  \label{ni90}
\end{equation}
Therefore, the probability amplitudes are 
\begin{eqnarray}
&&\Psi (1,0^{(\widehat{{\bf a}})};(+\tfrac 12)^{(\widehat{{\bf c}}%
_1)},(+\tfrac 12)^{(\widehat{{\bf c}}_2)})=\frac 1{\sqrt{2}}\sin \theta
[\sin \dfrac{\theta _1}2\sin \dfrac{\theta _2}2e^{i(\varphi -\varphi
_1-\varphi _2)}  \nonumber \\
&&-\cos \dfrac{\theta _1}2\cos \dfrac{\theta _2}2e^{-i\varphi }]  \nonumber
\\
&&\ \ +\frac 1{\sqrt{2}}\cos \theta [\cos \dfrac{\theta _1}2\sin \dfrac{%
\theta _2}2e^{-i\varphi _2}+\sin \dfrac{\theta _1}2\cos \dfrac{\theta _2}%
2e^{-i\varphi _1}],  \label{ni91}
\end{eqnarray}
\begin{eqnarray}
&&\Psi (1,0^{(\widehat{{\bf a}})};(+\tfrac 12)^{(\widehat{{\bf c}}%
_1)},(-\tfrac 12)^{(\widehat{{\bf c}}_2)})=\frac 1{\sqrt{2}}\sin \theta
[\sin \dfrac{\theta _1}2\cos \dfrac{\theta _2}2e^{i(\varphi -\varphi
_1-\varphi _2)}  \nonumber \\
&&+\cos \dfrac{\theta _1}2\sin \dfrac{\theta _2}2e^{-i\varphi }]  \nonumber
\\
&&\ \ \ +\frac 1{\sqrt{2}}\cos \theta [\cos \dfrac{\theta _1}2\cos \dfrac{%
\theta _2}2e^{-i\varphi _2}-\sin \dfrac{\theta _1}2\sin \dfrac{\theta _2}%
2e^{-i\varphi _1}],  \label{ni92}
\end{eqnarray}

\begin{eqnarray}
&&\Psi (1,0^{(\widehat{{\bf a}})};(-\tfrac 12)^{(\widehat{{\bf c}}%
_1)},(+\tfrac 12)^{(\widehat{{\bf c}}_2)})=\frac 1{\sqrt{2}}\sin \theta
[\cos \dfrac{\theta _1}2\sin \dfrac{\theta _2}2e^{i(\varphi -\varphi
_1-\varphi _2)}  \nonumber \\
&&+\sin \dfrac{\theta _1}2\cos \dfrac{\theta _2}2e^{-i\varphi }]  \nonumber
\\
&&\ \ \ \ +\frac 1{\sqrt{2}}\cos \theta [\cos \dfrac{\theta _1}2\cos \dfrac{%
\theta _2}2e^{-i\varphi _1}-\sin \dfrac{\theta _1}2\sin \dfrac{\theta _2}%
2e^{-i\varphi _2}]  \label{ni93}
\end{eqnarray}
and 
\begin{eqnarray}
&&\Psi (1,0^{(\widehat{{\bf a}})};(-\tfrac 12)^{(\widehat{{\bf c}}%
_1)},(-\tfrac 12)^{(\widehat{{\bf c}}_2)})=\frac 1{\sqrt{2}}\sin \theta
[\cos \dfrac{\theta _1}2\cos \dfrac{\theta _2}2e^{i(\varphi -\varphi
_1-\varphi _2)}  \nonumber \\
&&-\sin \dfrac{\theta _1}2\sin \dfrac{\theta _2}2e^{-i\varphi }]  \nonumber
\\
&&\ \ \ -\frac 1{\sqrt{2}}\cos \theta [\sin \dfrac{\theta _1}2\cos \dfrac{%
\theta _2}2e^{-i\varphi _2}+\cos \dfrac{\theta _1}2\sin \dfrac{\theta _2}%
2e^{-i\varphi _1}].  \label{ni94}
\end{eqnarray}

\subsubsection{The $M^{(\widehat{{\bf a}})}=-1$ State}

For $M^{(\widehat{{\bf a}})}=-1$, again the $\xi $'s remain the same, while
the $\chi $'s are [3] 
\begin{equation}
\chi (1,(-1)^{(\widehat{{\bf a}})};1,1^{(\widehat{{\bf k}})})=-\sin ^2\dfrac
\theta 2e^{-i\varphi },  \label{ni95}
\end{equation}
\begin{equation}
\chi (1,(-1)^{(\widehat{{\bf a}})};1,0^{(\widehat{{\bf k}})})=\frac 1{\sqrt{2%
}}\sin \theta  \label{ni96}
\end{equation}
and

\begin{equation}
\chi (1,(-1)^{(\widehat{{\bf a}})};1,(-1)^{(\widehat{{\bf k}})})=-\cos
^2\dfrac \theta 2e^{i\varphi }.  \label{ni97}
\end{equation}
Hence, the probability amplitudes are 
\begin{eqnarray}
&&\Psi (1,(-1)^{(\widehat{{\bf a}})};(+\tfrac 12)^{(\widehat{{\bf c}}%
_1)},(+\tfrac 12)^{(\widehat{{\bf c}}_2)})=-\sin ^2\dfrac \theta 2\cos 
\dfrac{\theta _1}2\cos \dfrac{\theta _2}2e^{-i\varphi }  \nonumber \\
&&-\cos ^2\dfrac \theta 2\sin \dfrac{\theta _1}2\sin \dfrac{\theta _2}%
2e^{i(\varphi -\varphi _1-\varphi _2)}  \nonumber \\
&&\ \ \ +\frac 12\sin \theta [\cos \dfrac{\theta _1}2\sin \frac{\theta _2}%
2e^{-i\varphi _2}+\sin \dfrac{\theta _1}2\cos \frac{\theta _2}2e^{-i\varphi
_1}],  \label{ni98}
\end{eqnarray}
\begin{eqnarray}
&&\Psi (1,(-1)^{(\widehat{{\bf a}})};(+\tfrac 12)^{(\widehat{{\bf c}}%
_1)},(-\tfrac 12)^{(\widehat{{\bf c}}_2)})=\sin ^2\dfrac \theta 2\cos \dfrac{%
\theta _1}2\sin \frac{\theta _2}2e^{-i\varphi }  \nonumber \\
&&-\cos ^2\dfrac \theta 2\sin \dfrac{\theta _1}2\cos \frac{\theta _2}%
2e^{i(\varphi -\varphi _1-\varphi _2)}  \nonumber \\
&&\ \ \ +\ \frac 12\sin \theta [\cos \dfrac{\theta _1}2\cos \frac{\theta _2}%
2e^{-i\varphi _2}-\sin \dfrac{\theta _1}2\sin \frac{\theta _2}2e^{-i\varphi
_1}],  \label{ni99}
\end{eqnarray}

\begin{eqnarray}
&&\Psi (1,(-1)^{(\widehat{{\bf a}})};(-\tfrac 12)^{(\widehat{{\bf c}}%
_1)},(+\tfrac 12)^{(\widehat{{\bf c}}_2)})=\sin ^2\dfrac \theta 2\sin \dfrac{%
\theta _1}2\cos \frac{\theta _2}2e^{-i\varphi }  \nonumber \\
&&-\cos ^2\dfrac \theta 2\cos \dfrac{\theta _1}2\sin \frac{\theta _2}%
2e^{i(\varphi -\varphi _1-\varphi _2)}  \nonumber \\
&&\ \ \ \ \ +\frac 12\sin \theta [\cos \dfrac{\theta _1}2\cos \frac{\theta _2%
}2e^{-i\varphi _1}-\sin \dfrac{\theta _1}2\sin \frac{\theta _2}2e^{-i\varphi
_2}]  \label{hu100}
\end{eqnarray}
and 
\begin{eqnarray}
&&\Psi (1,(-1)^{(\widehat{{\bf a}})};(-\tfrac 12)^{(\widehat{{\bf c}}%
_1)},(-\tfrac 12)^{(\widehat{{\bf c}}_2)})=-\sin ^2\dfrac \theta 2\sin 
\dfrac{\theta _1}2\sin \frac{\theta _2}2e^{-i\varphi }  \nonumber \\
&&-\cos ^2\dfrac \theta 2\cos \dfrac{\theta _1}2\cos \frac{\theta _2}%
2e^{i(\varphi -\varphi _1-\varphi _2)}  \nonumber \\
&&\ \ \ \ \ \ -\frac 12\sin \theta [\cos \dfrac{\theta _1}2\sin \frac{\theta
_2}2e^{-i\varphi _1}+\sin \dfrac{\theta _1}2\cos \frac{\theta _2}%
2e^{-i\varphi _2}].  \label{hu101}
\end{eqnarray}

\section{Probabilities}

\subsection{Singlet-State Probabilities}

The probabilities for the singlet state are

\begin{eqnarray}
&&P(0,0^{(\widehat{{\bf a}})};(+\tfrac 12)^{(\widehat{{\bf c}}_1)},(+\tfrac
12)^{(\widehat{{\bf c}}_2)})=\left| \Psi (0,0;(+\tfrac 12)^{(\widehat{{\bf c}%
}_1)},(+\tfrac 12)^{(\widehat{{\bf c}}_2)})\right| ^2  \nonumber \\
&=&\frac 12[\sin ^2\dfrac{(\theta _2-\theta _1)}2+\sin \theta _1\sin \theta
_2\sin ^2\dfrac{(\varphi _2-\varphi _1)}2],  \label{fi59}
\end{eqnarray}

\begin{eqnarray}
&&P(0,0^{(\widehat{{\bf a}})};(+\tfrac 12)^{(\widehat{{\bf c}}_1)},(-\tfrac
12)^{(\widehat{{\bf c}}_2)})=\frac 12[\cos ^2\dfrac{(\theta _2-\theta _1)}2 
\nonumber \\
&&-\sin \theta _1\sin \theta _2\sin ^2\dfrac{(\varphi _2-\varphi _1)}2],
\label{si60}
\end{eqnarray}

\begin{equation}
P(0,0^{(\widehat{{\bf a}})};(-\tfrac 12)^{(\widehat{{\bf c}}_1)},(+\tfrac
12)^{(\widehat{{\bf c}}_2)})=P(0,0^{(\widehat{{\bf a}})};(+\tfrac 12)^{(%
\widehat{{\bf c}}_1)},(-\tfrac 12)^{(\widehat{{\bf c}}_2)})  \label{si61}
\end{equation}
and 
\begin{equation}
P(0,0^{(\widehat{{\bf a}})};(-\tfrac 12)^{(\widehat{{\bf c}}_1)},(-\tfrac
12)^{(\widehat{{\bf c}}_2)})=P(0,0^{(\widehat{{\bf a}})};(+\tfrac 12)^{(%
\widehat{{\bf c}}_1)},(+\tfrac 12)^{(\widehat{{\bf c}}_2)}).  \label{si62}
\end{equation}

Now let $P(0,0^{(\widehat{{\bf a}})};(+\tfrac 12)^{(\widehat{{\bf c}}_1)})$
be the probability of finding the spin projection of subsystem $1$ up with
respect to the direction $\widehat{{\bf c}}_1$ irrespective of the
projection found for subsystem $2.$ Then

\begin{eqnarray}
&&P(0,0^{(\widehat{{\bf a}})};(+\tfrac 12)^{(\widehat{{\bf c}}_1)})=P(0,0^{(%
\widehat{{\bf a}})};(+\tfrac 12)^{(\widehat{{\bf c}}_1)},(+\tfrac 12)^{(%
\widehat{{\bf c}}_2)})  \nonumber \\
&&+P(0,0^{(\widehat{{\bf a}})};(+\tfrac 12)^{(\widehat{{\bf c}}_1)},(-\tfrac
12)^{(\widehat{{\bf c}}_2)})  \nonumber \\
&=&\frac 12.  \label{si62a}
\end{eqnarray}
$=$

Let $P(0,0^{(\widehat{{\bf a}})};(-\tfrac 12)^{(\widehat{{\bf c}}_1)})$ be
the probability of finding the spin projection of subsystem $1$ down with
respect to the direction $\widehat{{\bf c}}_1$ irrespective of the
projection found for subsystem $2.$ Then

\begin{eqnarray}
&&P(0,0^{(\widehat{{\bf a}})};(-\tfrac 12)^{(\widehat{{\bf c}}_1)})=P(0,0^{(%
\widehat{{\bf a}})};(-\tfrac 12)^{(\widehat{{\bf c}}_1)},(+\tfrac 12)^{(%
\widehat{{\bf c}}_2)})  \nonumber \\
&&+P(0,0^{(\widehat{{\bf a}})};(-\tfrac 12)^{(\widehat{{\bf c}}_1)},(-\tfrac
12)^{(\widehat{{\bf c}}_2)})  \nonumber \\
&=&\frac 12.  \label{si62b}
\end{eqnarray}

Similarly, 
\begin{equation}
P(0,0^{(\widehat{{\bf a}})};(+\tfrac 12)^{(\widehat{{\bf c}}_2)})=\frac 12
\label{si62c}
\end{equation}
and 
\begin{equation}
P(0,0^{(\widehat{{\bf a}})};(-\tfrac 12)^{(\widehat{{\bf c}}_2)})=\frac 12.
\label{si62d}
\end{equation}

\subsection{Triplet-State Probabilities}

\subsubsection{The $M^{(\widehat{{\bf a}})}=1$ State}

The probabilities corresponding to the amplitudes for the initial state
characterized by $s=1$, $M^{(\widehat{{\bf a}})}=1$ are

\begin{eqnarray}
&&P(1,1^{(\widehat{{\bf a}})};(+\tfrac 12)^{(\widehat{{\bf c}}_1)},(+\tfrac
12)^{(\widehat{{\bf c}}_2)})=\left| \Psi (1,1^{(\widehat{{\bf a}})};(+\tfrac
12)^{(\widehat{{\bf c}}_1)},(+\tfrac 12)^{(\widehat{{\bf c}}_2)})\right| ^2 
\nonumber \\
\ &=&\cos ^4\dfrac \theta 2\cos ^2\dfrac{\theta _1}2\cos ^2\frac{\theta _2}%
2+\sin ^4\dfrac \theta 2\sin ^2\dfrac{\theta _1}2\sin ^2\frac{\theta _2}2 
\nonumber \\
&&\ \ +\frac 14\sin ^2\theta \cos ^2\dfrac{\theta _1}2\sin ^2\frac{\theta _2}%
2+\frac 14\sin ^2\theta \sin ^2\dfrac{\theta _1}2\cos ^2\frac{\theta _2}2 
\nonumber \\
&&\ \ +\frac 14\sin ^2\theta \sin \theta _1\sin \theta _2\cos (\varphi
-\varphi _1)\cos (\varphi -\varphi _2)  \nonumber \\
&&\ \ +\frac 12\sin \theta \sin \theta _1(\cos ^2\dfrac \theta 2\cos ^2\frac{%
\theta _2}2+\sin ^2\dfrac \theta 2\sin ^2\frac{\theta _2}2)\cos (\varphi
_1-\varphi )  \nonumber \\
&&\ \ +\frac 12\sin \theta \sin \theta _2(\cos ^2\dfrac \theta 2\cos ^2%
\dfrac{\theta _1}2+\sin ^2\dfrac \theta 2\sin ^2\dfrac{\theta _1}2)\cos
(\varphi _2-\varphi ),  \nonumber \\
&&  \label{hu102}
\end{eqnarray}

\begin{eqnarray}
\ &&P(1,1^{(\widehat{{\bf a}})};(+\tfrac 12)^{(\widehat{{\bf c}}%
_1)},(-\tfrac 12)^{(\widehat{{\bf c}}_2)})=\cos ^4\dfrac \theta 2\cos ^2%
\dfrac{\theta _1}2\sin ^2\frac{\theta _2}2  \nonumber \\
&&+\sin ^4\dfrac \theta 2\sin ^2\dfrac{\theta _1}2\cos ^2\frac{\theta _2}2 
\nonumber \\
&&\ \ \ +\frac 14\sin ^2\theta \cos ^2\dfrac{\theta _1}2\cos ^2\frac{\theta
_2}2+\frac 14\sin ^2\theta \sin ^2\dfrac{\theta _1}2\sin ^2\frac{\theta _2}2
\nonumber \\
&&\ \ \ -\frac 14\sin ^2\theta \sin \theta _1\sin \theta _2\cos (\varphi
-\varphi _1)\cos (\varphi -\varphi _2)  \nonumber \\
&&\ \ \ +\frac 12\sin \theta \sin \theta _1(\cos ^2\dfrac \theta 2\sin ^2%
\frac{\theta _2}2+\sin ^2\dfrac \theta 2\cos ^2\frac{\theta _2}2)\cos
(\varphi _1-\varphi )  \nonumber \\
&&\ \ \ -\frac 12\sin \theta \sin \theta _2(\cos ^2\dfrac \theta 2\cos ^2%
\dfrac{\theta _1}2+\sin ^2\dfrac \theta 2\sin ^2\dfrac{\theta _1}2)\cos
(\varphi _2-\varphi ),  \nonumber \\
&&  \label{hu103}
\end{eqnarray}

\begin{eqnarray}
\ &&P(1,1^{(\widehat{{\bf a}})};(-\tfrac 12)^{(\widehat{{\bf c}}%
_1)},(+\tfrac 12)^{(\widehat{{\bf c}}_2)})=\cos ^4\dfrac \theta 2\sin ^2%
\dfrac{\theta _1}2\cos ^2\frac{\theta _2}2  \nonumber \\
&&+\sin ^4\dfrac \theta 2\cos ^2\dfrac{\theta _1}2\sin ^2\frac{\theta _2}2 
\nonumber \\
&&\ \ \ \ +\frac 14\sin ^2\theta \cos ^2\dfrac{\theta _1}2\cos ^2\frac{%
\theta _2}2+\frac 14\sin ^2\theta \sin ^2\dfrac{\theta _1}2\sin ^2\frac{%
\theta _2}2  \nonumber \\
&&\ \ \ \ -\frac 14\sin ^2\theta \sin \theta _1\sin \theta _2\cos (\varphi
-\varphi _1)\cos (\varphi -\varphi _2)  \nonumber \\
&&\ \ \ \ -\frac 12\sin \theta \sin \theta _1(\cos ^2\dfrac \theta 2\cos ^2%
\dfrac{\theta _2}2+\sin ^2\dfrac \theta 2\sin ^2\dfrac{\theta _2}2)\cos
(\varphi _1-\varphi )  \nonumber \\
&&\ \ \ \ +\frac 12\sin \theta \sin \theta _2(\cos ^2\dfrac \theta 2\sin ^2%
\dfrac{\theta _1}2+\sin ^2\dfrac \theta 2\cos ^2\dfrac{\theta _1}2)\cos
(\varphi _2-\varphi ),  \nonumber \\
&&  \label{hu104}
\end{eqnarray}

\begin{eqnarray}
&&\ P(1,1^{(\widehat{{\bf a}})};(-\tfrac 12)^{(\widehat{{\bf c}}%
_1)},(-\tfrac 12)^{(\widehat{{\bf c}}_2)})=\cos ^4\dfrac \theta 2\sin ^2%
\dfrac{\theta _1}2\sin ^2\dfrac{\theta _2}2  \nonumber \\
&&+\sin ^4\dfrac \theta 2\cos ^2\dfrac{\theta _1}2\cos ^2\dfrac{\theta _2}2 
\nonumber \\
&&\ \ \ \ +\frac 14\sin ^2\theta \sin ^2\dfrac{\theta _1}2\cos ^2\dfrac{%
\theta _2}2+\frac 14\sin ^2\theta \cos ^2\dfrac{\theta _1}2\sin ^2\dfrac{%
\theta _2}2  \nonumber \\
&&\ \ \ \ +\frac 14\sin ^2\theta \sin \theta _1\sin \theta _2\cos (\varphi
-\varphi _1)\cos (\varphi -\varphi _2)  \nonumber \\
&&\ \ \ \ -\frac 12\sin \theta \sin \theta _1(\cos ^2\dfrac \theta 2\sin ^2%
\frac{\theta _2}2+\sin ^2\dfrac \theta 2\cos ^2\frac{\theta _2}2)\cos
(\varphi _1-\varphi )  \nonumber \\
&&\ \ \ \ -\frac 12\sin \theta \sin \theta _2(\cos ^2\dfrac \theta 2\sin ^2%
\dfrac{\theta _1}2+\sin ^2\dfrac \theta 2\cos ^2\dfrac{\theta _1}2)\cos
(\varphi _2-\varphi ),  \nonumber \\
&&  \label{hu105}
\end{eqnarray}

These probabilities add to unity. We also find

\begin{eqnarray}
&&P(1,1^{(\widehat{{\bf a}})};(+\tfrac 12)^{(\widehat{{\bf c}}_1)})=P(1,1^{(%
\widehat{{\bf a}})};(+\tfrac 12)^{(\widehat{{\bf c}}_1)},(+\tfrac 12)^{(%
\widehat{{\bf c}}_2)})  \nonumber \\
&+P(1,1^{(\widehat{{\bf a}})}:&(+\tfrac 12)^{(\widehat{{\bf c}}_1)},(-\tfrac
12)^{(\widehat{{\bf c}}_2)})  \nonumber \\
\ &=&\cos ^2\dfrac \theta 2\cos ^2\dfrac{\theta _1}2+\sin ^2\dfrac \theta
2\sin ^2\dfrac{\theta _1}2  \nonumber \\
&&\ \ \ +\frac 12\sin \theta \sin \theta _1\cos (\varphi _1-\varphi ),
\label{hu106}
\end{eqnarray}
\begin{eqnarray}
&&P(1,1^{(\widehat{{\bf a}})};(-\tfrac 12)^{(\widehat{{\bf c}}_1)})=P(1,1^{(%
\widehat{{\bf a}})};(-\tfrac 12)^{(\widehat{{\bf c}}_1)},(+\tfrac 12)^{(%
\widehat{{\bf c}}_2)})  \nonumber \\
&&+P(1,1^{(\widehat{{\bf a}})};(-\tfrac 12)^{(\widehat{{\bf c}}_1)},(-\tfrac
12)^{(\widehat{{\bf c}}_2)})  \nonumber \\
\ &=&\cos ^2\dfrac \theta 2\sin ^2\dfrac{\theta _1}2+\sin ^2\dfrac \theta
2\cos ^2\dfrac{\theta _1}2  \nonumber \\
&&\ \ \ -\frac 12\sin \theta \sin \theta _1\cos (\varphi _1-\varphi ),
\label{hu107}
\end{eqnarray}
\begin{eqnarray}
&&P(1,1^{(\widehat{{\bf a}})};(+\tfrac 12)^{(\widehat{{\bf c}}_2)})=P(1,1^{(%
\widehat{{\bf a}})};(+\tfrac 12)^{(\widehat{{\bf c}}_1)},(+\tfrac 12)^{(%
\widehat{{\bf c}}_2)})  \nonumber \\
&&+P(1,1^{(\widehat{{\bf a}})};(-\tfrac 12)^{(\widehat{{\bf c}}_1)},(+\tfrac
12)^{(\widehat{{\bf c}}_2)})  \nonumber \\
\ &=&\cos ^2\dfrac \theta 2\cos ^2\dfrac{\theta _1}2+\sin ^2\dfrac \theta
2\sin ^2\dfrac{\theta _1}2  \nonumber \\
&&\ \ \ +\frac 12\sin \theta \sin \theta _2\cos (\varphi _2-\varphi )
\label{hu108}
\end{eqnarray}
and 
\begin{eqnarray}
&&P(1,1^{(\widehat{{\bf a}})};(-\tfrac 12)^{(\widehat{{\bf c}}_1)})=P(1,1^{(%
\widehat{{\bf a}})};(+\tfrac 12)^{(\widehat{{\bf c}}_1)},(-\tfrac 12)^{(%
\widehat{{\bf c}}_2)})  \nonumber \\
&&+P(1,1^{(\widehat{{\bf a}})};(-\tfrac 12)^{(\widehat{{\bf c}}_1)},(-\tfrac
12)^{(\widehat{{\bf c}}_2)})  \nonumber \\
\ &=&\cos ^2\dfrac \theta 2\sin ^2\frac{\theta _2}2+\sin ^2\dfrac \theta
2\cos ^2\frac{\theta _2}2  \nonumber \\
&&\ \ \ -\frac 12\sin \theta \sin \theta _2\cos (\varphi _2-\varphi ).
\label{hu109}
\end{eqnarray}

\paragraph{The $M^{(\widehat{{\bf a}})}=0$ State}

For this case, the probabilities are

\begin{eqnarray}
&&\ P(1,0^{(\widehat{{\bf a}})};(+\tfrac 12)^{(\widehat{{\bf c}}%
_1)},(+\tfrac 12)^{(\widehat{{\bf c}}_2)})=\frac 12\sin ^2\theta \sin ^2%
\dfrac{\theta _1}2\sin ^2\frac{\theta _2}2  \nonumber \\
&&+\frac 12\sin ^2\theta \cos ^2\dfrac{\theta _1}2\cos ^2\frac{\theta _2}2 
\nonumber \\
&&\ \ +\frac 12\cos ^2\theta \cos ^2\dfrac{\theta _1}2\sin ^2\frac{\theta _2}%
2+\frac 12\cos ^2\theta \sin ^2\dfrac{\theta _1}2\cos ^2\frac{\theta _2}2 
\nonumber \\
&&\ \ -\frac 12\sin ^2\theta \sin \theta _1\sin \theta _2\cos (\varphi
-\varphi _1)\cos (\varphi -\varphi _2)  \nonumber \\
&&\ \ +\frac 14\sin \theta _1\sin \theta _2\cos (\varphi _2-\varphi _1) 
\nonumber \\
&&\ \ -\frac 12\sin \theta \cos \theta \sin \theta _1\cos \theta _2\cos
(\varphi _1-\varphi )  \nonumber \\
&&\ \ -\frac 12\sin \theta \cos \theta \sin \theta _2\cos \theta _1\cos
(\varphi _2-\varphi ),  \label{hu110}
\end{eqnarray}
\begin{eqnarray}
\ &&P(1,0^{(\widehat{{\bf a}})};(+\tfrac 12)^{(\widehat{{\bf c}}%
_1)},(-\tfrac 12)^{(\widehat{{\bf c}}_2)})=\frac 12\sin ^2\theta \cos ^2%
\dfrac{\theta _1}2\sin ^2\frac{\theta _2}2  \nonumber \\
&&+\frac 12\sin ^2\theta \sin ^2\dfrac{\theta _1}2\cos ^2\frac{\theta _2}2 
\nonumber \\
&&\ \ \ +\frac 12\cos ^2\theta \cos ^2\dfrac{\theta _1}2\cos ^2\frac{\theta
_2}2+\frac 12\cos ^2\theta \sin ^2\dfrac{\theta _1}2\sin ^2\frac{\theta _2}2
\nonumber \\
&&\ \ \ +\frac 12\sin ^2\theta \sin \theta _1\sin \theta _2\cos (\varphi
-\varphi _1)\cos (\varphi -\varphi _2)  \nonumber \\
&&\ \ \ -\frac 14\sin \theta _1\sin \theta _2\cos (\varphi _2-\varphi _1) 
\nonumber \\
&&\ \ \ +\frac 12\sin \theta \cos \theta \sin \theta _1\cos \theta _2\cos
(\varphi _1-\varphi )  \nonumber \\
&&\ \ \ +\frac 12\sin \theta \cos \theta \sin \theta _2\cos \theta _1\cos
(\varphi _2-\varphi ),  \label{hu111}
\end{eqnarray}

\begin{equation}
\ P(1,0^{(\widehat{{\bf a}})};(-\tfrac 12)^{(\widehat{{\bf c}}_1)},(+\tfrac
12)^{(\widehat{{\bf c}}_2)})=P(1,0^{(\widehat{{\bf a}})};(+\tfrac 12)^{(%
\widehat{{\bf c}}_1)},(-\tfrac 12)^{(\widehat{{\bf c}}_2)})  \label{hu112}
\end{equation}
and

$\ $%
\begin{equation}
P(1,0^{(\widehat{{\bf a}})};(-\tfrac 12)^{(\widehat{{\bf c}}_1)},(-\tfrac
12)^{(\widehat{{\bf c}}_2)})=P(1,0^{(\widehat{{\bf a}})};(+\tfrac 12)^{(%
\widehat{{\bf c}}_1)},(+\tfrac 12)^{(\widehat{{\bf c}}_2)}).  \label{hu113}
\end{equation}

These probabilities sum to unity. Also, we have

\begin{eqnarray}
P(1,0^{(\widehat{{\bf a}})};(+\tfrac 12)^{(\widehat{{\bf c}}_1)}) &=&P(1,0^{(%
\widehat{{\bf a}})};(-\tfrac 12)^{(\widehat{{\bf c}}_1)})  \nonumber \\
&=&P(1,0^{(\widehat{{\bf a}})};(+\tfrac 12)^{(\widehat{{\bf c}}_2)}) 
\nonumber \\
&=&P(1,0^{(\widehat{{\bf a}})};(-\tfrac 12)^{(\widehat{{\bf c}}_2)})=\frac
12.  \label{hu114}
\end{eqnarray}

\subsubsection{The $M^{(\widehat{{\bf a}})}=-1$ State}

For this case, the probabilities are 
\begin{eqnarray}
\  &&P(1,(-1)^{(\widehat{{\bf a}})};(+\tfrac 12)^{(\widehat{{\bf c}}%
_1)},(+\tfrac 12)^{(\widehat{{\bf c}}_2)})=\sin ^4\dfrac \theta 2\cos ^2%
\dfrac{\theta _1}2\cos ^2\frac{\theta _2}2  \nonumber \\
&&+\cos ^4\dfrac \theta 2\sin ^2\dfrac{\theta _1}2\sin ^2\frac{\theta _2}2 
\nonumber \\
&&\ \ \ \ +\frac 14\sin ^2\theta \cos ^2\dfrac{\theta _1}2\sin ^2\frac{%
\theta _2}2+\frac 14\sin ^2\theta \sin ^2\dfrac{\theta _1}2\cos ^2\frac{%
\theta _2}2  \nonumber \\
&&\ \ +\frac 14\sin ^2\theta \sin \theta _1\sin \theta _2\cos (\varphi
-\varphi _1)\cos (\varphi -\varphi _2)  \nonumber \\
&&\ \ -\frac 12\sin \theta \sin \theta _1[\sin ^2\dfrac \theta 2\cos ^2\frac{%
\theta _2}2+\cos ^2\dfrac \theta 2\sin ^2\frac{\theta _2}2]\cos (\varphi
_1-\varphi )  \nonumber \\
&&\ \ -\frac 12\sin \theta \sin \theta _2[\sin ^2\dfrac \theta 2\cos ^2%
\dfrac{\theta _1}2+\cos ^2\dfrac \theta 2\sin ^2\dfrac{\theta _1}2]\cos
(\varphi _2-\varphi ).  \nonumber \\
&&  \label{hu115}
\end{eqnarray}
\begin{eqnarray}
\  &&P(1,(-1)^{(\widehat{{\bf a}})};(+\tfrac 12)^{(\widehat{{\bf c}}%
_1)},(-\tfrac 12)^{(\widehat{{\bf c}}_2)})=\sin ^4\dfrac \theta 2\cos ^2%
\dfrac{\theta _1}2\sin ^2\dfrac{\theta _2}2  \nonumber \\
&&+\cos ^4\dfrac \theta 2\sin ^2\dfrac{\theta _1}2\cos ^2\dfrac{\theta _2}2 
\nonumber \\
&&\ \ \ \ \ +\frac 14\sin ^2\theta \sin ^2\dfrac{\theta _1}2\sin ^2\dfrac{%
\theta _2}2+\frac 14\sin ^2\theta \cos ^2\dfrac{\theta _1}2\cos ^2\dfrac{%
\theta _2}2  \nonumber \\
&&\ \ -\frac 14\sin ^2\theta \sin \theta _1\sin \theta _2\cos (\varphi
-\varphi _1)\cos (\varphi -\varphi _2)  \nonumber \\
&&\ \ \ -\frac 12\sin \theta \sin \theta _1[\sin ^2\dfrac \theta 2\sin ^2%
\dfrac{\theta _2}2+\cos ^2\dfrac \theta 2\cos ^2\dfrac{\theta _2}2]\cos
(\varphi _1-\varphi )  \nonumber \\
&&\ \ \ +\frac 12\sin \theta \sin \theta _2[\sin ^2\dfrac \theta 2\cos ^2%
\dfrac{\theta _1}2+\cos ^2\dfrac \theta 2\sin ^2\dfrac{\theta _1}2]\cos
(\varphi _2-\varphi ),  \nonumber \\
&&  \label{hu116}
\end{eqnarray}
\begin{eqnarray}
\  &&P(1,(-1)^{(\widehat{{\bf a}})};(-\tfrac 12)^{(\widehat{{\bf c}}%
_1)},(+\tfrac 12)^{(\widehat{{\bf c}}_2)})=\sin ^4\dfrac \theta 2\sin ^2%
\dfrac{\theta _1}2\cos ^2\dfrac{\theta _2}2  \nonumber \\
&&+\cos ^4\dfrac \theta 2\cos ^2\dfrac{\theta _1}2\sin ^2\dfrac{\theta _2}2 
\nonumber \\
&&\ \ \ \ \ \ +\frac 14\sin ^2\theta \sin ^2\dfrac{\theta _1}2\sin ^2\dfrac{%
\theta _2}2+\frac 14\sin ^2\theta \cos ^2\dfrac{\theta _1}2\cos ^2\dfrac{%
\theta _2}2  \nonumber \\
&&\ \ \ -\frac 14\sin ^2\theta \sin \theta _1\sin \theta _2\cos (\varphi
-\varphi _1)\cos (\varphi -\varphi _2)  \nonumber \\
&&\ \ \ \ +\frac 12\sin \theta \sin \theta _1[\sin ^2\dfrac \theta 2\cos ^2%
\frac{\theta _2}2+\cos ^2\dfrac \theta 2\sin ^2\frac{\theta _2}2]\cos
(\varphi _1-\varphi )  \nonumber \\
&&\ \ \ \ -\frac 12\sin \theta \sin \theta _2[\sin ^2\dfrac \theta 2\sin ^2%
\dfrac{\theta _1}2+\cos ^2\dfrac \theta 2\cos ^2\dfrac{\theta _1}2]\cos
(\varphi _2-\varphi )  \nonumber \\
&&  \label{hu117}
\end{eqnarray}
and

\begin{eqnarray}
\ &&P(1,(-1)^{(\widehat{{\bf a}})};(-\tfrac 12)^{(\widehat{{\bf c}}%
_1)},(-\tfrac 12)^{(\widehat{{\bf c}}_2)})=\sin ^4\dfrac \theta 2\sin ^2%
\dfrac{\theta _1}2\sin ^2\frac{\theta _2}2  \nonumber \\
&&+\cos ^4\dfrac \theta 2\cos ^2\dfrac{\theta _1}2\cos ^2\frac{\theta _2}2 
\nonumber \\
&&\ \ \ \ \ \ +\frac 14\sin ^2\theta \cos ^2\dfrac{\theta _1}2\sin ^2\frac{%
\theta _2}2+\frac 14\sin ^2\theta \sin ^2\dfrac{\theta _1}2\cos ^2\frac{%
\theta _2}2  \nonumber \\
&&\ \ \ +\frac 14\sin ^2\theta \sin \theta _1\sin \theta _2\cos (\varphi
-\varphi _1)\cos (\varphi -\varphi _2)  \nonumber \\
&&\ \ \ \ +\frac 12\sin \theta \sin \theta _1[\sin ^2\dfrac \theta 2\sin ^2%
\frac{\theta _2}2+\cos ^2\dfrac \theta 2\cos ^2\frac{\theta _2}2]\cos
(\varphi _1-\varphi )  \nonumber \\
&&\ \ \ \ +\frac 12\sin \theta \sin \theta _2[\sin ^2\dfrac \theta 2\sin ^2%
\dfrac{\theta _1}2+\cos ^2\dfrac \theta 2\cos ^2\dfrac{\theta _1}2]\cos
(\varphi _2-\varphi ).  \nonumber \\
&&  \label{hu118}
\end{eqnarray}
These probabilities add up to unity. We also have

\begin{eqnarray}
P(1,(-1)^{(\widehat{{\bf a}})};(+\tfrac 12)^{(\widehat{{\bf c}}_1)}) &=&\sin
^2\dfrac \theta 2\cos ^2\dfrac{\theta _1}2+\cos ^2\dfrac \theta 2\sin ^2%
\dfrac{\theta _1}2  \nonumber \\
&&-\frac 12\sin \theta \sin \theta _1\cos (\varphi _1-\varphi ),
\label{hu119}
\end{eqnarray}

\begin{eqnarray}
P(1,(-1)^{(\widehat{{\bf a}})};(-\tfrac 12)^{(\widehat{{\bf c}}_1)}) &=&\sin
^2\dfrac \theta 2\sin ^2\dfrac{\theta _1}2+\cos ^2\dfrac \theta 2\cos ^2%
\dfrac{\theta _1}2  \nonumber \\
&&+\frac 12\sin \theta \sin \theta _1\cos (\varphi _1-\varphi ),
\label{hu120}
\end{eqnarray}
\begin{eqnarray}
P(1,(-1)^{(\widehat{{\bf a}})};(+\tfrac 12)^{(\widehat{{\bf c}}_2)}) &=&\sin
^2\dfrac \theta 2^2\cos \frac{\theta _2}2+\cos ^2\dfrac \theta 2\sin ^2\frac{%
\theta _2}2  \nonumber \\
&&-\frac 12\sin \theta \sin \theta _2\cos (\varphi _2-\varphi )
\label{hu121}
\end{eqnarray}
and 
\begin{eqnarray}
P(1,(-1)^{(\widehat{{\bf a}})};(-\tfrac 12)^{(\widehat{{\bf c}}_2)}) &=&\sin
^2\dfrac \theta 2\sin ^2\frac{\theta _2}2+\cos ^2\dfrac \theta 2\cos ^2\frac{%
\theta _2}2  \nonumber \\
&&+\frac 12\sin \theta \sin \theta _2\cos (\varphi _2-\varphi ).
\label{hu122}
\end{eqnarray}

\section{Clebsch-Gordan Coefficients}

In this paper, we have assumed that the Clebsch-Gordan coefficients are
probability amplitudes for obtaining, starting from a given state of the
compound system, certain special combinations of spin projections for joint
measurements on the subsystems . Such a measurement is a special case of the
general measurement which we have treated in this paper. In the general
case, we are interested in the probability amplitudes of joint measurements
of the spin projections of the subsystems with respect to arbitrary
directions. The special case that yields the Clebsch-Gordan coefficients
arises in the following way.

The vectors $\widehat{{\bf a}}$, $\widehat{{\bf c}}_1$ and $\widehat{{\bf c}}%
_2$ define the situation obtaining. While $\widehat{{\bf a}}$ gives the
initial direction of the compound spin, $\widehat{{\bf c}}_1$ and $\widehat{%
{\bf c}}_2$ give the directions along which we measure the spin projections
of subsystems $1$ and $2$ respectively. The probability amplitudes for
measurements of these projections are given by the Clebsch-Gordan
coefficients if $\widehat{{\bf c}}_1$ defines the direction of spin ${\bf S}%
_1$ and $\widehat{{\bf c}}_2$ that of spin ${\bf S}_2$ in such a way that
the spin addition condition 
\begin{equation}
{\bf S}={\bf S}_1+{\bf S}_2  \label{hu400}
\end{equation}
is satisfied. But this is not all. In addition, $\widehat{{\bf a}}$ must be
oriented to point in the $z$ direction.

This immediately suggests a generalization of the Clebsch-Gordan

coefficients. If the direction $\widehat{{\bf a}}$ is made arbitrary, but
with the directions of $\widehat{{\bf c}}_1$ and $\widehat{{\bf c}}_2$ still
such that the condition Eq. (\ref{hu400}) holds, then we obtain joint
probability amplitudes which are generalized forms of the Clebsch-Gordan
coefficients.

For $s=0$ or $s=1$, as in this paper, the generalized Clebsch-Gordan
coefficients can be deduced from the joint probability amplitudes if we set $%
\widehat{{\bf c}}_1=\widehat{{\bf c}}_2=\widehat{{\bf a}}.$ If in addition,
we set $\widehat{{\bf a}}=\widehat{{\bf k}}$, so that $\theta =\varphi =0$,
we obtain the standard Clebsch-Gordan coefficients. Though the systems in
this paper are such that the vectors $\widehat{{\bf a}}$, $\widehat{{\bf c}}%
_1$ and $\widehat{{\bf c}}_2$ are collinear for the situation that yields
the Clebsch-Gordan coefficients, this will clearly not be the case for an
arbitrary system. The exact mutual orientations of the vectors will have to
be deduced on a system-by-system basis. However, in all cases, the
generalized Clebsch-Gordan coefficients can be obtained from the standard
ones by a rotation that carries the vector $\widehat{{\bf k}}$ into the
vector $\widehat{{\bf a}}$. Also, the probabilities resulting from
generalized Clebsch-Gordan coefficients will always equal those resulting
from the standard Clebsch-Gordan coefficients.

We denote the generalized Clebsch-Gordan coefficients by $%
C(s_1s_2s;m_1m_2M)_{\text{gen}}$. We obtain them from Eqs. (\ref{fi54})-(\ref
{fi57}), Eqs. (\ref{ei84}) -(\ref{ei87}), Eqs. (\ref{ni91}) - (\ref{ni94})
and Eqs. (\ref{ni98}) - (\ref{hu101}) by setting $\theta _1=\theta _2=\theta 
$ and $\varphi _1=\varphi _2=\varphi .$

For the singlet state, they are

\begin{equation}
C(\tfrac 12\tfrac 120;\tfrac 12\tfrac 120)_{\text{gen}}=\Psi (0,0^{(\widehat{%
{\bf a}})};(+\tfrac 12)^{(\widehat{{\bf a}})},(+\tfrac 12)^{(\widehat{{\bf a}%
})})=0,  \label{hu401}
\end{equation}

\begin{equation}
C(\tfrac 12\tfrac 120;\tfrac 12,-\tfrac 120)_{\text{gen}}=\Psi (0,0^{(%
\widehat{{\bf a}})};(+\tfrac 12)^{(\widehat{{\bf a}})},(-\tfrac 12)^{(%
\widehat{{\bf a}})})=\frac 1{\sqrt{2}}e^{-i\varphi },  \label{hu402}
\end{equation}

\begin{equation}
C(\tfrac 12\tfrac 120;-\tfrac 12\tfrac 120)_{\text{gen}}=\Psi (0,0^{(%
\widehat{{\bf a}})};(-\tfrac 12)^{(\widehat{{\bf a}})},(+\tfrac 12)^{(%
\widehat{{\bf a}})})=-\frac 1{\sqrt{2}}e^{-i\varphi }  \label{hu403}
\end{equation}
and 
\begin{equation}
C(\tfrac 12\tfrac 120;-\tfrac 12,-\tfrac 12,0)_{\text{gen}}=\Psi (0,0^{(%
\widehat{{\bf a}})};(-\tfrac 12)^{(\widehat{{\bf a}})},(-\tfrac 12)^{(%
\widehat{{\bf a}})})=0.  \label{hu404}
\end{equation}

For the case $s=1$, the generalized Clebsch-Gordan coefficients for $M^{(%
\widehat{{\bf a}})}=1$ are

\begin{equation}
C(\tfrac 12\tfrac 121;\tfrac 12\tfrac 121)_{\text{gen}}=\Psi (1,1^{(\widehat{%
{\bf a}})};(+\tfrac 12)^{(\widehat{{\bf a}})},(+\tfrac 12)^{(\widehat{{\bf a}%
})})=e^{-i\varphi },  \label{hu405}
\end{equation}
\begin{equation}
C(\tfrac 12\tfrac 121;\tfrac 12,-\tfrac 121)_{\text{gen}}=\Psi (1,1^{(%
\widehat{{\bf a}})};(+\tfrac 12)^{(\widehat{{\bf a}})},(-\tfrac 12)^{(%
\widehat{{\bf a}})})=0,  \label{hu406}
\end{equation}
\begin{equation}
C(\tfrac 12\tfrac 121;-\tfrac 12\tfrac 121)_{\text{gen}}=\Psi (1,1^{(%
\widehat{{\bf a}})};(-\tfrac 12)^{(\widehat{{\bf a}})},(+\tfrac 12)^{(%
\widehat{{\bf a}})})=0  \label{hu407}
\end{equation}
and 
\begin{equation}
C(\tfrac 12\tfrac 121;-\tfrac 12,-\tfrac 12,1)_{\text{gen}}=\Psi (1,1^{(%
\widehat{{\bf a}})};(-\tfrac 12)^{(\widehat{{\bf a}})},(-\tfrac 12)^{(%
\widehat{{\bf a}})})=0.  \label{hu408}
\end{equation}

For $s=1$, $M^{(\widehat{{\bf a}})}=0$, they are

\begin{equation}
C(\tfrac 12\tfrac 121;\tfrac 12\tfrac 120)_{\text{gen}}=\Psi (1,0^{(\widehat{%
{\bf a}})};(+\tfrac 12)^{(\widehat{{\bf a}})},(+\tfrac 12)^{(\widehat{{\bf a}%
})})=e^{-i\varphi },  \label{hu409}
\end{equation}
\begin{equation}
C(\tfrac 12\tfrac 121;\tfrac 12,-\tfrac 120)_{\text{gen}}=\Psi (1,0^{(%
\widehat{{\bf a}})};(+\tfrac 12)^{(\widehat{{\bf a}})},(-\tfrac 12)^{(%
\widehat{{\bf a}})})=\frac 1{\sqrt{2}}e^{-i\varphi },  \label{hu410}
\end{equation}
\begin{equation}
C(\tfrac 12\tfrac 121;-\tfrac 12\tfrac 120)_{\text{gen}}=\Psi (1,0^{(%
\widehat{{\bf a}})};(-\tfrac 12)^{(\widehat{{\bf a}})},(+\tfrac 12)^{(%
\widehat{{\bf a}})})=\frac 1{\sqrt{2}}e^{-i\varphi }  \label{hu411}
\end{equation}
and 
\begin{equation}
C(\tfrac 12\tfrac 121;-\tfrac 12,-\tfrac 120)_{\text{gen}}=\Psi (1,0^{(%
\widehat{{\bf a}})};(-\tfrac 12)^{(\widehat{{\bf a}})},(-\tfrac 12)^{(%
\widehat{{\bf a}})})=0.  \label{hu412}
\end{equation}

Finally, for $s=1,$ $M^{(\widehat{{\bf a}})}=-1$, the generalized
Clebsch-Gordan coefficients are

\begin{equation}
C(\tfrac 12\tfrac 121;\tfrac 12\tfrac 12-1)_{\text{gen}}=\Psi (1,(-1)^{(%
\widehat{{\bf a}})};(+\tfrac 12)^{(\widehat{{\bf a}})},(+\tfrac 12)^{(%
\widehat{{\bf a}})})=0,  \label{hu413}
\end{equation}
\begin{equation}
C(\tfrac 12\tfrac 121;\tfrac 12,-\tfrac 12,-1)_{\text{gen}}=\Psi (1,(-1)^{(%
\widehat{{\bf a}})};(+\tfrac 12)^{(\widehat{{\bf a}})},(-\tfrac 12)^{(%
\widehat{{\bf a}})})=0,  \label{hu414}
\end{equation}
\begin{equation}
C(\tfrac 12\tfrac 121;-\tfrac 12\tfrac 12,-1)_{\text{gen}}=\Psi (1,(-1)^{(%
\widehat{{\bf a}})};(-\tfrac 12)^{(\widehat{{\bf a}})},(+\tfrac 12)^{(%
\widehat{{\bf a}})})=0  \label{hu415}
\end{equation}
and 
\begin{equation}
C(\tfrac 12\tfrac 121;-\tfrac 12,-\tfrac 12,-1)_{\text{gen}}=\Psi (1,(-1)^{(%
\widehat{{\bf a}})};(-\tfrac 12)^{(\widehat{{\bf a}})},(-\tfrac 12)^{(%
\widehat{{\bf a}})})=-e^{-i\varphi }.  \label{hu416}
\end{equation}

Thus, we see that we do indeed obtain quantities that we may justly term
generalized Clebsch-Gordan coefficients. The simple appearance of these
expressions in this case is probably due to the fact that ${\bf S}_1$, ${\bf %
S}_2$ and ${\bf S}$ always lie along one line in the cases dealt with here.
More complicated expressions ought to arise from those cases where the three
angular momenta actually form a triangle. But this needs to be investigated
fully by actually obtaining the probability amplitudes for such cases.

\section{Matrix Mechanics}

\subsection{Introductory Remarks}

We obtain a matrix description from a probability-amplitude description by
means of the expansion Eq. (\ref{one1}) of the probability amplitudes. In
order to obtain the triplet-state probability amplitudes $\Psi (1,1^{(%
\widehat{{\bf a}})};(m_1)_u^{(\widehat{{\bf c}}_1)},(m_2)_v^{(\widehat{{\bf c%
}}_2)})$, we have had to utilize two expansions, which means that we can
have at least two matrix descriptions of these probability amplitudes and of
the expectation values we can calculate through them. In order to obtain the
probability amplitudes $\Psi (0,0^{(\widehat{{\bf a}})};(m_1)_u^{(\widehat{%
{\bf c}}_1)},(m_2)_v^{(\widehat{{\bf c}}_2)})$ for the singlet state, we
have employed only one expansion, which gives only one way of casting
expectation values involving this state into matrix form. We start off with
this case.

\subsection{Vectors and Operators for the Singlet State}

For this case, the expansion of the probability amplitudes connecting the
initial state with the possible final states is

\begin{eqnarray}
&&\Psi (0,0^{(\widehat{{\bf a}})};(m_1)_u^{(\widehat{{\bf c}}_1)},(m_2)_v^{(%
\widehat{{\bf c}}_2)})=\sum_{\alpha ,\alpha ^{\prime }}\eta (0,0^{(\widehat{%
{\bf k}})};(m_1)_\alpha ^{(\widehat{{\bf k}})},(m_2)_{\alpha ^{\prime }}^{(%
\widehat{{\bf k}})})  \nonumber \\
&&\times \psi ((m_1)_\alpha ^{(\widehat{{\bf k}})},(m_2)_{\alpha ^{\prime
}}^{(\widehat{{\bf k}})};(m_1)_u^{(\widehat{{\bf c}}_1)},(m_2)_v^{(\widehat{%
{\bf c}}_2)}).  \label{hu123}
\end{eqnarray}

The expectation value of $R=R((m_1)^{(\widehat{{\bf c}}_1)},(m_2)^{(\widehat{%
{\bf c}}_2)})$ is

\begin{eqnarray}
&&\left\langle R((m_1)^{(\widehat{{\bf c}}_1)},(m_2)^{(\widehat{{\bf c}}%
_2)})\right\rangle =\sum_{u,v}\left| \Psi (0,0^{(\widehat{{\bf a}}%
)};(m_1)_u^{(\widehat{{\bf c}}_1)},(m_2)_v^{(\widehat{{\bf c}}_2)})\right| ^2
\nonumber \\
&&\times R((m_1)_u^{(\widehat{{\bf c}}_1)},(m_2)_v^{(\widehat{{\bf c}}_2)}).
\label{hu124}
\end{eqnarray}

In order to facilitate the transition to matrix mechanics, we define a new
observable corresponding to combined values of $((m_1)^{(\widehat{{\bf k}}%
)},(m_2)^{(\widehat{{\bf k}})}).$ This is denoted by $B$. Hence $B_\gamma
=((m_1)^{(\widehat{{\bf k}})},(m_2)^{(\widehat{{\bf k}})})_\gamma $, where $%
\gamma =1,2,3,4.$ Hence we have

\begin{equation}
B_1=(m_1)_1^{(\widehat{{\bf k}})},(m_2)_1^{(\widehat{{\bf k}})})=((+\frac
12)^{(\widehat{{\bf k}})},(+\frac 12)^{(\widehat{{\bf k}})})),  \label{fi53h}
\end{equation}
\begin{equation}
B_2=(m_1)_1^{(\widehat{{\bf k}})},(m_2)_2^{(\widehat{{\bf k}})})=((+\frac
12)^{(\widehat{{\bf k}})},(-\frac 12)^{(\widehat{{\bf k}})})),  \label{fi53i}
\end{equation}
\begin{equation}
B_3=(m_1)_2^{(\widehat{{\bf k}})},(m_2)_1^{(\widehat{{\bf k}})})=((-\frac
12)^{(\widehat{{\bf k}})},(+\frac 12)^{(\widehat{{\bf k}})}))  \label{fi53j}
\end{equation}
and 
\begin{equation}
B_4=(m_1)_2^{(\widehat{{\bf k}})},(m_2)_2^{(\widehat{{\bf k}})})=((-\frac
12)^{(\widehat{{\bf k}})},(-\frac 12)^{(\widehat{{\bf k}})})).  \label{fi53k}
\end{equation}

Thus, the probability amplitude Eq. (\ref{hu123}) becomes

\[
\Psi ^{*}(0,0^{(\widehat{{\bf a}})};(m_1)_u^{(\widehat{{\bf c}}%
_1)},(m_2)_v^{(\widehat{{\bf c}}_2)})=\sum_\gamma \eta ^{*}(0,0^{(\widehat{%
{\bf k}})};B_\gamma )\psi ^{*}(B_\gamma ;(m_1)_u^{(\widehat{{\bf c}}%
_1)},(m_2)_v^{(\widehat{{\bf c}}_2)}) 
\]
and

\[
\Psi (0,0^{(\widehat{{\bf a}})};(m_1)_u^{(\widehat{{\bf c}}_1)},(m_2)_v^{(%
\widehat{{\bf c}}_2)})=\sum_{\gamma ^{\prime }}\eta (0,0^{(\widehat{{\bf k}}%
)};B_{\gamma ^{\prime }})\psi (B_{\gamma ^{\prime }};(m_1)_u^{(\widehat{{\bf %
c}}_1)},(m_2)_v^{(\widehat{{\bf c}}_2)}) 
\]

The expectation value becomes 
\begin{equation}
\left\langle R((m_1)^{(\widehat{{\bf c}}_1)},(m_2)^{(\widehat{{\bf c}}%
_2)})\right\rangle =\sum_{\gamma ,\gamma ^{^{\prime }}}\eta ^{*}(0,0^{(%
\widehat{{\bf k}})};B_\gamma )R_{\gamma \gamma ^{^{\prime }}}\eta (0,0^{(%
\widehat{{\bf k}})};B_{\gamma ^{\prime }}),  \label{hu126}
\end{equation}
where

\begin{eqnarray}
&&R_{\gamma \gamma ^{^{\prime }}}=\sum_{u,v=1}^2\psi ^{*}(B_\gamma
;(m_1)_u^{(\widehat{{\bf c}}_1)},(m_2)_v^{(\widehat{{\bf c}}_2)})R((m_1)_u^{(%
\widehat{{\bf c}}_1)},(m_2)_v^{(\widehat{{\bf c}}_2)})  \nonumber \\
&&\times \psi (B_{\gamma ^{\prime }};(m_1)_u^{(\widehat{{\bf c}}%
_1)},(m_2)_v^{(\widehat{{\bf c}}_2)}).  \label{hu127}
\end{eqnarray}

Therefore 
\begin{eqnarray}
&&\left\langle R((m_1)^{(\widehat{{\bf c}}_1)},(m_2)^{(\widehat{{\bf c}}%
_2)}))\right\rangle =[\Psi (0,0^{(\widehat{{\bf a}})};(m_1)^{(\widehat{{\bf c%
}}_1)},(m_2)^{(\widehat{{\bf c}}_2)})]^{\dagger }[R]  \nonumber \\
&&\times [\Psi (0,0^{(\widehat{{\bf a}})};(m_1)^{(\widehat{{\bf c}}%
_1)},(m_2)^{(\widehat{{\bf c}}_2)})],  \label{hu128}
\end{eqnarray}
where

\begin{eqnarray}
\lbrack \Psi (0,0^{(\widehat{{\bf a}})};(m_1)^{(\widehat{{\bf c}}%
_1)},(m_2)^{(\widehat{{\bf c}}_2)})] &=&\left( 
\begin{array}{c}
\eta (0,0^{(\widehat{{\bf k}})};B_1) \\ 
\eta (0,0^{(\widehat{{\bf k}})};B_2) \\ 
\eta (0,0^{(\widehat{{\bf k}})};B_3) \\ 
\eta (0,0^{(\widehat{{\bf k}})};B_4)
\end{array}
\right)  \nonumber \\
&=&\left( 
\begin{array}{c}
\eta (0,0^{(\widehat{{\bf k}})};(+\frac 12)^{(\widehat{{\bf k}})},(+\frac
12)^{(\widehat{{\bf k}})}) \\ 
\eta (0,0^{(\widehat{{\bf k}})};(+\frac 12)^{(\widehat{{\bf k}})},(-\frac
12)^{(\widehat{{\bf k}})}) \\ 
\eta (0,0^{(\widehat{{\bf k}})};(-\frac 12)^{(\widehat{{\bf k}})},(+\frac
12)^{(\widehat{{\bf k}})}) \\ 
\eta (0,0^{(\widehat{{\bf k}})};(-\frac 12)^{(\widehat{{\bf k}})},(-\frac
12)^{(\widehat{{\bf k}})})
\end{array}
\right)  \label{hu129}
\end{eqnarray}
and 
\begin{equation}
\lbrack R]=\left( 
\begin{array}{cccc}
R_{11} & R_{12} & R_{13} & R_{14} \\ 
R_{21} & R_{22} & R_{23} & R_{24} \\ 
R_{31} & R_{32} & R_{33} & R_{34} \\ 
R_{41} & R_{42} & R_{43} & R_{44}
\end{array}
\right) .  \label{hu130}
\end{equation}

As the $\eta $'s are the Clebsch-Gordan coefficients, Eqs. (\ref{fi53c}) - (%
\ref{fi53f}), we have

\begin{equation}
\lbrack \Psi (0,0^{(\widehat{{\bf a}})};(m_1)^{(\widehat{{\bf c}}%
_1)},(m_2)^{(\widehat{{\bf c}}_2)})]=\left( 
\begin{array}{c}
0 \\ 
\frac 1{\sqrt{2}} \\ 
-\frac 1{\sqrt{2}} \\ 
0
\end{array}
\right) .  \label{hu131}
\end{equation}

The elements of $[R]$ are as follows.

\begin{eqnarray}
R_{11} &=&\left| \psi ((+\tfrac 12)^{(\widehat{{\bf k}})},(+\tfrac 12)^{(%
\widehat{{\bf k}})};(+\tfrac 12)^{(\widehat{{\bf c}}_1)},(+\tfrac 12)^{(%
\widehat{{\bf c}}_2)})\right| ^2R((+\tfrac 12)^{(\widehat{{\bf c}}%
_1)},(+\tfrac 12)^{(\widehat{{\bf c}}_2)})  \nonumber \\
&&+\left| \psi ((+\tfrac 12)^{(\widehat{{\bf k}})},(+\tfrac 12)^{(\widehat{%
{\bf k}})};(+\tfrac 12)^{(\widehat{{\bf c}}_1)},(-\tfrac 12)^{(\widehat{{\bf %
c}}_2)})\right| ^2R((+\tfrac 12)^{(\widehat{{\bf c}}_1)},(-\tfrac 12)^{(%
\widehat{{\bf c}}_2)})  \nonumber \\
&&+\left| \psi ((+\tfrac 12)^{(\widehat{{\bf k}})},(+\tfrac 12)^{(\widehat{%
{\bf k}})};(-\tfrac 12)^{(\widehat{{\bf c}}_1)},(+\tfrac 12)^{(\widehat{{\bf %
c}}_2)})\right| ^2R((-\tfrac 12)^{(\widehat{{\bf c}}_1)},(+\tfrac 12)^{(%
\widehat{{\bf c}}_2)})  \nonumber \\
&&+\left| \psi ((+\tfrac 12)^{(\widehat{{\bf k}})},(+\tfrac 12)^{(\widehat{%
{\bf k}})};(-\tfrac 12)^{(\widehat{{\bf c}}_1)},(-\tfrac 12)^{(\widehat{{\bf %
c}}_2)})\right| ^2R((-\tfrac 12)^{(\widehat{{\bf c}}_1)},(-\tfrac 12)^{(%
\widehat{{\bf c}}_2)})  \nonumber \\
&=&\cos ^2\dfrac{\theta _1}2\cos ^2\dfrac{\theta _2}2R((+\tfrac 12)^{(%
\widehat{{\bf c}}_1)},(+\tfrac 12)^{(\widehat{{\bf c}}_2)})  \nonumber \\
&&+\cos ^2\dfrac{\theta _1}2\sin ^2\dfrac{\theta _2}2R((+\tfrac 12)^{(%
\widehat{{\bf c}}_1)},(-\tfrac 12)^{(\widehat{{\bf c}}_2)})  \nonumber \\
&&+\sin ^2\dfrac{\theta _1}2\cos ^2\dfrac{\theta _2}2R((-\tfrac 12)^{(%
\widehat{{\bf c}}_1)},(+\tfrac 12)^{(\widehat{{\bf c}}_2)})  \nonumber \\
&&+\sin ^2\dfrac{\theta _1}2\sin ^2\dfrac{\theta _2}2R((-\tfrac 12)^{(%
\widehat{{\bf c}}_1)},(-\tfrac 12)^{(\widehat{{\bf c}}_2)}),  \label{hu132}
\end{eqnarray}

By the same token, 
\begin{eqnarray}
R_{12} &=&\psi ^{*}((+\tfrac 12)^{(\widehat{{\bf k}})},(+\tfrac 12)^{(%
\widehat{{\bf k}})};(+\tfrac 12)^{(\widehat{{\bf c}}_1)},(+\tfrac 12)^{(%
\widehat{{\bf c}}_2)})  \nonumber \\
&&\times \psi ((+\tfrac 12)^{(\widehat{{\bf k}})},(-\tfrac 12)^{(\widehat{%
{\bf k}})};(+\tfrac 12)^{(\widehat{{\bf c}}_1)},(+\tfrac 12)^{(\widehat{{\bf %
c}}_2)})R((+\tfrac 12)^{(\widehat{{\bf c}}_1)},(+\tfrac 12)^{(\widehat{{\bf c%
}}_2)})  \nonumber \\
&&+\psi ^{*}((+\tfrac 12)^{(\widehat{{\bf k}})},(+\tfrac 12)^{(\widehat{{\bf %
k}})};(+\tfrac 12)^{(\widehat{{\bf c}}_1)},(-\tfrac 12)^{(\widehat{{\bf c}}%
_2)})  \nonumber \\
&&\times \psi ((+\tfrac 12)^{(\widehat{{\bf k}})},(-\tfrac 12)^{(\widehat{%
{\bf k}})};(+\tfrac 12)^{(\widehat{{\bf c}}_1)},(-\tfrac 12)^{(\widehat{{\bf %
c}}_2)})R((+\tfrac 12)^{(\widehat{{\bf c}}_1)},(-\tfrac 12)^{(\widehat{{\bf c%
}}_2)})  \nonumber \\
&&+\psi ^{*}((+\tfrac 12)^{(\widehat{{\bf k}})},(+\tfrac 12)^{(\widehat{{\bf %
k}})};(-\tfrac 12)^{(\widehat{{\bf c}}_1)},(+\tfrac 12)^{(\widehat{{\bf c}}%
_2)})  \nonumber \\
&&\times \psi ((+\tfrac 12)^{(\widehat{{\bf k}})},(-\tfrac 12)^{(\widehat{%
{\bf k}})};(-\tfrac 12)^{(\widehat{{\bf c}}_1)},(+\tfrac 12)^{(\widehat{{\bf %
c}}_2)})R((-\tfrac 12)^{(\widehat{{\bf c}}_1)},(+\tfrac 12)^{(\widehat{{\bf c%
}}_2)})  \nonumber \\
&&+\psi ^{*}((+\tfrac 12)^{(\widehat{{\bf k}})},(+\tfrac 12)^{(\widehat{{\bf %
k}})};(-\tfrac 12)^{(\widehat{{\bf c}}_1)},(-\tfrac 12)^{(\widehat{{\bf c}}%
_2)})  \nonumber \\
&&\times \psi ((+\tfrac 12)^{(\widehat{{\bf k}})},(-\tfrac 12)^{(\widehat{%
{\bf k}})};(-\tfrac 12)^{(\widehat{{\bf c}}_1)},(-\tfrac 12)^{(\widehat{{\bf %
c}}_2)})R((-\tfrac 12)^{(\widehat{{\bf c}}_1)},(-\tfrac 12)^{(\widehat{{\bf c%
}}_2)}).  \nonumber \\
&&  \label{hu133}
\end{eqnarray}

Thus, 
\begin{eqnarray}
R_{12} &=&\frac 12\cos ^2\dfrac{\theta _1}2\sin \theta _2e^{-i\varphi
_2}[R((+\tfrac 12)^{(\widehat{{\bf c}}_1)},(+\tfrac 12)^{(\widehat{{\bf c}}%
_2)})-R((+\tfrac 12)^{(\widehat{{\bf c}}_1)},(-\tfrac 12)^{(\widehat{{\bf c}}%
_2)})]  \nonumber \\
&&+\frac 12\sin ^2\dfrac{\theta _1}2\sin \theta _2e^{-i\varphi
_2}[R((-\tfrac 12)^{(\widehat{{\bf c}}_1)},(+\tfrac 12)^{(\widehat{{\bf c}}%
_2)})-R((-\tfrac 12)^{(\widehat{{\bf c}}_1)},(-\tfrac 12)^{(\widehat{{\bf c}}%
_2)})],  \nonumber \\
&&  \label{hu133z}
\end{eqnarray}

\begin{eqnarray}
R_{13} &=&\frac 12\cos ^2\dfrac{\theta _2}2\sin \theta _1e^{-i\varphi
_1}[R((+\tfrac 12)^{(\widehat{{\bf c}}_1)},(+\tfrac 12)^{(\widehat{{\bf c}}%
_2)})-R((-\tfrac 12)^{(\widehat{{\bf c}}_1)},(+\tfrac 12)^{(\widehat{{\bf c}}%
_2)})]  \nonumber \\
&&+\frac 12\sin ^2\dfrac{\theta _2}2\sin \theta _1e^{-i\varphi
_1}[R((+\tfrac 12)^{(\widehat{{\bf c}}_1)},(-\tfrac 12)^{(\widehat{{\bf c}}%
_2)})-R((-\tfrac 12)^{(\widehat{{\bf c}}_1)},(-\tfrac 12)^{(\widehat{{\bf c}}%
_2)})],  \nonumber \\
&&  \label{hu134}
\end{eqnarray}

\begin{eqnarray}
R_{14} &=&\frac 14\sin \theta _1\sin \theta _2e^{-i(\varphi _1+\varphi
_2)}[R((+\tfrac 12)^{(\widehat{{\bf c}}_1)},(+\tfrac 12)^{(\widehat{{\bf c}}%
_2)})-R((+\tfrac 12)^{(\widehat{{\bf c}}_1)},(-\tfrac 12)^{(\widehat{{\bf c}}%
_2)})  \nonumber \\
&&-R((-\tfrac 12)^{(\widehat{{\bf c}}_1)},(+\tfrac 12)^{(\widehat{{\bf c}}%
_2)})+R((-\tfrac 12)^{(\widehat{{\bf c}}_1)},(-\tfrac 12)^{(\widehat{{\bf c}}%
_2)})],  \label{hu135}
\end{eqnarray}

\begin{equation}
R_{21}=R_{12}^{*},  \label{hu136}
\end{equation}

\begin{eqnarray}
R_{22} &=&\cos ^2\dfrac{\theta _1}2\sin ^2\dfrac{\theta _2}2R((+\tfrac 12)^{(%
\widehat{{\bf c}}_1)},(+\tfrac 12)^{(\widehat{{\bf c}}_2)})  \nonumber \\
&&+\cos ^2\dfrac{\theta _1}2\cos ^2\dfrac{\theta _2}2R((+\tfrac 12)^{(%
\widehat{{\bf c}}_1)},(-\tfrac 12)^{(\widehat{{\bf c}}_2)})  \nonumber \\
&&+\sin ^2\dfrac{\theta _1}2\sin ^2\dfrac{\theta _2}2R((-\tfrac 12)^{(%
\widehat{{\bf c}}_1)},(+\tfrac 12)^{(\widehat{{\bf c}}_2)})  \nonumber \\
&&+\cos ^2\dfrac{\theta _2}2\sin ^2\dfrac{\theta _1}2R((-\tfrac 12)^{(%
\widehat{{\bf c}}_1)},(-\tfrac 12)^{(\widehat{{\bf c}}_2)}),  \label{hu137}
\end{eqnarray}

\begin{eqnarray}
R_{23} &=&\frac 14\sin \theta _1\sin \theta _2e^{i(\varphi _2-\varphi
_1)}[R((+\tfrac 12)^{(\widehat{{\bf c}}_1)},(+\tfrac 12)^{(\widehat{{\bf c}}%
_2)})-R((+\tfrac 12)^{(\widehat{{\bf c}}_1)},(-\tfrac 12)^{(\widehat{{\bf c}}%
_2)})  \nonumber \\
&&-R((-\tfrac 12)^{(\widehat{{\bf c}}_1)},(+\tfrac 12)^{(\widehat{{\bf c}}%
_2)})+R((-\tfrac 12)^{(\widehat{{\bf c}}_1)},(-\tfrac 12)^{(\widehat{{\bf c}}%
_2)})],  \label{hu138}
\end{eqnarray}

\begin{eqnarray}
R_{24} &=&\frac 12\sin ^2\dfrac{\theta _2}2\sin \theta _1e^{-i\varphi
_1}[R((+\tfrac 12)^{(\widehat{{\bf c}}_1)},(+\tfrac 12)^{(\widehat{{\bf c}}%
_2)})-R((-\tfrac 12)^{(\widehat{{\bf c}}_1)},(+\tfrac 12)^{(\widehat{{\bf c}}%
_2)})]  \nonumber \\
&&+\frac 12\cos ^2\dfrac{\theta _2}2\sin \theta _1e^{-i\varphi
_1}[R((+\tfrac 12)^{(\widehat{{\bf c}}_1)},(-\tfrac 12)^{(\widehat{{\bf c}}%
_2)})-R((-\tfrac 12)^{(\widehat{{\bf c}}_1)},(-\tfrac 12)^{(\widehat{{\bf c}}%
_2)})],  \nonumber \\
&&  \label{hu139}
\end{eqnarray}

\begin{equation}
R_{31}=R_{13}^{*},  \label{hu140}
\end{equation}

\begin{equation}
R_{32}=R_{23}^{*},  \label{hu141}
\end{equation}

\begin{eqnarray}
R_{33} &=&\sin ^2\dfrac{\theta _1}2\cos ^2\dfrac{\theta _2}2R((+\tfrac 12)^{(%
\widehat{{\bf c}}_1)},(+\tfrac 12)^{(\widehat{{\bf c}}_2)})  \nonumber \\
&&+\sin ^2\dfrac{\theta _1}2\sin ^2\dfrac{\theta _2}2R((+\tfrac 12)^{(%
\widehat{{\bf c}}_1)},(-\tfrac 12)^{(\widehat{{\bf c}}_2)})  \nonumber \\
&&+\cos ^2\dfrac{\theta _1}2\cos ^2\dfrac{\theta _2}2R((-\tfrac 12)^{(%
\widehat{{\bf c}}_1)},(+\tfrac 12)^{(\widehat{{\bf c}}_2)})  \nonumber \\
&&+\cos ^2\dfrac{\theta _1}2\sin ^2\dfrac{\theta _2}2R((-\tfrac 12)^{(%
\widehat{{\bf c}}_1)},(-\tfrac 12)^{(\widehat{{\bf c}}_2)})  \label{hu142}
\end{eqnarray}

\begin{eqnarray}
R_{34} &=&\frac 12\sin ^2\frac{\theta _1}2\sin \theta _2e^{-i\varphi
_2}[R((+\tfrac 12)^{(\widehat{{\bf c}}_1)},(+\tfrac 12)^{(\widehat{{\bf c}}%
_2)})-R((+\tfrac 12)^{(\widehat{{\bf c}}_1)},(-\tfrac 12)^{(\widehat{{\bf c}}%
_2)})]  \nonumber \\
&&+\frac 12\cos ^2\frac{\theta _1}2\sin \theta _2e^{-i\varphi _2}[R((-\tfrac
12)^{(\widehat{{\bf c}}_1)},(+\tfrac 12)^{(\widehat{{\bf c}}_2)})-R((-\tfrac
12)^{(\widehat{{\bf c}}_1)},(-\tfrac 12)^{(\widehat{{\bf c}}_2)})]  \nonumber
\\
&&  \label{hu143}
\end{eqnarray}

\begin{equation}
R_{41}=R_{14}^{*},  \label{hu144}
\end{equation}

\begin{equation}
R_{42}=R_{24}^{*},  \label{hu145}
\end{equation}

\begin{equation}
R_{43}=R_{34}^{*}  \label{hu146}
\end{equation}
and

\begin{eqnarray}
R_{44} &=&\sin ^2\dfrac{\theta _1}2\sin ^2\dfrac{\theta _2}2R((+\tfrac 12)^{(%
\widehat{{\bf c}}_1)},(+\tfrac 12)^{(\widehat{{\bf c}}_2)})  \nonumber \\
&&+\sin ^2\dfrac{\theta _1}2\cos ^2\dfrac{\theta _2}2R((+\tfrac 12)^{(%
\widehat{{\bf c}}_1)},(-\tfrac 12)^{(\widehat{{\bf c}}_2)})  \nonumber \\
&&+\cos ^2\dfrac{\theta _1}2\sin ^2\dfrac{\theta _2}2R((-\tfrac 12)^{(%
\widehat{{\bf c}}_1)},(+\tfrac 12)^{(\widehat{{\bf c}}_2)})  \nonumber \\
&&+\cos ^2\dfrac{\theta _1}2\cos ^2\dfrac{\theta _2}2R((-\tfrac 12)^{(%
\widehat{{\bf c}}_1)},(-\tfrac 12)^{(\widehat{{\bf c}}_2)}).  \label{hu147}
\end{eqnarray}

The formulas given here correspond to the general results given in Section
3.2.

\subsection{The Triplet State}

\subsubsection{ Vectors and Operators in Four-Dimensional Treatment}

As observed above, the case $s=1$ admits of two different matrix
representations because the calculation of the amplitudes involves two
different expansions. We shall give both.

The probability amplitudes Eqs. (\ref{tw21}) can be written in the form

\begin{eqnarray}
&&\Psi (1,M_i^{(\widehat{{\bf a}})};(m_1)_u^{(\widehat{{\bf c}}%
_1)},(m_2)_v^{(\widehat{{\bf c}}_2)})=\sum_{\alpha ,\alpha ^{\prime }}\beta
(1,M_i^{(\widehat{{\bf a}})};(m_1)_\alpha ^{(\widehat{{\bf k}}%
)},(m_2)_{\alpha ^{\prime }}^{(\widehat{{\bf k}})})  \nonumber \\
&&\times \psi ((m_1)_\alpha ^{(\widehat{{\bf k}})},(m_2)_{\alpha ^{\prime
}}^{(\widehat{{\bf k}})};(m_1)_u^{(\widehat{{\bf c}}_1)},(m_2)_v^{(\widehat{%
{\bf c}}_2)}),  \label{hu148}
\end{eqnarray}
where 
\begin{eqnarray}
&&\beta (1,M_i^{(\widehat{{\bf a}})};(m_1)_\alpha ^{(\widehat{{\bf k}}%
)},(m_2)_{\alpha ^{\prime }}^{(\widehat{{\bf k}})})=\sum_s\chi (1,M_i^{(%
\widehat{{\bf a}})};1,M_s^{(\widehat{{\bf k}})})  \nonumber \\
&&\times \eta (1,M_s^{(\widehat{{\bf k}})};(m_1)_\alpha ^{(\widehat{{\bf k}}%
)},(m_2)_{\alpha ^{\prime }}^{(\widehat{{\bf k}})}).  \label{hu149}
\end{eqnarray}

Alternatively the probability amplitudes can be written in the form 
\begin{eqnarray}
&&\Psi (1,M_i^{(\widehat{{\bf a}})};(m_1)_u^{(\widehat{{\bf c}}%
_1)},(m_2)_v^{(\widehat{{\bf c}}_2)})=\sum_r\chi (1,M_i^{(\widehat{{\bf a}}%
)};1,M_r^{(\widehat{{\bf k}})})  \nonumber \\
&&\times \xi (1,M_r^{(\widehat{{\bf k}})};(m_1)_u^{(\widehat{{\bf c}}%
_1)},(m_2)_v^{(\widehat{{\bf c}}_2)}),  \label{hu150}
\end{eqnarray}
where 
\begin{eqnarray}
&&\xi (1,M_r^{(\widehat{{\bf k}})};(m_1)_u^{(\widehat{{\bf c}}_1)},(m_2)_v^{(%
\widehat{{\bf c}}_2)})=\dsum\limits_{\alpha ,\alpha ^{\prime }}\eta (1,M_r^{(%
\widehat{{\bf k}})};(m_1)_\alpha ^{(\widehat{{\bf k}})},(m_2)_{\alpha
^{\prime }}^{(\widehat{{\bf k}})})  \nonumber \\
&&\times \psi ((m_1)_\alpha ^{(\widehat{{\bf k}})},(m_2)_{\alpha ^{\prime
}}^{(\widehat{{\bf k}})};(m_1)_u^{(\widehat{{\bf c}}_1)},(m_2)_v^{(\widehat{%
{\bf c}}_2)}).  \label{hu151}
\end{eqnarray}

If we start off with the choice Eq. (\ref{hu148}), then the operator is a $%
4\times 4$ matrix and the vector is a $4\times 1$ column. The expectation
value of $R((m_1)^{(\widehat{{\bf c}}_1)},(m_2)^{(\widehat{{\bf c}}_2)})$ is
given by 
\begin{eqnarray}
&&\left\langle R((m_1)^{(\widehat{{\bf c}}_1)},(m_2)^{(\widehat{{\bf c}}%
_2)})\right\rangle =[\Psi (1,M_i^{(\widehat{{\bf a}})};(m_1)^{(\widehat{{\bf %
c}}_1)}(m_2)^{(\widehat{{\bf c}}_2)})]^{\dagger }[R]  \nonumber \\
&&\times [\Psi (1,M_i^{(\widehat{{\bf a}})};(m_1)^{(\widehat{{\bf c}}%
_1)}(m_2)^{(\widehat{{\bf c}}_2)})]\text{,}  \label{hu152}
\end{eqnarray}
where

\begin{equation}
\lbrack \Psi (1,M_i^{(\widehat{{\bf a}})};(m_1)^{(\widehat{{\bf c}}%
_1)}(m_2)^{(\widehat{{\bf c}}_2)})]=\left( 
\begin{array}{c}
\beta (1,M_i^{(\widehat{{\bf a}})};(+\tfrac 12)^{(\widehat{{\bf k}}%
)},(+\tfrac 12)^{(\widehat{{\bf k}})}) \\ 
\beta (1,M_i^{(\widehat{{\bf a}})};(+\tfrac 12)^{(\widehat{{\bf k}}%
)},(-\tfrac 12)^{(\widehat{{\bf k}})}) \\ 
\beta (1,M_i^{(\widehat{{\bf a}})};(-\tfrac 12)^{(\widehat{{\bf k}}%
)},(+\tfrac 12)^{(\widehat{{\bf k}})}) \\ 
\beta (1,M_i^{(\widehat{{\bf a}})};(-\tfrac 12)^{(\widehat{{\bf k}}%
)},(-\tfrac 12)^{(\widehat{{\bf k}})})
\end{array}
\right) .  \label{hu153}
\end{equation}

The operator is the same as was found for the singlet state and so its
elements are given by Eqs. (\ref{hu132}) - (\ref{hu147}).

The elements of the vectors are known. We consider first the case $M^{(%
\widehat{{\bf a}})}=1.$ We have from Eqs. (\ref{si65}) - (\ref{si67})
combined with Eqs. (\ref{si69}) - (\ref{se72}), Eqs. (\ref{se74}) - (\ref
{se77}) and Eqs. (\ref{se79}) - (\ref{ei82}) 
\begin{eqnarray}
\lbrack \Psi (1,1^{(\widehat{{\bf a}})};(m_1)^{(\widehat{{\bf c}}_1)}(m_2)^{(%
\widehat{{\bf c}}_2)})] &=&\left( 
\begin{array}{c}
\beta (1,1^{(\widehat{{\bf a}})};(+\tfrac 12)^{(\widehat{{\bf k}})},(+\tfrac
12)^{(\widehat{{\bf k}})}) \\ 
\beta (1,1^{(\widehat{{\bf a}})};(+\tfrac 12)^{(\widehat{{\bf k}})},(-\tfrac
12)^{(\widehat{{\bf k}})}) \\ 
\beta (1,1^{(\widehat{{\bf a}})};(-\tfrac 12)^{(\widehat{{\bf k}})},(+\tfrac
12)^{(\widehat{{\bf k}})}) \\ 
\beta (1,1^{(\widehat{{\bf a}})};(-\tfrac 12)^{(\widehat{{\bf k}})},(-\tfrac
12)^{(\widehat{{\bf k}})})
\end{array}
\right)  \nonumber \\
&=&\left( 
\begin{array}{c}
\cos ^2\dfrac \theta 2e^{-i\varphi } \\ 
\dfrac 12\sin \theta \\ 
\dfrac 12\sin \theta \\ 
\sin ^2\dfrac \theta 2e^{i\varphi }
\end{array}
\right) .  \label{hu154}
\end{eqnarray}

For $M^{(\widehat{{\bf a}})}=0,$ we use Eqs. (\ref{ei88}) - (\ref{ni90})
together with Eqs. (\ref{si69}) - (\ref{se72}), Eqs. (\ref{se74}) - (\ref
{se77}) and Eqs. (\ref{se79}) - (\ref{ei82}) to obtain 
\begin{eqnarray}
\lbrack \Psi (1,0^{(\widehat{{\bf a}})};(m_1)^{(\widehat{{\bf c}}_1)}(m_2)^{(%
\widehat{{\bf c}}_2)})] &=&\left( 
\begin{array}{c}
\beta (1,0^{(\widehat{{\bf a}})};(+\tfrac 12)^{(\widehat{{\bf k}})},(+\tfrac
12)^{(\widehat{{\bf k}})}) \\ 
\beta (1,0^{(\widehat{{\bf a}})};(+\tfrac 12)^{(\widehat{{\bf k}})},(-\tfrac
12)^{(\widehat{{\bf k}})}) \\ 
\beta (1,0^{(\widehat{{\bf a}})};(-\tfrac 12)^{(\widehat{{\bf k}})},(+\tfrac
12)^{(\widehat{{\bf k}})}) \\ 
\beta (1,0^{(\widehat{{\bf a}})};(-\tfrac 12)^{(\widehat{{\bf k}})},(-\tfrac
12)^{(\widehat{{\bf k}})})
\end{array}
\right)  \nonumber \\
&=&\left( 
\begin{array}{c}
-\frac 1{\sqrt{2}}\sin \theta e^{-i\varphi } \\ 
\frac 1{\sqrt{2}}\cos \theta \\ 
\frac 1{\sqrt{2}}\cos \theta \\ 
\frac 1{\sqrt{2}}\sin \theta e^{-i\varphi }
\end{array}
\right) .  \label{hu155}
\end{eqnarray}

Finally for or $M^{(\widehat{{\bf a}})}=-1$, we have combined Eqs. (\ref
{ni95}) - (\ref{ni97}) with Eqs. (\ref{si69}) - (\ref{se72}), Eqs. (\ref
{se74}) - (\ref{se77}) and Eqs. (\ref{se79}) - (\ref{ei82}) to get 
\begin{eqnarray}
\lbrack \Psi (1,(-1)^{(\widehat{{\bf a}})};(m_1)^{(\widehat{{\bf c}}%
_1)}(m_2)^{(\widehat{{\bf c}}_2)})] &=&\left( 
\begin{array}{c}
\beta (1,(-1)^{(\widehat{{\bf a}})};(+\tfrac 12)^{(\widehat{{\bf k}}%
)},(+\tfrac 12)^{(\widehat{{\bf k}})}) \\ 
\beta (1,(-1)^{(\widehat{{\bf a}})};(+\tfrac 12)^{(\widehat{{\bf k}}%
)},(-\tfrac 12)^{(\widehat{{\bf k}})}) \\ 
\beta (1,(-1)^{(\widehat{{\bf a}})};(-\tfrac 12)^{(\widehat{{\bf k}}%
)},(+\tfrac 12)^{(\widehat{{\bf k}})}) \\ 
\beta (1,(-1)^{(\widehat{{\bf a}})};(-\tfrac 12)^{(\widehat{{\bf k}}%
)},(-\tfrac 12)^{(\widehat{{\bf k}})})
\end{array}
\right)  \nonumber \\
&=&\left( 
\begin{array}{c}
-\sin ^2\dfrac \theta 2e^{-i\varphi } \\ 
\dfrac 12\sin \theta \\ 
\dfrac 12\sin \theta \\ 
-\cos ^2\dfrac \theta 2e^{i\varphi }
\end{array}
\right) .  \label{hu156}
\end{eqnarray}

These then are the matrix forms of the various states. The three vectors are
mutually orthogonal. As expected, each one is normalized to unity. We
reiterate the basic result that the elements of the vectors are probability
amplitudes. They are probability amplitudes for getting spin projections
along the $z$ axis starting from states of specified projection of the total
spin along the arbitrary vector $\widehat{{\bf a}}$.

\subsubsection{Vectors and Operators in Three-Dimensional Treatment}

If we choose to write the probability amplitude as in Eq. (\ref{hu150}), the
expectation value of $R((m_1)^{(\widehat{{\bf c}}_1)},(m_2)^{(\widehat{{\bf c%
}}_2)})$ is 
\begin{eqnarray}
&&\left\langle R((m_1)^{(\widehat{{\bf c}}_1)},(m_2)^{(\widehat{{\bf c}}%
_2)})\right\rangle =[\Psi (1,M_i^{(\widehat{{\bf a}})};(m_1)^{(\widehat{{\bf %
c}}_1)}(m_2)^{(\widehat{{\bf c}}_2)})]^{\dagger }[R]  \nonumber \\
&&\times [\Psi (1,M_i^{(\widehat{{\bf a}})};(m_1)^{(\widehat{{\bf c}}%
_1)}(m_2)^{(\widehat{{\bf c}}_2)})],  \label{hu159}
\end{eqnarray}
with the matrix states being

\begin{equation}
\lbrack \Psi (1,M_i^{(\widehat{{\bf a}})};(m_1)^{(\widehat{{\bf c}}%
_1)}(m_2)^{(\widehat{{\bf c}}_2)})]=\left( 
\begin{array}{c}
\chi (1,M_i^{(\widehat{{\bf a}})};1,1^{(\widehat{{\bf k}})}) \\ 
\chi (1,M_i^{(\widehat{{\bf a}})};1,0^{(\widehat{{\bf k}})}) \\ 
\chi (1,M_i^{(\widehat{{\bf a}})};1,(-1)^{(\widehat{{\bf k}})})
\end{array}
\right) ,  \label{hu160}
\end{equation}
and the operator $[R]$ being a $3\times 3$ matrix whose elements are given by

\begin{eqnarray}
&&R_{pp^{\prime }}=\dsum_{u,v}\xi ^{*}(1,M_p^{(\widehat{{\bf k}})};(m_1)_u^{(%
\widehat{{\bf c}}_1)},(m_2)_v^{(\widehat{{\bf c}}_2)})R((m_1)_u^{(\widehat{%
{\bf c}}_1)},(m_2)_v^{(\widehat{{\bf c}}_2)})  \nonumber \\
&&\times \xi (1,M_{p^{\prime }}^{(\widehat{{\bf k}})};(m_1)_u^{(\widehat{%
{\bf c}}_1)},(m_2)_v^{(\widehat{{\bf c}}_2)}).  \label{hu161}
\end{eqnarray}

The states $[\Psi (1,M_i^{(\widehat{{\bf a}})};(m_1)^{(\widehat{{\bf c}}%
_1)}(m_2)^{(\widehat{{\bf c}}_2)})]$ are known because their elements are
given by Eqs. (\ref{si65}) - (\ref{si67}), Eqs. (\ref{ei88}) - (\ref{ni90})
and Eqs. (\ref{ni95}) - (\ref{ni97}). They are:

\begin{equation}
\lbrack \Psi (1,1^{(\widehat{{\bf a}})};(m_1)^{(\widehat{{\bf c}}_1)}(m_2)^{(%
\widehat{{\bf c}}_2)})]=\left( 
\begin{array}{c}
\cos ^2\dfrac \theta 2e^{-i\varphi } \\ 
\frac 1{\sqrt{2}}\sin \theta \\ 
\sin ^2\dfrac \theta 2e^{i\varphi }
\end{array}
\right) ,  \label{hu162}
\end{equation}

\begin{equation}
\lbrack \Psi (1,0^{(\widehat{{\bf a}})};(m_1)^{(\widehat{{\bf c}}_1)}(m_2)^{(%
\widehat{{\bf c}}_2)})]=\left( 
\begin{array}{c}
-\frac 1{\sqrt{2}}\sin \theta e^{-i\varphi } \\ 
\cos \theta \\ 
\frac 1{\sqrt{2}}\sin \theta e^{-i\varphi }
\end{array}
\right)  \label{hu163}
\end{equation}
and 
\begin{equation}
\lbrack \Psi (1,(-1)^{(\widehat{{\bf a}})};(m_1)^{(\widehat{{\bf c}}%
_1)}(m_2)^{(\widehat{{\bf c}}_2)})]=\left( 
\begin{array}{c}
-\sin ^2\dfrac \theta 2e^{-i\varphi } \\ 
\frac 1{\sqrt{2}}\sin \theta \\ 
-\cos ^2\dfrac \theta 2e^{i\varphi }
\end{array}
\right) .  \label{hu164}
\end{equation}

Using Eq. (\ref{hu151}) for the definition of $\xi (1,1,M_p^{(\widehat{{\bf k%
}})};(m_1)_u^{(\widehat{{\bf c}}_1)},(m_2)_v^{(\widehat{{\bf c}}_2)})$, we
find that

\begin{eqnarray}
&&\xi (1,1^{(\widehat{{\bf k}})};(+\tfrac 12)^{(\widehat{{\bf c}}%
_1)},(+\tfrac 12)^{(\widehat{{\bf c}}_2)})  \nonumber \\
&=&\eta (1,1^{(\widehat{{\bf k}})};(+\tfrac 12)^{(\widehat{{\bf k}}%
)},(+\tfrac 12)^{(\widehat{{\bf k}})})\psi ((+\tfrac 12)^{(\widehat{{\bf k}}%
)},(+\tfrac 12)^{(\widehat{{\bf k}})};(+\tfrac 12)^{(\widehat{{\bf c}}%
_1)},(+\tfrac 12)^{(\widehat{{\bf c}}_2)})  \nonumber \\
&&\ +\eta (1,1^{(\widehat{{\bf k}})};(+\tfrac 12)^{(\widehat{{\bf k}}%
)},(-\tfrac 12)^{(\widehat{{\bf k}})})\psi ((+\tfrac 12)^{(\widehat{{\bf k}}%
)},(-\tfrac 12)^{(\widehat{{\bf k}})};(+\tfrac 12)^{(\widehat{{\bf c}}%
_1)},(+\tfrac 12)^{(\widehat{{\bf c}}_2)})  \nonumber \\
&&\ \ +\eta (1,1^{(\widehat{{\bf k}})};(-\tfrac 12)^{(\widehat{{\bf k}}%
)},(-\tfrac 12)^{(\widehat{{\bf k}})})\psi ((-\tfrac 12)^{(\widehat{{\bf k}}%
)},(-\tfrac 12)^{(\widehat{{\bf k}})};(+\tfrac 12)^{(\widehat{{\bf c}}%
_1)},(+\tfrac 12)^{(\widehat{{\bf c}}_2)})  \nonumber \\
&&\ +\eta (1,1^{(\widehat{{\bf k}})};(-\tfrac 12)^{(\widehat{{\bf k}}%
)},(-\tfrac 12)^{(\widehat{{\bf k}})})\psi ((-\tfrac 12)^{(\widehat{{\bf k}}%
)},(-\tfrac 12)^{(\widehat{{\bf k}})};(+\tfrac 12)^{(\widehat{{\bf c}}%
_1)},(+\tfrac 12)^{(\widehat{{\bf c}}_2)})  \nonumber \\
\ &=&\psi ((+\tfrac 12)^{(\widehat{{\bf k}})},(+\tfrac 12)^{(\widehat{{\bf k}%
})};(+\tfrac 12)^{(\widehat{{\bf c}}_1)},(+\tfrac 12)^{(\widehat{{\bf c}}%
_2)})=\cos \frac{\theta _1}2\cos \frac{\theta _2}2.  \label{hu165}
\end{eqnarray}
by virtue of the values of the Clebsch-Gordan coefficients $\eta $ given in
Eqs. (\ref{si69}) - (\ref{se72}), and the expressions for the $\psi $'s
given in Eqs. (\ref{th37}) - (\ref{fi53}).

Similarly, 
\begin{eqnarray}
&&\xi (1,1^{(\widehat{{\bf k}})};(+\tfrac 12)^{(\widehat{{\bf c}}%
_1)},(-\tfrac 12)^{(\widehat{{\bf c}}_2)})=\psi ((+\tfrac 12)^{(\widehat{%
{\bf k}})},(+\tfrac 12)^{(\widehat{{\bf k}})};(+\tfrac 12)^{(\widehat{{\bf c}%
}_1)},(-\tfrac 12)^{(\widehat{{\bf c}}_2)})  \nonumber \\
&=&-\cos \frac{\theta _1}2\sin \frac{\theta _2}2,  \label{hu166}
\end{eqnarray}

\begin{eqnarray}
&&\xi (1,1^{(\widehat{{\bf k}})};(-\tfrac 12)^{(\widehat{{\bf c}}%
_1)},(+\tfrac 12)^{(\widehat{{\bf c}}_2)})=\psi ((+\tfrac 12)^{(\widehat{%
{\bf k}})},(+\tfrac 12)^{(\widehat{{\bf k}})};(-\tfrac 12)^{(\widehat{{\bf c}%
}_1)},(+\tfrac 12)^{(\widehat{{\bf c}}_2)})  \nonumber \\
&=&-\sin \frac{\theta _1}2\cos \frac{\theta _2}2,  \label{hu167}
\end{eqnarray}

\begin{eqnarray}
&&\xi (1,1^{(\widehat{{\bf k}})};(-\tfrac 12)^{(\widehat{{\bf c}}%
_1)},(-\tfrac 12)^{(\widehat{{\bf c}}_2)})=\psi ((+\tfrac 12)^{(\widehat{%
{\bf k}})},(+\tfrac 12)^{(\widehat{{\bf k}})};(-\tfrac 12)^{(\widehat{{\bf c}%
}_1)},(-\tfrac 12)^{(\widehat{{\bf c}}_2)})  \nonumber \\
&=&\sin \frac{\theta _1}2\sin \frac{\theta _2}2,  \label{hu168}
\end{eqnarray}

\begin{eqnarray}
&&\xi (1,0^{(\widehat{{\bf k}})};(+\tfrac 12)^{(\widehat{{\bf c}}%
_1)},(+\tfrac 12)^{(\widehat{{\bf c}}_2)})=\frac 1{\sqrt{2}}[\psi ((+\tfrac
12)^{(\widehat{{\bf k}})},(-\tfrac 12)^{(\widehat{{\bf k}})};(+\tfrac 12)^{(%
\widehat{{\bf c}}_1)},(+\tfrac 12)^{(\widehat{{\bf c}}_2)})  \nonumber \\
&&+\psi ((-\tfrac 12)^{(\widehat{{\bf k}})},(+\tfrac 12)^{(\widehat{{\bf k}}%
)};(+\tfrac 12)^{(\widehat{{\bf c}}_1)},(+\tfrac 12)^{(\widehat{{\bf c}}%
_2)})]  \nonumber \\
&=&\frac 1{\sqrt{2}}[\sin \frac{\theta _1}2\cos \frac{\theta _2}%
2e^{-i\varphi _1}+\cos \frac{\theta _1}2\sin \frac{\theta _2}2e^{-i\varphi
_2}],  \label{hu169}
\end{eqnarray}

\begin{eqnarray}
&&\xi (1,0^{(\widehat{{\bf k}})};(+\tfrac 12)^{(\widehat{{\bf c}}%
_1)},(-\tfrac 12)^{(\widehat{{\bf c}}_2)})=\frac 1{\sqrt{2}}[\psi ((+\tfrac
12)^{(\widehat{{\bf k}})},(-\tfrac 12)^{(\widehat{{\bf k}})};(+\tfrac 12)^{(%
\widehat{{\bf c}}_1)},(-\tfrac 12)^{(\widehat{{\bf c}}_2)})  \nonumber \\
&&+\psi ((-\tfrac 12)^{(\widehat{{\bf k}})},(+\tfrac 12)^{(\widehat{{\bf k}}%
)};(+\tfrac 12)^{(\widehat{{\bf c}}_1)},(-\tfrac 12)^{(\widehat{{\bf c}}%
_2)})]  \nonumber \\
\ &=&\frac 1{\sqrt{2}}[\cos \frac{\theta _1}2\cos \frac{\theta _2}%
2e^{-i\varphi _2}-\sin \frac{\theta _1}2\sin \frac{\theta _2}2e^{-i\varphi
_1}],  \label{hu170}
\end{eqnarray}
\begin{eqnarray}
&&\xi (1,0^{(\widehat{{\bf k}})};(-\tfrac 12)^{(\widehat{{\bf c}}%
_1)},(+\tfrac 12)^{(\widehat{{\bf c}}_2)})=\frac 1{\sqrt{2}}[\psi ((+\tfrac
12)^{(\widehat{{\bf k}})},(-\tfrac 12)^{(\widehat{{\bf k}})};(-\tfrac 12)^{(%
\widehat{{\bf c}}_1)},(+\tfrac 12))  \nonumber \\
&&+\psi ((-\tfrac 12)^{(\widehat{{\bf k}})},(+\tfrac 12)^{(\widehat{{\bf k}}%
)};(-\tfrac 12)^{(\widehat{{\bf c}}_1)},(+\tfrac 12))]  \nonumber \\
\ &=&\frac 1{\sqrt{2}}[\cos \frac{\theta _1}2\cos \frac{\theta _2}%
2e^{-i\varphi _1}-\sin \frac{\theta _1}2\sin \frac{\theta _2}2e^{-i\varphi
_2}],  \label{hu171}
\end{eqnarray}
\begin{eqnarray}
&&\xi (1,0^{(\widehat{{\bf k}})};(-\tfrac 12)^{(\widehat{{\bf c}}%
_1)},(-\tfrac 12))=\frac 1{\sqrt{2}}[\psi ((+\tfrac 12)^{(\widehat{{\bf k}}%
)},(-\tfrac 12)^{(\widehat{{\bf k}})};(-\tfrac 12)^{(\widehat{{\bf c}}%
_1)},(-\tfrac 12))  \nonumber \\
&&+\psi ((-\tfrac 12)^{(\widehat{{\bf k}})},(+\tfrac 12)^{(\widehat{{\bf k}}%
)};(-\tfrac 12)^{(\widehat{{\bf c}}_1)},(-\tfrac 12))]  \nonumber \\
\ &=&-\frac 1{\sqrt{2}}[\cos \frac{\theta _1}2\sin \frac{\theta _2}%
2e^{-i\varphi _1}+\sin \frac{\theta _1}2\cos \frac{\theta _2}2e^{-i\varphi
_2}],  \label{hu172}
\end{eqnarray}
\begin{eqnarray}
&&\xi (1,(-1)^{(\widehat{{\bf k}})};(+\tfrac 12)^{(\widehat{{\bf c}}%
_1)},(+\tfrac 12))=\psi (-k,-k;(+\tfrac 12)^{(\widehat{{\bf c}}_1)},(+\tfrac
12))  \nonumber \\
&=&\sin \frac{\theta _1}2\sin \frac{\theta _2}2e^{-i(\varphi _1+\varphi _2)},
\label{hu173}
\end{eqnarray}
\begin{eqnarray}
&&\xi (1,(-1)^{(\widehat{{\bf k}})};(+\tfrac 12)^{(\widehat{{\bf c}}%
_1)},(-\tfrac 12))=\psi ((-\tfrac 12)^{(\widehat{{\bf k}})},(-\tfrac 12)^{(%
\widehat{{\bf k}})};(+\tfrac 12)^{(\widehat{{\bf c}}_1)},(-\tfrac 12)) 
\nonumber \\
&=&\sin \frac{\theta _1}2\cos \frac{\theta _2}2e^{-i(\varphi _1+\varphi _2)},
\label{hu174}
\end{eqnarray}
\begin{eqnarray}
&&\xi (1,(-1)^{(\widehat{{\bf k}})};(-\tfrac 12)^{(\widehat{{\bf c}}%
_1)},(+\tfrac 12))=\psi ((-\tfrac 12)^{(\widehat{{\bf k}})},(-\tfrac 12)^{(%
\widehat{{\bf k}})};(-\tfrac 12)^{(\widehat{{\bf c}}_1)},(+\tfrac 12)) 
\nonumber \\
&=&\cos \frac{\theta _1}2\sin \frac{\theta _2}2e^{-i(\varphi _1+\varphi _2)}
\label{hu175}
\end{eqnarray}
and 
\begin{eqnarray}
&&\xi (1,(-1)^{(\widehat{{\bf k}})};(-\tfrac 12)^{(\widehat{{\bf c}}%
_1)},(-\tfrac 12))=\psi ((-\tfrac 12)^{(\widehat{{\bf k}})},(-\tfrac 12)^{(%
\widehat{{\bf k}})};(-\tfrac 12)^{(\widehat{{\bf c}}_1)},(-\tfrac 12)) 
\nonumber \\
&=&\cos \frac{\theta _1}2\cos \frac{\theta _2}2e^{-i(\varphi _1+\varphi _2)}.
\label{hu176}
\end{eqnarray}
The elements $R_{pp^{\prime }}$ of $[R]$ are

\begin{eqnarray}
R_{11} &=&\left| \xi (1,1^{(\widehat{{\bf k}})};(+\tfrac 12)^{(\widehat{{\bf %
c}}_1)},(+\tfrac 12)^{(\widehat{{\bf c}}_2)})\right| ^2R((+\tfrac 12)^{(%
\widehat{{\bf c}}_1)},(+\tfrac 12)^{(\widehat{{\bf c}}_2)})  \nonumber \\
&&+\left| \xi (1,1^{(\widehat{{\bf k}})};(+\tfrac 12)^{(\widehat{{\bf c}}%
_1)},(-\tfrac 12)^{(\widehat{{\bf c}}_2)})\right| ^2R((+\tfrac 12)^{(%
\widehat{{\bf c}}_1)},(-\tfrac 12)^{(\widehat{{\bf c}}_2)})  \nonumber \\
&&+\left| \xi (1,1^{(\widehat{{\bf k}})};(-\tfrac 12)^{(\widehat{{\bf c}}%
_1)},(+\tfrac 12)^{(\widehat{{\bf c}}_2)})\right| ^2R((-\tfrac 12)^{(%
\widehat{{\bf c}}_1)},(+\tfrac 12)^{(\widehat{{\bf c}}_2)})  \nonumber \\
&&+\left| \xi (1,1^{(\widehat{{\bf k}})};(-\tfrac 12)^{(\widehat{{\bf c}}%
_1)},(-\tfrac 12)^{(\widehat{{\bf c}}_2)})\right| ^2R((-\tfrac 12)^{(%
\widehat{{\bf c}}_1)},(-\tfrac 12)^{(\widehat{{\bf c}}_2)})  \nonumber \\
&=&\cos ^2\frac{\theta _1}2\cos ^2\frac{\theta _2}2R((+\tfrac 12)^{(\widehat{%
{\bf c}}_1)},(+\tfrac 12)^{(\widehat{{\bf c}}_2)})  \nonumber \\
&&+\cos ^2\frac{\theta _1}2\sin ^2\frac{\theta _2}2R((+\tfrac 12)^{(\widehat{%
{\bf c}}_1)},(-\tfrac 12)^{(\widehat{{\bf c}}_2)})  \nonumber \\
&&\sin ^2\frac{\theta _1}2\cos ^2\frac{\theta _2}2R((-\tfrac 12)^{(\widehat{%
{\bf c}}_1)},(+\tfrac 12)^{(\widehat{{\bf c}}_2)})  \nonumber \\
&&+\sin ^2\frac{\theta _1}2\sin ^2\frac{\theta _2}2R((-\tfrac 12)^{(\widehat{%
{\bf c}}_1)},(-\tfrac 12)^{(\widehat{{\bf c}}_2)}).  \label{hu177}
\end{eqnarray}
Similarly, 
\begin{eqnarray}
\ &&R_{12}=\xi ^{*}(1,1^{(\widehat{{\bf k}})};(+\tfrac 12)^{(\widehat{{\bf c}%
}_1)},(+\tfrac 12)^{(\widehat{{\bf c}}_2)})  \nonumber \\
&&\ \times \xi (1,0^{(\widehat{{\bf k}})};(+\tfrac 12)^{(\widehat{{\bf c}}%
_1)},(+\tfrac 12)^{(\widehat{{\bf c}}_2)})R((+\tfrac 12)^{(\widehat{{\bf c}}%
_1)},(+\tfrac 12)^{(\widehat{{\bf c}}_2)})  \nonumber \\
&&\ +\xi ^{*}(1,1^{(\widehat{{\bf k}})};(+\tfrac 12)^{(\widehat{{\bf c}}%
_1)},(-\tfrac 12)^{(\widehat{{\bf c}}_2)})\xi (1,0^{(\widehat{{\bf k}}%
)};(+\tfrac 12)^{(\widehat{{\bf c}}_1)},(-\tfrac 12)^{(\widehat{{\bf c}}_2)})
\nonumber \\
&&\ \times R((+\tfrac 12)^{(\widehat{{\bf c}}_1)},(-\tfrac 12)^{(\widehat{%
{\bf c}}_2)})  \nonumber \\
&&\ +\xi ^{*}(1,1^{(\widehat{{\bf k}})};(-\tfrac 12)^{(\widehat{{\bf c}}%
_1)},(+\tfrac 12)^{(\widehat{{\bf c}}_2)})\xi (1,0^{(\widehat{{\bf k}}%
)};(-\tfrac 12)^{(\widehat{{\bf c}}_1)},(+\tfrac 12)^{(\widehat{{\bf c}}_2)})
\nonumber \\
&&\ \times R((-\tfrac 12)^{(\widehat{{\bf c}}_1)},(+\tfrac 12)^{(\widehat{%
{\bf c}}_2)})  \nonumber \\
&&\ +\xi ^{*}(1,1^{(\widehat{{\bf k}})};(-\tfrac 12)^{(\widehat{{\bf c}}%
_1)},(-\tfrac 12)^{(\widehat{{\bf c}}_2)})\xi (1,0^{(\widehat{{\bf k}}%
)};(-\tfrac 12)^{(\widehat{{\bf c}}_1)},(-\tfrac 12)^{(\widehat{{\bf c}}_2)})
\nonumber \\
&&\ \times R((-\tfrac 12)^{(\widehat{{\bf c}}_1)},(-\tfrac 12)^{(\widehat{%
{\bf c}}_2)})  \label{hu177a}
\end{eqnarray}

Thus, 
\begin{eqnarray}
R_{12} &=&\frac 1{\sqrt{2}}\cos \frac{\theta _1}2\cos \frac{\theta _2}2[\sin 
\frac{\theta _1}2\cos \frac{\theta _2}2e^{-i\varphi _1}+\cos \frac{\theta _1}%
2\sin \frac{\theta _2}2e^{-i\varphi _2}]  \nonumber \\
&&\times R((+\tfrac 12)^{(\widehat{{\bf c}}_1)},(+\tfrac 12)^{(\widehat{{\bf %
c}}_2)})  \nonumber \\
&&+\frac 1{\sqrt{2}}\cos \frac{\theta _1}2\sin \frac{\theta _2}2[\sin \frac{%
\theta _1}2\sin \frac{\theta _2}2e^{-i\varphi _1}-\cos \frac{\theta _1}2\cos 
\frac{\theta _2}2e^{-i\varphi _2}]  \nonumber \\
&&\times R((+\tfrac 12)^{(\widehat{{\bf c}}_1)},(-\tfrac 12)^{(\widehat{{\bf %
c}}_2)})  \nonumber \\
&&+\frac 1{\sqrt{2}}\sin \frac{\theta _1}2\cos \frac{\theta _2}2[\sin \frac{%
\theta _1}2\sin \frac{\theta _2}2e^{-i\varphi _2}-\cos \frac{\theta _1}2\cos 
\frac{\theta _2}2e^{-i\varphi _1}]  \nonumber \\
&&\times R((-\tfrac 12)^{(\widehat{{\bf c}}_1)},(+\tfrac 12)^{(\widehat{{\bf %
c}}_2)})  \nonumber \\
&&-\frac 1{\sqrt{2}}\sin \frac{\theta _1}2\sin \frac{\theta _2}2[\cos \frac{%
\theta _1}2\sin \frac{\theta _2}2e^{-i\varphi _1}+\sin \frac{\theta _1}2\cos 
\frac{\theta _2}2e^{-i\varphi _2}]  \nonumber \\
&&\times R((-\tfrac 12)^{(\widehat{{\bf c}}_1)},(-\tfrac 12)^{(\widehat{{\bf %
c}}_2)}),  \nonumber \\
&&  \label{hu178}
\end{eqnarray}

In the same way, 
\begin{eqnarray}
R_{13} &=&\frac 14\sin \theta _1\sin \theta _2e^{-i(\varphi _1+\varphi
_2)}[R((+\tfrac 12)^{(\widehat{{\bf c}}_1)},(+\tfrac 12)^{(\widehat{{\bf c}}%
_2)})-R((+\tfrac 12)^{(\widehat{{\bf c}}_1)},(-\tfrac 12)^{(\widehat{{\bf c}}%
_2)})  \nonumber \\
&&-R((-\tfrac 12)^{(\widehat{{\bf c}}_1)},(+\tfrac 12)^{(\widehat{{\bf c}}%
_2)})+R((-\tfrac 12)^{(\widehat{{\bf c}}_1)},(-\tfrac 12)^{(\widehat{{\bf c}}%
_2)})],  \label{hu179}
\end{eqnarray}

\begin{equation}
R_{21}=R_{12}^{*},  \label{hu180}
\end{equation}

\begin{eqnarray}
R_{22} &=&\frac 12[\cos ^2\frac{\theta _1}2\sin ^2\frac{\theta _2}2+\sin ^2%
\frac{\theta _1}2\cos ^2\frac{\theta _2}2  \nonumber \\
&&+\frac 12\sin \theta _1\sin \theta _2\cos (\varphi _1-\varphi
_2)]R((+\tfrac 12)^{(\widehat{{\bf c}}_1)},(+\tfrac 12)^{(\widehat{{\bf c}}%
_2)})  \nonumber \\
&&+\frac 12[\sin ^2\frac{\theta _1}2\sin ^2\frac{\theta _2}2+\cos ^2\frac{%
\theta _1}2\cos ^2\frac{\theta _2}2  \nonumber \\
&&-\frac 12\sin \theta _1\sin \theta _2\cos (\varphi _1-\varphi
_2)]R((+\tfrac 12)^{(\widehat{{\bf c}}_1)},(-\tfrac 12)^{(\widehat{{\bf c}}%
_2)})  \nonumber \\
&&+\frac 12[\sin ^2\frac{\theta _1}2\sin ^2\frac{\theta _2}2+\cos ^2\frac{%
\theta _1}2\cos ^2\frac{\theta _2}2  \nonumber \\
&&-\frac 12\sin \theta _1\sin \theta _2\cos (\varphi _1-\varphi
_2)]R((-\tfrac 12)^{(\widehat{{\bf c}}_1)},(+\tfrac 12)^{(\widehat{{\bf c}}%
_2)})  \nonumber \\
&&+\frac 12[\sin ^2\frac{\theta _1}2\sin ^2\frac{\theta _2}2+\cos ^2\frac{%
\theta _1}2\cos ^2\frac{\theta _2}2  \nonumber \\
&&+\frac 12\sin \theta _1\sin \theta _2\cos (\varphi _1-\varphi
_2)]R((-\tfrac 12)^{(\widehat{{\bf c}}_1)},(-\tfrac 12)^{(\widehat{{\bf c}}%
_2)}),  \label{hu181}
\end{eqnarray}

\begin{eqnarray}
R_{23} &=&\frac 1{\sqrt{2}}\sin \frac{\theta _1}2\sin \frac{\theta _2}2[\cos 
\frac{\theta _1}2\sin \frac{\theta _2}2e^{-i\varphi _1}  \nonumber \\
&&+\sin \frac{\theta _1}2\cos \frac{\theta _2}2e^{-i\varphi _2}]R((+\tfrac
12)^{(\widehat{{\bf c}}_1)},(+\tfrac 12)^{(\widehat{{\bf c}}_2)})  \nonumber
\\
&&+\frac 1{\sqrt{2}}\sin \frac{\theta _1}2\cos \frac{\theta _2}2[\cos \frac{%
\theta _1}2\cos \frac{\theta _2}2e^{-i\varphi _1}  \nonumber \\
&&-\sin \frac{\theta _1}2\sin \frac{\theta _2}2e^{-i\varphi _2}]R((+\tfrac
12)^{(\widehat{{\bf c}}_1)},(-\tfrac 12)^{(\widehat{{\bf c}}_2)})  \nonumber
\\
&&+\frac 1{\sqrt{2}}\cos \frac{\theta _1}2\sin \frac{\theta _2}2[\cos \frac{%
\theta _1}2\cos \frac{\theta _2}2e^{-i\varphi _2}  \nonumber \\
&&-\sin \frac{\theta _1}2\sin \frac{\theta _2}2e^{-i\varphi _1}]R((-\tfrac
12)^{(\widehat{{\bf c}}_1)},(+\tfrac 12)^{(\widehat{{\bf c}}_2)})  \nonumber
\\
&&-\frac 1{\sqrt{2}}\cos \frac{\theta _1}2\cos \frac{\theta _2}2[\sin \frac{%
\theta _1}2\cos \frac{\theta _2}2e^{-i\varphi _1}  \nonumber \\
&&+\cos \frac{\theta _1}2\sin \frac{\theta _2}2e^{-i\varphi _2}]R((-\tfrac
12)^{(\widehat{{\bf c}}_1)},(-\tfrac 12)^{(\widehat{{\bf c}}_2)}),  \nonumber
\\
&&  \label{hu182}
\end{eqnarray}

\begin{equation}
R_{31}=R_{13}^{*},  \label{hu183}
\end{equation}
\begin{equation}
R_{32}=R_{23}^{*}  \label{hu184}
\end{equation}
and

\begin{eqnarray}
R_{33} &=&\sin ^2\frac{\theta _1}2\sin ^2\frac{\theta _2}2R((+\tfrac 12)^{(%
\widehat{{\bf c}}_1)},(+\tfrac 12)^{(\widehat{{\bf c}}_2)})  \nonumber \\
&&+\sin ^2\frac{\theta _1}2\cos ^2\frac{\theta _2}2R((+\tfrac 12)^{(\widehat{%
{\bf c}}_1)},(-\tfrac 12)^{(\widehat{{\bf c}}_2)}) \\
&&+\cos ^2\frac{\theta _1}2\sin ^2\frac{\theta _2}2R((-\tfrac 12)^{(\widehat{%
{\bf c}}_1)},(+\tfrac 12)^{(\widehat{{\bf c}}_2)})  \nonumber \\
&&+\cos ^2\frac{\theta _1}2\cos ^2\frac{\theta _2}2R((-\tfrac 12)^{(\widehat{%
{\bf c}}_1)},(-\tfrac 12)^{(\widehat{{\bf c}}_2)}).  \label{hu185}
\end{eqnarray}

Thus, this form of the matrix representation of the triplet state consists
of $3\times 1$ columns for the state and $3\times 3$ matrices for the
operators.

\section{Applications to Measurements on Entangled Systems}

\subsection{Singlet State}

We have given the general form for the operator of any quantity which is a
function of the final spin projections of subsystems $1$ and $2$. In this
section, we shall see how this general operator can be used to calculate
expectation values for joint measurements on a system such as singlet or a
triplet state. These measurements reveal correlations which play a very
important role in the debate over the interpretation of quantum theory [12].
The most interesting such system is the singlet state.

To use our formulas to study the correlations in the results of measurements
on the singlet state, we start by investigating the behaviour of the
probability amplitudes. For one thing, we ought to find that for $\widehat{%
{\bf c}}_1=\widehat{{\bf c}}_2$ the probability amplitudes for finding the
two spin projections parallel should vanish. Indeed, when we set $\theta
_1=\theta _2$ and $\varphi _1=\varphi _2$ so that $\widehat{{\bf c}}_1=%
\widehat{{\bf c}}_2=\widehat{{\bf c}}$ in Eqs. (\ref{fi54}) - (\ref{fi57}),
we find that 
\begin{equation}
P(0,0^{(\widehat{{\bf a}})};(+\tfrac 12)^{(\widehat{{\bf c}}_1)},(+\tfrac
12)^{(\widehat{{\bf c}}_2)})=P(0,0^{(\widehat{{\bf a}})};(-\tfrac 12)^{(%
\widehat{{\bf c}}_1)},(-\tfrac 12)^{(\widehat{{\bf c}}_2)})=0  \label{hu185a}
\end{equation}
but 
\begin{equation}
P(0,0^{(\widehat{{\bf a}})};(+\tfrac 12)^{(\widehat{{\bf c}}_1)},(-\tfrac
12)^{(\widehat{{\bf c}}_2)})=P(0,0^{(\widehat{{\bf a}})};(-\tfrac 12)^{(%
\widehat{{\bf c}}_1)},(+\tfrac 12)^{(\widehat{{\bf c}}_2)})=\frac 12.
\label{hu185b}
\end{equation}

Thus, whenever one subspin is found up, the other will be found down with
certainty. The present approach gives a reason for this correlation. Since
the angles defining the initial direction $\widehat{{\bf a}}$ are absent
from the expressions for the probability amplitudes, the singlet state is
symmetric with respect to the coordinate system. Thus, whatever vector $%
\widehat{{\bf c}}$ we choose in order to make our spin projection
measurements takes on the role of the vector $\widehat{{\bf a}}$ along which
the two spin-$1/2$ projections are assumed to lie. To put it another way,
the absence of the angles for $\widehat{{\bf a}}$ from the probability
amplitudes means that the state does not recognise the original direction
along which lie the two spins added to give zero. When one measurement is
made, the direction with respect to which it is done assumes the role of the
vector along which the two spins lie anti-parallel. Hence a measurement of
the other spin projection along the same vector finds it anti-parallel to
the other. Thus, the correlation is a symmetry effect.

We expect similar results to occur whenever the spin of the composite system
is zero. In such cases, the expansion of the probability amplitudes will
involve only one summation, as in Eq. (\ref{fi53b}) owing to the fact that
the summation over the spin projections of the compound spin vanishes. There
will then be no correlation between the original orientations of the
constituent subsystems and the directions of the vectors $\widehat{{\bf c}}=%
\widehat{{\bf c}}_1=\widehat{{\bf c}}_2.$

In order to compute the expectation values for the entangled system, we
consider the observable $R$ defined by assigning the value $+1$ ($-1$) when
the spin projection of subsystem $1$ or subsystem $2$ is up (down) with
respect to $\widehat{{\bf c}}_1$ or $\widehat{{\bf c}}_2$ respectively. If
these values are then multiplied to give an observable describing joint
measurements on the subsystems, the four possible outcomes are

\begin{equation}
R((+\tfrac 12)^{(\widehat{{\bf c}}_1)},(+\tfrac 12)^{(\widehat{{\bf c}}%
_2)})=R((-\tfrac 12)^{(\widehat{{\bf c}}_1)},(-\tfrac 12)^{(\widehat{{\bf c}}%
_2)})=1  \label{hu186a}
\end{equation}
and 
\begin{equation}
R((+\tfrac 12)^{(\widehat{{\bf c}}_1)},(-\tfrac 12)^{(\widehat{{\bf c}}%
_2)})=R((-\tfrac 12)^{(\widehat{{\bf c}}_1)},(+\tfrac 12)^{(\widehat{{\bf c}}%
_2)})=-1.  \label{hu186b}
\end{equation}

Hence the operator $[R]$ has the elements

\begin{equation}
R_{11}=\cos \theta _1\cos \theta _2,  \label{hu186}
\end{equation}

\begin{equation}
R_{12}=\cos \theta _1\sin \theta _2e^{-i\varphi _2},  \label{hu187}
\end{equation}
\begin{equation}
R_{13}=\sin \theta _1\cos \theta _2e^{-i\varphi _1},  \label{hu188}
\end{equation}

\begin{equation}
R_{14}=\sin \theta _1\sin \theta _2e^{-i(\varphi _1+\varphi _2)},
\label{hu189}
\end{equation}
\begin{equation}
R_{21}=R_{12}^{*},  \label{hu189a}
\end{equation}

\begin{equation}
R_{22}=-\cos \theta _1\cos \theta _2,  \label{hu190}
\end{equation}

\begin{equation}
R_{23}=\sin \theta _1\sin \theta _2e^{i(\varphi _2-\varphi _1)},
\label{hu191}
\end{equation}
\begin{equation}
R_{24}=-\sin \theta _1\cos \theta _2e^{-i\varphi _1},  \label{hu192}
\end{equation}
\begin{equation}
R_{31}=R_{13}^{*},  \label{hu192a}
\end{equation}
\begin{equation}
R_{32}=R_{23}^{*},  \label{hu192b}
\end{equation}
\begin{equation}
R_{33}=-\cos \theta _1\cos \theta _2,  \label{hu193}
\end{equation}

\begin{equation}
R_{34}=-\sin \theta _2\cos \theta _1e^{-i\varphi _2},  \label{hu194}
\end{equation}
\begin{equation}
R_{41}=R_{14}^{*},  \label{hu194a}
\end{equation}
\begin{equation}
R_{42}=R_{24}^{*},  \label{hu194b}
\end{equation}
\begin{equation}
R_{43}=R_{34}^{*}  \label{hu194c}
\end{equation}
and 
\begin{equation}
R_{44}=\cos \theta _1\cos \theta _2.  \label{hu195}
\end{equation}

Using this operator and the state, Eq. (\ref{hu131}), the expectation value
of $R$ is found to be

\begin{eqnarray}
\left\langle R\right\rangle &=&-[\cos (\theta _2-\theta _1)-2\sin \theta
_1\sin \theta _2\sin ^2\dfrac{(\varphi _2-\varphi _1)}2]  \nonumber \\
\ &=&-\widehat{{\bf c}}_1\cdot \widehat{{\bf c}}_2.  \label{hu195b}
\end{eqnarray}

\subsection{Triplet State}

As we have seen, there are two matrix representations for the triplet state.
The four-dimensional representation has the same operator as the
singlet-state case, but with the states being given by Eqs. (\ref{hu154}), (%
\ref{hu155}) and (\ref{hu156}) for the cases $M=1,0$ and $-1$ respectively.

For the three-dimensional treatment, the elements of the operator $[R]$ are
obtained by putting 
\begin{equation}
R((+\tfrac 12)^{(\widehat{{\bf c}}_1)},(+\tfrac 12)^{(\widehat{{\bf c}}%
_2)})=R((-\tfrac 12)^{(\widehat{{\bf c}}_1)},(-\tfrac 12)^{(\widehat{{\bf c}}%
_2)})=1  \label{hu195c}
\end{equation}
and 
\begin{equation}
R((+\tfrac 12)^{(\widehat{{\bf c}}_1)},(-\tfrac 12)^{(\widehat{{\bf c}}%
_2)})=R((-\tfrac 12)^{(\widehat{{\bf c}}_1)},(+\tfrac 12)^{(\widehat{{\bf c}}%
_2)})=-1  \label{hu195d}
\end{equation}
in Eqs. (\ref{hu177}) - (\ref{hu185}), and are

\begin{equation}
R_{11}=-\cos \theta _1\cos \theta _2,  \label{hu196}
\end{equation}

\begin{equation}
R_{12}=\frac 1{\sqrt{2}}[\sin \theta _1\cos \theta _2e^{-i\varphi _1}+\cos
\theta _1\sin \theta _2e^{-i\varphi _2}],  \label{hu197}
\end{equation}
\begin{equation}
R_{13}=\sin \theta _1\sin \theta _2e^{-i(\varphi _1+\varphi _2)},
\label{hu198}
\end{equation}
\begin{equation}
R_{21}=R_{12}^{*},  \label{hu198a}
\end{equation}

\begin{equation}
R_{22}=-\cos \theta _1\cos \theta _2+\sin \theta _1\sin \theta _2\cos
(\varphi _1-\varphi _2),  \label{hu199}
\end{equation}
\begin{equation}
R_{23}=-\frac 1{\sqrt{2}}[\sin \theta _1\cos \theta _2e^{-i\varphi _1}+\cos
\theta _1\sin \theta _2e^{-i\varphi _2}]  \label{hu200}
\end{equation}
\begin{equation}
R_{31}=R_{13}^{*},  \label{hu200a}
\end{equation}
\begin{equation}
R_{32}=R_{23}^{*}  \label{hu200b}
\end{equation}
and

\begin{equation}
R_{33}=\cos \theta _1\cos \theta _2.  \label{hu201}
\end{equation}

We can calculate the expectation value by using the probability amplitudes
directly, or by the matrix-mechanics approach, in which case we have a
choice between the three- and the four- dimensional representations.
Whatever method we use, the expectation values are 
\begin{eqnarray}
\left\langle R\right\rangle _{M^{(\widehat{{\bf a}})}=1} &=&\left\langle
R\right\rangle _{M^{(\widehat{{\bf a}})}=-1}=\cos ^2\theta \cos \theta
_1\cos \theta _2  \nonumber \\
&&+\sin ^2\theta \sin \theta _1\sin \theta _2\cos (\varphi -\varphi _2)\cos
(\varphi -\varphi _1)  \nonumber \\
&&+\sin \theta \sin \theta _1\cos \theta \cos \theta _2\cos (\varphi
-\varphi _1)  \nonumber \\
&&+\sin \theta \sin \theta _2\cos \theta \cos \theta _1\cos (\varphi
-\varphi _2)  \label{hu201a}
\end{eqnarray}
and 
\begin{eqnarray}
\left\langle R\right\rangle _{M^{(\widehat{{\bf a}})}=0} &=&-\cos 2\theta
\cos \theta _1\cos \theta _2+\sin \theta _1\sin \theta _2\cos (\varphi
_2-\varphi _1)  \nonumber \\
&&\ -2\sin ^2\theta \sin \theta _1\sin \theta _2\cos (\varphi -\varphi
_2)\cos (\varphi -\varphi _1)  \nonumber \\
&&\ -\sin 2\theta \sin \theta _1\cos \theta _2\cos (\varphi -\varphi _1) 
\nonumber \\
&&\ -\sin 2\theta \sin \theta _2\cos \theta _1\cos (\varphi -\varphi _2).
\label{hu201b}
\end{eqnarray}

The standard expectation values correspond to setting $\theta =\varphi =0,$
so that the compound spin is initially along the $z$ axis. In this limit, we
find

\begin{equation}
\left\langle R\right\rangle _{M^{(\widehat{{\bf a}})}=1}=\left\langle
R\right\rangle _{M^{(\widehat{{\bf a}})}=-1}=\cos \theta _1\cos \theta _2
\label{hu202}
\end{equation}
and

\begin{equation}
\left\langle R\right\rangle _{M^{(\widehat{{\bf a}})}=0}=-\cos \theta _1\cos
\theta _2+\sin \theta _1\sin \theta _2\cos (\varphi _2-\varphi _1).
\label{hu203}
\end{equation}

If in addition $\widehat{{\bf c}}_2=\widehat{{\bf c}}_1$, then 
\begin{equation}
\left\langle R\right\rangle _{M^{(\widehat{{\bf a}})}=1}=\left\langle
R\right\rangle _{M^{(\widehat{{\bf a}})}=-1}=\cos ^2\theta _1  \label{hu204}
\end{equation}
and 
\begin{equation}
\left\langle R\right\rangle _{M^{(\widehat{{\bf a}})}=0}=-\cos 2\theta _1.
\label{hu205}
\end{equation}

Further applications are easily realized by plugging the appropriate values $%
R((m_1)^{(\widehat{{\bf c}}_1)},(m_2)^{(\widehat{{\bf c}}_2)})$ into the
formulas Eqs. (\ref{hu177}) - (\ref{hu185}) if we select the
three-dimensional treatment or into Eqs. (\ref{hu132}) - (\ref{hu147}) if we
prefer the four-dimensional treatment.

\section{Discussion and Conclusion}

In this paper, we have approached the treatment of systems of compounded
angular momentum from a new direction. We have given a general theory for
their treatment which can be used on all such systems. This theory has been
specialized to the case of the addition of two spins of value $1/2$ each to
obtain total spin $0$ and $1$. Thus, for the singlet and triplet states thus
resulting, we have derived the probability amplitudes for obtaining by
measurement all possible combinations of the spin projections of the
subsystems along arbitrary directions.

We have obtained new matrix treatments of these systems from first
principles. Thus, we have given the matrix representations of the ${\bf S}=0$
and ${\bf S}=1$ states and given the general form of the matrix operator
corresponding to measurements on these systems. For the ${\bf S}=1$ system,
we have found two matrix representations. It is evident that in a general
case where a probability amplitude is obtained indirectly through several
expansions, it is possible to use any of those expansions to define a matrix
representation. Thus, a general system has as many matrix representations as
there are possible intermediate expansions of its probability amplitudes. In
fact from this point of view, wave mechanics is merely matrix mechanics of $%
N=1$ dimensions. We can see this from the wave-mechanics expectation value
of the quantity $R(C)$:

\begin{equation}
\left\langle R\right\rangle =\sum_j\left| \Psi (A_i;C_j)\right| ^2R(C_j).
\label{hu210}
\end{equation}
This can be expressed as

\begin{eqnarray}
\left\langle R\right\rangle &=&(1)\left( \sum_j\left| \Psi (A_i;C_j)\right|
^2R(C_j)\right) (1)  \nonumber \\
&=&[\Psi (A_i)]^{\dagger }[R][\Psi (A_i)],  \label{hu211}
\end{eqnarray}
where 
\begin{equation}
\lbrack \Psi (A_i)]=(1)  \label{hu212}
\end{equation}
and

\begin{equation}
\lbrack R]=\left( \sum_j\left| \Psi (A_i,C_j)\right| ^2R(C_j)\right) .
\label{hu213}
\end{equation}

In our work on isolated spin-$1/2$ systems[1,2,4], we saw that changing the
phase of the probability amplitudes alters the forms of the spin operators
and vectors. In the present work, we have seen that not all choices of phase
for the spin-$1/2$ probability amplitudes are equivalent. It must be
realized that there are several choices of phase possible for the
probability amplitudes for the isolated spin-1 system as well; the choice
that appears in Ref. [3], and which has been adopted for use here, was
arbitrarily picked. In fact, the total number of phase choices for the spin-$%
1$ case is greater than for the spin-$1/2$ case. It is certain that the
choice of spin-1 probability amplitude phase has a bearing on the choice of
spin-$1/2$ phase to be combined with it in order to obtain the treatment of
the compound system. If in our derivation of the spin quantities for spin 1
we had used a different choice of phase for the probability amplitudes, we
would certainly have been obliged to use a different phase for the spin-$1/2$
probability amplitudes in order to obtain the results for the compound
system. We then would have found different forms for the compound-system
probability amplitudes, the matrix states and the matrix operators. Thus, we
see that the results we have presented here are not unique, but are only one
set in a family of such results. But as we have seen here, it is not true
that we can combine any spin-$1/2$ probability amplitudes with just any spin-%
$1$ probability amplitudes. Some combinations will not lead to acceptable
results. The rule for determining the correct combination of spin-$1$ and
spin-$1/2$ phase choices is that in the appropriate limit, the
compound-state probability amplitudes should reduce to the Clebsch-Gordan
coefficients. The matter of just how to match phase choices for spin-$1/2$
and spin-$1$ is of considerable interest in its own right and deserves
deeper investigation. It is possible that the phase choice combinations
which are not acceptable in the present work will be found appropriate in
other investigations where both spin-$1/2$ and spin-$1$ probability
amplitudes both enter.

According to the present approach, the Clebsch-Gordan coefficients can be
generalized and we have given the generalized forms for the singlet and
triplet states. For these cases, the forms are not very different from the
standard forms. However, the generalized Clebsch-Gordan coefficients are
themselves functions of the phase combinations deriving from the spin-$1/2$
and spin-$1$ phase choices. It is possible that even for the singlet and
triplet cases, a more involved dependence of the generalized Clebsch-Gordan
coefficients on the angles results from a different combination of phases
from that used here.

We have applied the results obtained to one practical problem - the issue of
correlations in measurements on entangled systems. We have gained a useful
insight into the systems with the observation that the correlations in the
results for the singlet state are due to the probability amplitudes not
being functions of the angles that define the initial direction of the
compounded spin. Here this direction is the direction along which the
subspins which add up to zero lie. We deduce that whenever we compound any
number of spins in order to get a total spin of zero, the probability
amplitudes we obtain will be independent of the directions of the spins
which are added to yield zero. Therefore in all such cases, we should
observe correlations in the results of spin-projection measurements on the
systems.

The fact that we have been able to do all this proves the soundness of
Land\'e's ideas, which underlie our method. The method has the advantage of
great generality and clarity and should prove useful in other departments of
physics. It is probably possible to use this approach to obtain the matrix
treatment from a probability amplitude basis for other systems which are
currently described only by matrices.

The extension of the present ideas to more complex systems is
straightforward, if more tedious. A system that suggests itself as being but
one step removed in complexity from the one treated here involves adding the
spins of spin-$1$ and spin-$1/2$ subsystems to obtain or a spin half or a
spin $3/2$ system. For the former case, we have all the tools we need
because we already have the individual treatments for the spin-$1$ and spin-$%
1/2$ subsystems and thereby also for the spin-$1/2$ compound system. But if
we wish to compound to get spin-$3/2$, then we need to apply the general
method we have devised in previous papers in order to obtain the probability
amplitudes for this case. Strictly speaking, we can get away with using the
standard generalized probability amplitudes. Whatever, both these cases
promise to be interesting because they ought to yield generalized
Clebsch-Gordan coefficients with a less simple dependence on the angles than
the ones for the cases in the present paper.

It is very striking that the matrix treatment we have derived here is
completely different from the standard treatment which involves $2\times 2$
operators for each subsystem directly multiplied to give the operator for
the compound system. The transition from one treatment to the other deserves
investigation.

\section{References}

1. Mweene H. V., ''Derivation of Spin Vectors and Operators From First
Principles'', quant-ph/9905012

2. Mweene H. V., ''Generalized Spin-1/2 Operators and Their Eigenvectors'',
quant-ph/9906002

3. Mweene H. V., ''Vectors and Operators for Spin 1 Derived From First
Principles'', quant-ph/9906043

4. Mweene H. V., ''Alternative Forms of Generalized Vectors and Operators
for Spin 1/2'', quant-ph/9907031

5. Mweene H. V., ''Spin Description and Calculations in the Land\'e
Interpretation of Quantum Mechanics'', quant-ph/9907033

6. Land\'e A., ''From Dualism To Unity in Quantum Physics'', Cambridge
University Press, 1960.

7. Land\'e A., ''New Foundations of Quantum Mechanics'', Cambridge
University Press, 1965.

8. Land\'e A., ''Foundations of Quantum Theory,'' Yale University Press,
1955.

9. Land\'e A., ''Quantum Mechanics in a New Key,'' Exposition Press, 1973.

10. Rose M. E., ''Elementary Theory of Angular Momentum'', John Wiley and
Sons, Inc. (New York), 1957

11.Bransden and Joachain, ''Introduction to Quantum Mechanics'', Longman
Scientific \& Technical, 1989.

12. See for example, Bell J. S., Physics {\bf 1} (1964), 195

\end{document}